\documentclass[11pt,a4paper]{article}
\usepackage{jcappub, slashed} 
\usepackage{bm,graphicx,amssymb}
\usepackage{enumerate}
\usepackage[italicdiff]{physics}

\begin{document}
\title{Perturbative limits on axion-SU(2) gauge dynamics during inflation from the energy density of spin-2 particles}
\author[a]{Koji Ishiwata,}
\author[b,c,d]{Eiichiro Komatsu}
\affiliation[a]{Institute for Theoretical Physics, Kanazawa University, Kanazawa 920-1192, Japan}
\affiliation[b]{Max-Planck-Institute for Astrophysics, Karl-Schwarzschild-Str. 1, 85741 Garching, Germany}
\affiliation[c]{Ludwig-Maximilians-Universität München, Schellingstr. 4, 80799 München, Germany}
\affiliation[d]{Kavli Institute for the Physics and Mathematics of the Universe (Kavli IPMU, WPI), UTIAS, The University of Tokyo, Chiba, 277-8583, Japan} 
\emailAdd{ishiwata@hep.s.kanazawa-u.ac.jp}

\abstract{
We investigate the conditions under which the perturbative treatment of the backreaction of spin-2 particles on the dynamics of an axion-SU(2) gauge field system breaks down during cosmic inflation. This condition is based on the ratio of the energy density of spin-2 particles from the SU(2) gauge field to that of the background field. The perturbative treatment breaks down when this ratio exceeds unity. We show that this occurs within a parameter space nearly identical to the strong backreaction regime identified in previous studies. However, in some cases, the ratio exceeds unity even before the system enters the strong backreaction regime. Our results suggest that attempts to study the strong backreaction regime using perturbation theory are necessarily limited. Reliable calculations require non-perturbative treatments, such as three-dimensional lattice simulations.
}

\maketitle
 
\section{Introduction}
\label{sec:intro}    
A pseudo Nambu-Goldstone boson, such as an `axionlike' field $\chi$, plays a profound role in building models of cosmic inflation~\cite{Freese:1990rb}. Rich new phenomenology arises when $\chi$ is coupled to Abelian~\cite{Anber:2009ua} and non-Abelian~\cite{Maleknejad:2011jw,Adshead:2012kp} gauge fields via the Chern-Simons term, $\chi F\tilde F$. In particular, the motion of $\chi$ leads to the copious production of gauge-field particles. These particles can source observable scalar- and tensor-mode fluctuations that are parity-violating and non-Gaussian (see Refs.~\cite{Maleknejad:2012fw,Komatsu:2022nvu} for reviews and references therein).

However, the produced particles backreact on the background axion and gauge-field dynamics, which could spoil the successful phenomenology of these models~\cite{Anber:2009ua,Barnaby:2011qe,Dimastrogiovanni:2016fuu,Fujita:2017jwq}. For this reason, the strong backreaction regime of the axion-U(1) \cite{Cheng:2015oqa,Notari:2016npn,DallAgata:2019yrr,Domcke:2020zez,Peloso:2022ovc,Gorbar:2021rlt,Durrer:2023rhc,vonEckardstein:2023gwk} and axion-SU(2) \cite{Maleknejad:2018nxz,Ishiwata:2021yne,Iarygina:2023mtj,Dimastrogiovanni:2025snj} models is an active area of research today. A comprehensive review of the former can be found in Ref.\,\cite{Barbon:2025wjl}. In this paper, we focus on the latter.

All of the aforementioned studies on the backreaction of axion-SU(2) models are based on perturbation theory. Specifically, the gauge field is decomposed into a background and linear perturbations. Then, the backreaction on the background evolution of the $\chi$ and gauge fields is calculated by averaging terms that are quadratic in the linear perturbations. But is this approach valid in the strong backreaction regime? We address this issue by calculating the energy density of spin-2 perturbations of the gauge field and comparing it to the background gauge-field energy density. For perturbation theory to be valid, the former must be smaller than the latter.

Ref.\,\cite{Dimastrogiovanni:2024lzj} studied perturbativity limits by including spin-2 perturbations at second order (one-loop contributions). By requiring the one-loop contributions to be smaller than the tree-level contributions, they obtained perturbative limits on the viable parameter space of axion-SU(2) gauge field dynamics. In contrast, our approach, which is based on energy density, is conceptually different. If the energy density of spin-2 particles exceeds the background energy density, perturbation theory breaks down at all orders.

Throughout this paper, we take $M_{\rm pl}=(8\pi G)^{-1/2}=1$ and use the Friedmann-Lema\^{i}tre-Robertson-Walker (FLRW) metric tensor, $g_{\mu\nu}={\rm diag}(-1,a^2(t),a^2(t),a^2(t))$, where $t$ and $a(t)$ are the physical time and the scale factor, respectively. The determinant of the metric tensor is denoted as $g$. In our convention, the Einstein's field equations are given by $G_{\mu\nu}+T_{\mu\nu}=0$, where $G_{\mu\nu}$ and $T_{\mu\nu}$ are the Einstein tensor and the energy-momentum tensor, respectively.

\section{Axion-SU(2) gauge model}
\label{sec:model}    

\subsection{The action and Euler-Lagrange equation}

In this paper, we work with the spectator axion-SU(2) gauge field
model~\cite{Dimastrogiovanni:2016fuu}, which is given by the action
\begin{align}
  S=\int d^4x \sqrt{-g}\left[\frac{1}{2}R
    +{\cal L}_{\phi}
    +{\cal L}_{\chi}
    +{\cal L}_{A}+{\cal L}_{\rm CS}
    \right]\,,
    \label{eq:action}
\end{align}
where $R$ is the Ricci scalar and 
\begin{align}
  {\cal L}_{\phi}&=
  -\frac{1}{2}g^{\mu\nu}\partial_\mu \phi\partial_\nu \phi
  -V_\phi(\phi)\,,\\
  {\cal L}_\chi&=
  -\frac{1}{2}g^{\mu\nu}\partial_\mu \chi\partial_\nu \chi-V_\chi(\chi)\,,
  \\
  {\cal L}_A &=
  -\frac{1}{4}F^a_{\mu\nu}F^{a\,\mu\nu}\,,
  \\
  {\cal L}_{\rm CS} &=
    -\frac{\lambda }{4f}\chi F^a_{\mu\nu}\tilde{F}^{a\,\mu\nu}
    \,.
\end{align}
Here, $\phi$ and $\chi$ are the inflaton and spectator axion fields with potentials $V_{\phi}(\phi)$ and $V_{\chi}(\chi)$, respectively.  ${\cal L}_{\rm CS}$ is the
Chern-Simons (CS) term, where the axion decay constant and the dimensionless coupling
constant are given by $f$ and $\lambda$, respectively.  $F_{\mu
  \nu}^a$ ($a=1,2,3$) is the field-strength tensor of the SU(2) gauge
field, and we define
$\tilde{F}^{a\,\mu\nu}=\epsilon^{\mu\nu\rho\sigma}F^a_{\rho\sigma}/(2\sqrt{-g})$
with the Levi-Civita symbol satisfying $\epsilon^{0123}=1$. We define the gauge covariant
derivative as $\partial_\mu -i g_A A_\mu^aT^a$, where $g_A$ and $T^a$
are the gauge coupling constant and the generator of the SU(2),
respectively. Here, we do not distinguish between the upper and lower Latin indices $(a,b,c)$ and the repeated indices are summed regardless of their location.  The field strength is given by
$F_{\mu\nu}^a=\partial_\mu A_\nu^a-\partial_\nu A_\mu^a+g_A
\epsilon^{abc}A_{\mu}^bA_{\nu}^c$. Our convention is the same as in
Refs.\,\cite{Maleknejad:2018nxz,Ishiwata:2021yne}, while in Ref.\,\cite{Dimastrogiovanni:2016fuu} the
opposite sign convention is used for the
Chern–Simons term and the interaction term in the gauge-covariant
derivative. They are related to each other via the replacement, 
$\lambda \to -\lambda$ and $g_A\to -g_A$.

The Euler-Lagrange equations are 
\begin{align}
   &\nabla_\mu (\partial^\mu \phi)
  -V'_\phi(\phi)=0\,,
  \label{eq:ELeq_phi}
  \\
   &\nabla_\mu (\partial^\mu \chi)
  -V'_\chi(\chi)-\frac{\lambda}{4f}F^a_{\mu\nu}\tilde{F}^{a\,\mu\nu}=0\,,
  \label{eq:ELeq_chi}\\
  &\left[D_\mu {\cal F}^{\mu\nu}
    \right]^a
  \equiv \nabla_\mu {\cal F}^{a\,\mu\nu} 
-ig_A(T^b)_{ac} A^b_\mu{\cal F}^{c\,\mu\nu} 
  =0\,, 
   \label{eq:ELeq_A}
\end{align}
where $\nabla_\mu$ is the covariant derivative,
$V'_\phi(\phi)=\dv*{V_\phi}{\phi}$,
$V'_\chi(\chi)=\dv*{V_\chi}{\chi}$, and ${\cal F}^{a\,\mu\nu}\equiv
F^{a\,\mu\nu}+\frac{\lambda}{f}\chi \tilde{F}^{a\,\mu\nu}$.
$i(T^b)_{ac}=\epsilon^{bac}$ is used for the adjoint representation. 

The total energy momentum tensor,
$T_{\mu\nu}=T^\phi_{\mu\nu}+T^\chi_{\mu\nu}+T^A_{\mu\nu}$, is 
conserved:
\begin{align}
  \nabla_\mu T^\mu_{~\nu}=
  \frac{1}{\sqrt{-g}}\partial_\mu(\sqrt{-g}T^\mu_{~\nu})-
  \frac{1}{2}\partial_\nu g_{\alpha\beta} T^{\alpha\beta} = 0\,,
\end{align}
where
\begin{align}
  T^\phi_{\mu\nu}&=
  -\partial_\mu \phi\partial_\nu \phi-g_{\mu\nu}{\cal L}_\phi\,,
  \\
  \label{eq:Tmunuchi}
  T^\chi_{\mu\nu}&=
  -\partial_\mu \chi\partial_\nu \chi-g_{\mu\nu}{\cal L}_\chi\,,
  \\
  \label{eq:TmunuA}
  T^A_{\mu\nu}&=
  -F^a_{\mu\alpha}F^{a\,\alpha}_\nu-g_{\mu\nu}{\cal L}_A\,.
\end{align}
Since the inflaton sector is decoupled, $\nabla_\mu
T^{\phi\mu}_{~~\nu}=0$ holds.  The energy density of the inflaton,
axion, and gauge fields are respectively given by $\rho_X=T^{X
  0}_{~~~0}$ ($X=\phi,\chi,A$) and evaluated in Appendix~\ref{sec:T_munu}. The Hubble parameter is
given by $H^2=\sum_X\rho_X/3$.

\subsection{Spectator axion and gauge fields}
\label{sec:spectator}

In our study, $\rho_\phi$ dominates the total
energy density during inflation,
\begin{align}
  \rho_\phi \gg \rho_\chi,\,\rho_A\,,~~ -\dot{H}/H^2\ll 1\,,
\end{align}
where dots denote time derivatives.  Hereafter we will disregard the inhomogeneity in the inflaton and axion fields. Namely, we
take
\begin{align}
  \phi=\phi(t)\,,~~~\chi=\chi(t)\,.
  \label{eq:homogeneous}
\end{align}
Then, the time derivative of $H$ is given by
\begin{align}
  -\frac{\dot{H}}{H^2}=\frac{\dot{\phi}^2}{2H^2}+\frac{\dot{\chi}^2}{2H^2}
  +\frac{2\rho_A}{3H^2}\,.
  \label{eq:Hdot}
\end{align}
We will use this expression to check that $-\dot{H}/H^2\ll 1$ is
satisfied. In what follows, we will neglect this term unless stated
otherwise.

Unlike in Refs.\,\cite{Dimastrogiovanni:2024xvc,Dimastrogiovanni:2025snj}, Eq.\,\eqref{eq:Hdot} does not contain the backreaction term, ${\cal
  J}_A$. The time derivative of $H$ is given by the sum of the energy density, $\rho_X$, and pressure, $P_X$, of $X=(\phi, \chi, A)$ as
 $\dot{H}=-\sum_X(\rho_X+P_X)/2$. Using $\rho_\phi+P_\phi=\dot{\phi}^2$, $\rho_\chi+P_\chi=\dot{\chi}^2$, and $\rho_A+P_A=4\rho_A/3$, we obtain Eq.\,\eqref{eq:Hdot} without invoking the equations of motion and, therefore, without ${\cal J}_A$. See Appendix~\ref{sec:Hdot} for more details.

Under this background configuration, $A^a_\mu$
acquires an isotropic and homogeneous background solution
$\bar{A}^a_\mu$~\cite{Maleknejad:2011jw,Maleknejad:2011sq}:
\begin{align}
  \bar{A}^a_0=0\,,~~~\bar{A}^a_i=a(t)Q(t)\delta^a_i\,.
\end{align}
This is an attractor solution~\cite{Maleknejad:2013npa,Wolfson:2020fqz,Wolfson:2021fya}.
In addition, the perturbation around the background
solution, $\delta A^a_i$, contains scalar, vector, and tensor modes. The tensor mode is
especially important and affects the dynamics of the
homogeneous solution via backreaction~\cite{Dimastrogiovanni:2016fuu,Fujita:2017jwq}. Writing the
tensor mode as $\delta A^a_i=B_{ai}\,(=B_{ia})$, and plugging
Eq.\,\eqref{eq:homogeneous} and
\begin{align}
  A^a_0=0\,,~~~A^a_i = \bar{A}^a_i +\delta A^a_i\,,
\end{align}
into Eqs.\,\eqref{eq:ELeq_phi} and \eqref{eq:ELeq_chi}, we
get
\begin{align}
  \ddot{\phi} + 3H \dot{\phi} + V'_\phi(\phi)=0\,,
  \\
  \ddot{\chi} + 3H \dot{\chi} + V'_\chi(\chi)
  +\frac{3g_A\lambda}{fa^3}(aQ)^2\partial_t(aQ)
  -{\cal P}_\chi=0\,,
  \label{eq:ELeq_chi_2}
  \\
  \frac{\partial_t^2(aQ)}{a}+\frac{H\partial_t(aQ)}{a}
  +2g_A^2Q^3-\frac{g_A\lambda}{f} \dot{\chi}Q^2
  +{\cal J}_A=0\,.
  \label{eq:ELeq_A_2}
\end{align}
Here, we use the FLRW metric, and $\partial_t= \partial/\partial t$.
For simplicity, the $t$ dependence of the fields and the scale factor is omitted.
Eq.\,\eqref{eq:ELeq_A_2} is obtained from the $\nu=i$ component of
Eq.~\eqref{eq:ELeq_A} contracted with $\delta^{a}_i$. Hereafter we
will ignore the dynamics of $\phi$, since it does not affect the other
fields.

${\cal P}_\chi$ and ${\cal J}_A$ are the contributions from the
perturbation at ${\cal
  O}(B_{ij}^2)$~\cite{Maleknejad:2018nxz,Ishiwata:2021yne},
\begin{align}
\label{eq:Pchi}
  {\cal P}_\chi&=\frac{\lambda}{2fa^3}\left[
    2\epsilon^{iqp}\partial_i B_{pj}\dot{B}_{qj}
    +\partial_t(g_A aQB_{ij}^2)\right]\,,\\
    \label{eq:JA}
  {\cal J}_A&=\frac{g_A}{3a^2}
  \left[\frac{1}{a}\epsilon^{iqp}\partial_iB_{jp} B_{jq}
    +\frac{\lambda}{2f}\dot{\chi}B_{ij}^2
    \right]\,,
\end{align}
where the Latin indices are 1,2,3 and the Levi-Civita symbol satisfies $\epsilon^{123}=1$. We do not distinguish between upper and lower indices and repeated indices are summed regardless of their location. We omit the
symbol of the vacuum expectation value $\expval{\cdots}$ on the
right-hand side for notational simplicity.  

We write the symmetric tensor,
$B_{ij}$, in terms of the helicity eigenstates, $T_L$ (left) and $T_R$ (right), of the spin-2 fields that are canonically normalized. For
$\vec{k}=(0,0,|k|)$, for instance, $T_{L,R}=B_{11}\pm i
B_{12}$. The sign $\pm$ corresponds to the left- ($+$) and right-handed ($-$) field. 
Then, $T_{L,R}$ are mixed with the helicity eigenstates of gravitational waves, $\psi_{L,R}$,
i.e., the tensor modes of the metric perturbation. For the same
$\vec{k}$, they are related to the metric perturbations of the tensor
modes as $\psi_{L,R}=a(h_+\pm ih_\times)/2$, where $\delta
g_{11}=a^2h_+$ and $\delta g_{12}=a^2h_\times$ .  

The linear equations of
motion for $T_{L,R}$ and $\psi_{L,R}$ in the wavenumber space are given
by ignoring the non-linear terms,
\begin{align}
  &\partial^2_\tau T_{L,R}+\Omega^2_TT_{L,R}
  =
  -2\frac{\delta_{Q1}}{\tau}\partial_\tau \psi_{L,R}
  +\frac{2}{\tau^2}\left[\sqrt{\epsilon_{Q_{B}}}(m_Q\mp k\tau)
    -(\delta_{Q1}+\delta_{Q2})
      \right]\psi_{L,R} \,,
   \label{eq:T_1}\\
   &\partial^2_\tau \psi_{L,R}+\Omega_\psi^2\psi_{L,R}
   =2\frac{\delta_{Q1}}{\tau}\partial_\tau T_{L,R}
    +\frac{2\sqrt{\epsilon_{Q_B}}}{\tau^2}
    \left[m_Q\mp k\tau \right]T_{L,R} \,,
    \label{eq:psi_1}
\end{align}
where $\tau=\int dt/a(t)=-1/(aH)$ is the conformal time and 
\begin{align}
  &\Omega_T^2=k^2+\frac{2}{\tau^2}\left\{m_Q\xi\mp k\tau(m_Q+\xi)\right\}\,,
  ~~\Omega_\psi^2 =k^2-\frac{2}{\tau^2}\,,\\
  &m_Q=\frac{g_AQ}{H}\,,~~\xi=\frac{\lambda \dot{\chi}}{2fH}\,,~~
  {\cal Q}\equiv aQ\,,
  \\
  \label{eq:miscvariables}
  &\delta_{Q1}=H\tau^2 \partial_\tau {\cal Q}\,,~~
  \delta_{Q2}=H\tau^3 \partial^2_\tau {\cal Q}\,,~~
  \epsilon_{Q_E}=\frac{(\dot{Q}+HQ)^2}{H^2}\,,~~
  \epsilon_{Q_B}=\frac{g_A^2Q^4}{H^2}\,.
\end{align}
Throughout this paper, we use the same notation, $T_{L,R}$ and $\psi_{L,R}$, for the
coordinate-space and momentum-space representations. In
Eqs.\,\eqref{eq:T_1} and \eqref{eq:psi_1}, the sign $\mp$ corresponds
to ``L'' and ``R'' of the subscript. 

In the above equations, $\delta_{Q1}$ and $\delta_{Q2}$ are new variables, 
which have not appeared in previous studies. By rewriting them as
\begin{align}
  \delta_{Q1}=\frac{\dot{Q}+HQ}{H}\,,~~~
  \delta_{Q2}=-\frac{2QH^2+3H\dot{Q}+\ddot{Q}}{H^2}\,,
\end{align}
they are related to the slow-roll parameter,
$\sqrt{\epsilon_{Q_E}}$, as
\begin{align}
  \delta_{Q1}&= \sqrt{\epsilon_{Q_E}} ~~~~~~~~{\rm if}~\dot{Q}+HQ>0\,,
  \\
  \delta_{Q1}+\delta_{Q2}&\simeq - \sqrt{\epsilon_{Q_E}}
  ~~~~~~{\rm if}~ \dot{Q}\simeq 0~{\rm and}~\ddot{Q}\simeq 0\,.
\end{align}
With this approximation, Eqs.\,\eqref{eq:T_1} and \eqref{eq:psi_1}
agree with those in the literature (e.g., Ref.\,\cite{Dimastrogiovanni:2016fuu}).  In this paper, we solve the differential equations given in Eqs.\,\eqref{eq:T_1} and \eqref{eq:psi_1} without this approximation.  

The
energy density of the gauge field at ${\cal O}(B_{ij}^2)$ is given by
$\rho_A=\bar{\rho}_A+\delta \rho_{A}$, where
\begin{align}
  \bar{\rho}_A &= \frac{3}{2}H^2(\epsilon_{Q_E}+\epsilon_{Q_B})\,,\\
  \delta\rho_A &=
  \frac{1}{2a^2}
  \left[\dot{B}_{ij}^2-
    \frac{1}{a^2}B_{ij}\nabla^2B_{ij}+\frac{2g_AQ}{a}
    \epsilon^{iqp}B_{jq}\partial_iB_{jp}
    \right]\,.
\end{align}
The derivation is given in Eqs.\,\eqref{eq:bar_rho_A} and \eqref{eq:delta_rho_A} in
Appendix~\ref{sec:T_munu}. When $\delta\rho_A>\bar\rho_A$, the gauge-field sector becomes non-perturbative due to self-interaction. The background-perturbation split is no longer valid and $B_{ij}$ cannot be treated as a perturbation. Therefore, we cannot write the equations of motion for $Q$ and $B_{ij}$ separately, as in Eqs.~\eqref{eq:ELeq_A_2} and \eqref{eq:T_1}, because higher-order terms in $B_{ij}$ due to self-interaction cannot be ignored. For this reason, we use the threshold, $|\delta\rho_A|/\bar\rho_A=1$, to define the perturbative limit.

Our criterion for the perturbative limit does not apply to axion-U(1) models. The equations of motion for the U(1) gauge field do not contain non-linear terms when the axion is homogeneous due to the absence of self-interaction. The perturbative limit of the axion-U(1) system comes from the interaction between the inhomogeneous axion and gauge fields \cite{Ferreira:2015omg}, which is not captured in our work.

Using the mode functions, $T_h$, we write ${\cal P}_\chi$ [Eq.~\eqref{eq:Pchi}], ${\cal
  J}_A$ [Eq.~\eqref{eq:JA}], and $\delta \rho_A$ in terms of the wavenumber
integral,
\begin{align}
\label{eq:Pchiint}
  {\cal P}_\chi &= \frac{\lambda}{2fa^3} \pdv{t}
  \sum_{h=L,R}\int \frac{d^3k}{(2\pi)^3}\left[c_hk+aHm_Q
    \right]|T_h|^2
  \\
  \label{eq:JAint}
  {\cal J}_A &= \frac{g_A}{3a^3}
  \sum_{h=L,R} \int  \frac{d^3k}{(2\pi)^3}
  \left[c_hk
    +aH\xi
    \right]|T_h|^2 \,,
  \\
  \label{eq:rhoAint}
  \delta \rho_A&=
  \frac{1}{2a^2}
  \sum_{h=L,R} \int  \frac{d^3k}{(2\pi)^3}
  \left[|\dot{T}_h|^2+\left(\frac{k^2}{a^2}
+2m_QH\frac{c_hk}{a}\right)|T_h|^2
    \right]\,,
\end{align}
where $c_h=1\,(-1)$ for the left (right)-handed field. 

\section{Numerical method}

We solve the integro-differential equation using \texttt{ODE} 
solvers provided in
\texttt{Julia}~\cite{Bezanson:2014pyv}. We introduce $y=-\ln(-\tau)=\ln aH$, which is
effectively the number of $e$-folds during inflation, as time coordinates. Using $\dv*{t}=H\dv*{y}$
and $\dv*{\tau}=-\tau^{-1}\dv*{y}$, the equations of motion for the
axion and the background gauge field become
\begin{align}
  &\tilde{\chi}''+3\tilde{\chi}'
  -\frac{\beta(\chi)}{\lambda}
  =
  -\frac{3}{\lambda\kappa}m_Q^2(m_Q+m_Q')+\tilde{{\cal P}}_\chi\,,
\label{eq:ELeq_chi_3}\\
&m_Q''+3m'_Q+2m_Q(1
+m^2_Q)=
  \lambda m_Q^2 \tilde{\chi}'-\tilde{{\cal J}}_A\,,
  \label{eq:ELeq_A_3}
\end{align}
where
\begin{align}
\label{eq:misc}
  \tilde{\chi}\equiv \frac{\chi}{f}\,,~~
  \kappa \equiv \left(\frac{g_Af}{\lambda H}\right)^2\,,~~
  \beta (\chi)\equiv -\frac{\lambda V'_{\chi}(\chi)}{H^2 f}\,,~~
  \tilde{{\cal P}}_\chi = \frac{1}{fH^2}{\cal P}_\chi\,,~~
  \tilde{{\cal J}}_A = \frac{g_A}{H^3}{\cal J}_A\,.
\end{align}
Here, the primes denote the derivatives with respect to $y$, unless they act on $V_\chi$ and $V_\phi$.
In our numerical
study, we take the potential of the axion as \cite{Freese:1990rb}
\begin{align}
  V_\chi=\mu^4(1+\cos\tilde{\chi})\,.
\end{align}
Similarly, the linear equations of motion for the tensor modes are
\begin{align}
    & T''_{L,R}+T'_{L,R}+\tau^2\Omega_T^2T_{L,R}
  =
  2\delta_{Q1}\psi'_{L,R}
  +2\left[\sqrt{\epsilon_{Q_{B}}}(m_Q\mp k\tau)
    -(\delta_{Q1}+\delta_{Q2})
      \right]\psi_{L,R} \,,
   \label{eq:T_2}
    \\
   & \psi''_{L,R}+\psi'_{L,R}+\tau^2\Omega_\psi^2\psi_{L,R}
    =
    -2\delta_{Q1} T'_{L,R}
    +2\sqrt{\epsilon_{Q_B}}
    \left[m_Q\mp k\tau \right]T_{L,R} \,,
    \label{eq:psi_2}
\end{align}
where $\tau=-e^{-y}$, $\delta_{Q1}=H(m_Q+m'_Q)/g_A$, and
$\delta_{Q2}=-H(m_Q''+3m_Q'+2m_Q)/g_A$.

We adopt the method given in
Ref.\,\cite{Fujita:2018ndp}, in which the
tensor modes are decomposed into intrinsic and sourced
perturbations. See Appendix~\ref{app:ELeq_tensor} for
details. We numerically solve Eqs.\,\eqref{eq:ELeq_chi_3},
\eqref{eq:ELeq_A_3}, \eqref{eq:T_2}, and \eqref{eq:psi_2} in the range
$y=[0,25]$.  Eqs.\,\eqref{eq:T_2} and \eqref{eq:psi_2} are solved for
each wavenumber $k$ and integrated to obtain $\tilde{{\cal P}}_\chi$,
$\tilde{J}_A$, and $\delta \rho_A$ using Eqs.~\eqref{eq:Pchiint}, \eqref{eq:JAint}, and \eqref{eq:rhoAint}, respectively. The wavenumber space is
$k=[10^{-3},10^{12}]$ in units of $aH$ at $y=0$, but we set a cutoff wavenumber $k_{\rm cut}$
to regularize the integrals as follows. The integrands include the contribution of the Bunch-Davies vacuum, which diverges at high $k$. This should be excluded from the integral,
since we are calculating the backreaction mainly due to the instability 
of $T_{R,L}$ caused by the dynamics of the background. 
To this end, we regularize the integral with a momentum
cutoff. The momentum cutoff is given by one of the roots of
$\Omega_T^2=0$~\cite{Maleknejad:2018nxz},
\begin{align}
  k_{\rm cut}=aH\left[|m_Q+\xi|+\sqrt{m_Q^2+\xi^2}\right]\,.
  \label{eq:kcut}
\end{align}
Here, $aH=e^y$, $m_Q$ and $\xi$ depend on $y$ and
we put the wavenumber upper bound at each $y$ in the integral. This regularization scheme properly accounts for the backreaction. See Appendix~\ref{app:cutoff} for more details.

The system has five parameters, $\lambda$, $g_A$, $\mu$, $f$, and
$H$. To investigate how these parameter affect the dynamics, we
utilize the known stationary solutions.  When $\tilde{\chi}''=m_Q''=0$
in Eqs.\,\eqref{eq:ELeq_chi_3} and \eqref{eq:ELeq_A_3}, then the
equations of motion for $\tilde{\chi}$ and $m_Q$ are diagonalized as
\begin{align}
  m'_Q&=
  \frac{
    \kappa \beta(\chi)m_Q^2 -3m_Q^5
    -6\kappa m_Q(1+m_Q^2)
    +\kappa \lambda m_Q^2\tilde{\cal P}_\chi-3\kappa\tilde{\cal J}_A}
       {3(3\kappa + m_Q^4)}\,,
  \label{eq:dxiA}\\
  \tilde{\chi}'&=
  \frac{ \kappa \beta(\chi)-m_Q^3+2m_Q^5
         +\kappa \lambda \tilde{\cal P}_\chi+m_Q^2 \tilde{\cal J}_A}
            {\lambda(3\kappa + m_Q^4)}\,.
  \label{eq:dchi}
\end{align}
The stationary solutions, $m_Q^{\rm st}$, given by $m_Q'=0$ without the
backreaction terms, can be classified depending on the value of
$\kappa$~\cite{Ishiwata:2021yne}, i.e.,
\begin{align}
  {\rm (a)}~&  m_Q^{\rm st}\simeq
  \left[\frac{\kappa \beta(\chi)}{3}\right]^{1/3}
  \label{eq:mQ_a}\,,
  \\
  {\rm (b)}~& m_Q^{\rm st}\simeq \frac{1}{12}\left[
    \beta(\chi)+\sqrt{\beta(\chi)^2-144}\right]
  \label{eq:mQ_b}\,,
\end{align}
for $\kappa \ll 1$ and $\kappa \gg 1$, respectively.  We will refer to them as
`case (a)' and `case (b)' in the following discussions. The case (a) and (b) solutions
were found in Refs.\,\cite{Adshead:2012kp} and \cite{Ishiwata:2021yne}, respectively. 

We use the stationary solutions and
$\kappa$ to set the five parameters. However, the dynamics are
independent of a parameter $\mu^4/H^2$ when
$\dot{H}$ is neglected.\footnote{The factor of $\mu^4/H^2$ in $\beta$ of
Eq.\,\eqref{eq:misc} is given by $m_Q^{\rm st}$ and $\kappa$.
The dependence of $\tilde{\cal P}_\chi$ and $\tilde{\cal J}_A$ on $H$ is
absorbed by $y$ as $aH=e^y$.} With this knowledge, we take the
following four parameters as input:
\begin{align}
  \lambda\,,~~g_A\,, ~~\kappa\,,~~m_{Q}^{\rm in}\,.
\end{align}
Then, using $m_{Q}^{\rm in}=m_Q^{\rm st}$ with
Eqs.\,\eqref{eq:mQ_a} and \eqref{eq:mQ_b} for $\kappa<1$ and
$\kappa>1$, respectively, and $\tilde{\chi}=0.5\pi$, the remaining
parameters are determined.

We set initial conditions for the background field values
as $m_Q^{\rm in}=m_Q^{\rm
  st}(\tilde{\chi}_i)$ where $\tilde{\chi}_i\equiv\tilde{\chi}(y=0)=0.3\pi$. 
For the derivative of the field, 
we consider two different initial conditions:
\begin{enumerate}[(I)]
  \item $m'_Q$ and $\tilde{\chi}'$ at $y=0$ are given by
    Eqs.\,\eqref{eq:dxiA} and \eqref{eq:dchi}\,,
  \item $m'_Q=\tilde{\chi}'=0$ at $y=0$\,.
\end{enumerate}
We set $\lambda=10^2$ and take
$\kappa=10^{-2}$ and $10^{-1}$ for case (a), and
$\kappa =10$ and $10^2$ for case (b).

\section{Results}
\label{sec:results}
\subsection{Criterion for the perturbative limit}
Using the numerical method described in the previous section, we solve the
integro-differential equations of the system. We focus on
the validity of the perturbative expansion of the gauge field.
As discussed in Section~\ref{sec:spectator}, the perturbative calculation is invalid if
\begin{align}
  \bar{\rho}_A < \delta \rho_A\,,
\end{align}
where the dependence on $y$ is implicit. This condition is independent of
$H$, as $\delta \rho_A/\bar{\rho}_A$ is a function of
$aH=e^y$. 

Additionally, we set a criterion for the `strong
backreaction' by comparing the backreaction 
terms to
the largest terms in the differential equations given in Eqs.~\eqref{eq:ELeq_chi_3}
and \eqref{eq:ELeq_A_3}:
\begin{align}
  \tilde{\cal J}_A>2m_Q(1+m_Q^2)\,~~{\rm or}~~
  \tilde{\cal P}_\chi>\beta/\lambda\,,
  \label{eq:strong_BR}
\end{align}
where again the dependence on $y$ is implicit. If this condition is satisfied, we call it
`strong backreaction.'

\subsection{Solutions for case (a): \texorpdfstring{$\kappa=0.01$}{kappa=0.01}}
\label{sec:solultionA}

\begin{figure}
 \begin{center}
   \includegraphics[width=7.5cm]{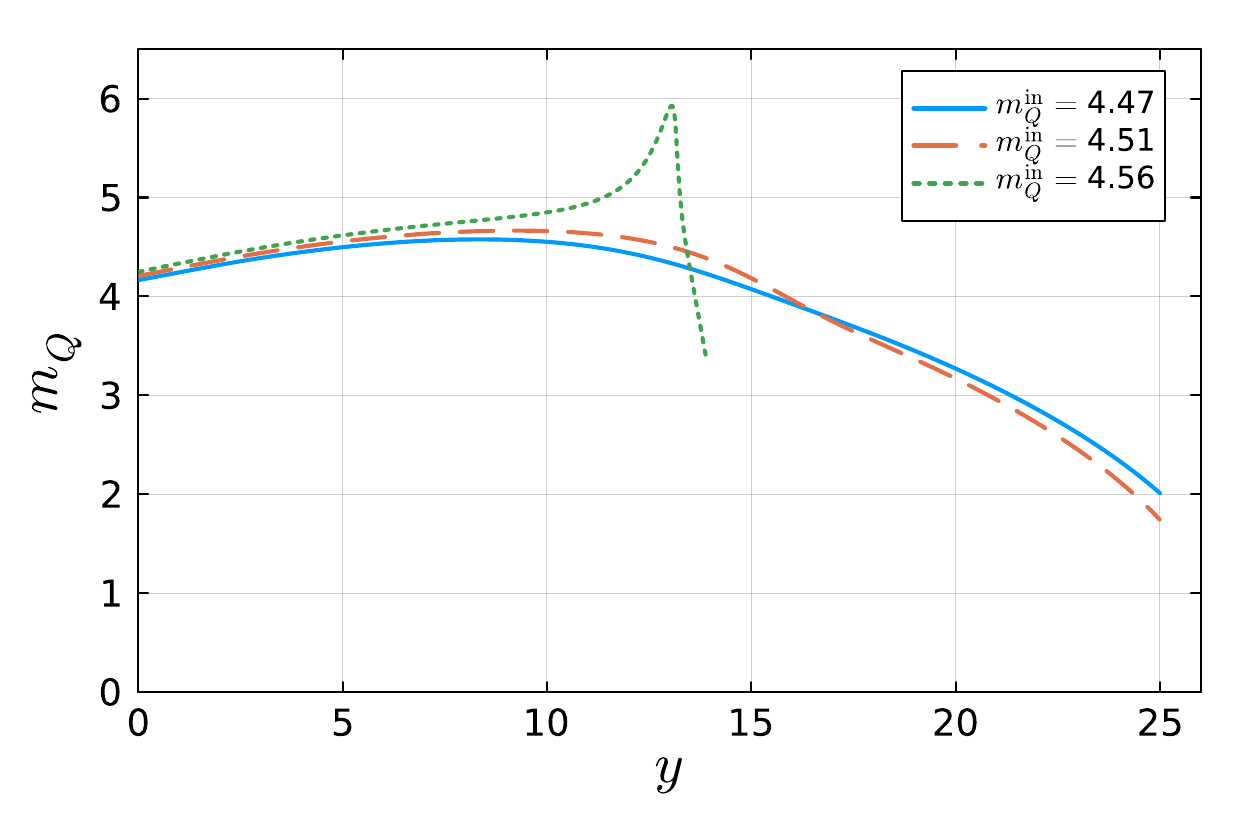}
   \includegraphics[width=7.5cm]{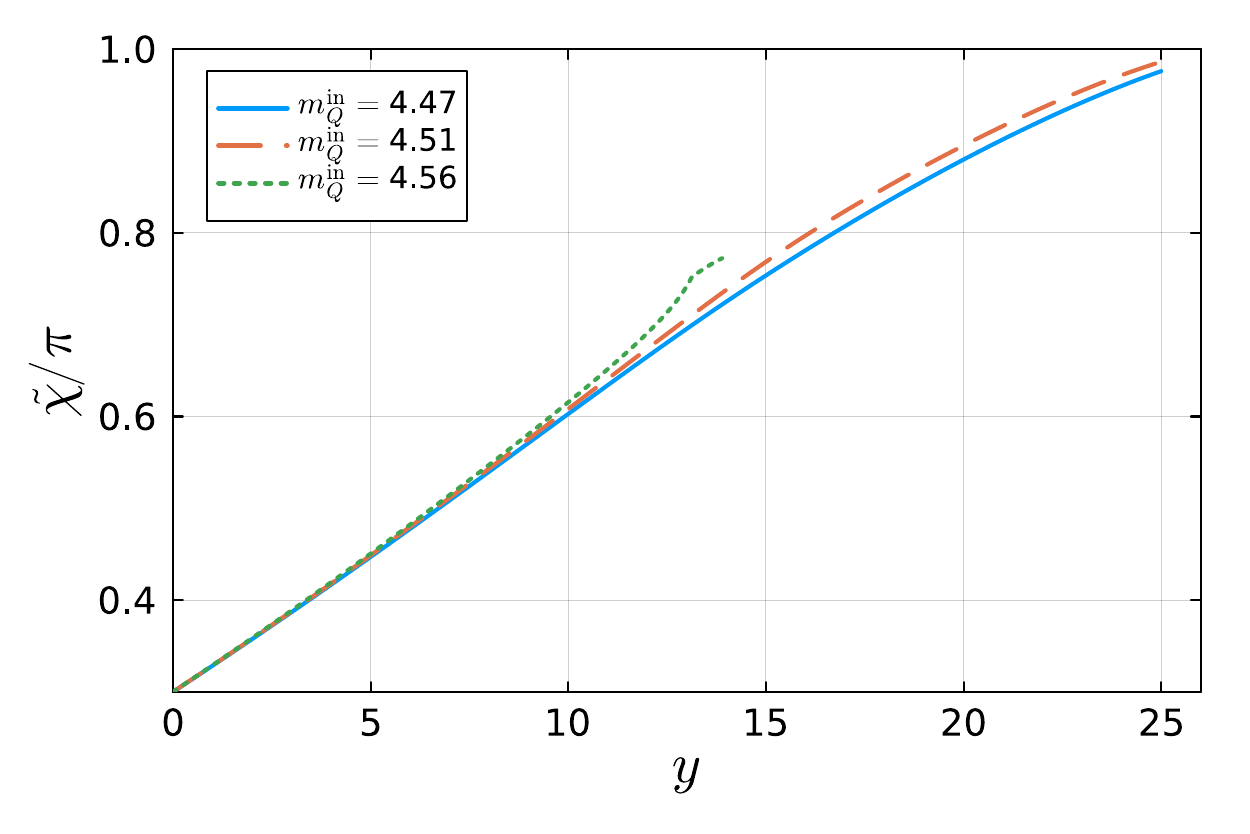}
   \includegraphics[width=7.5cm]{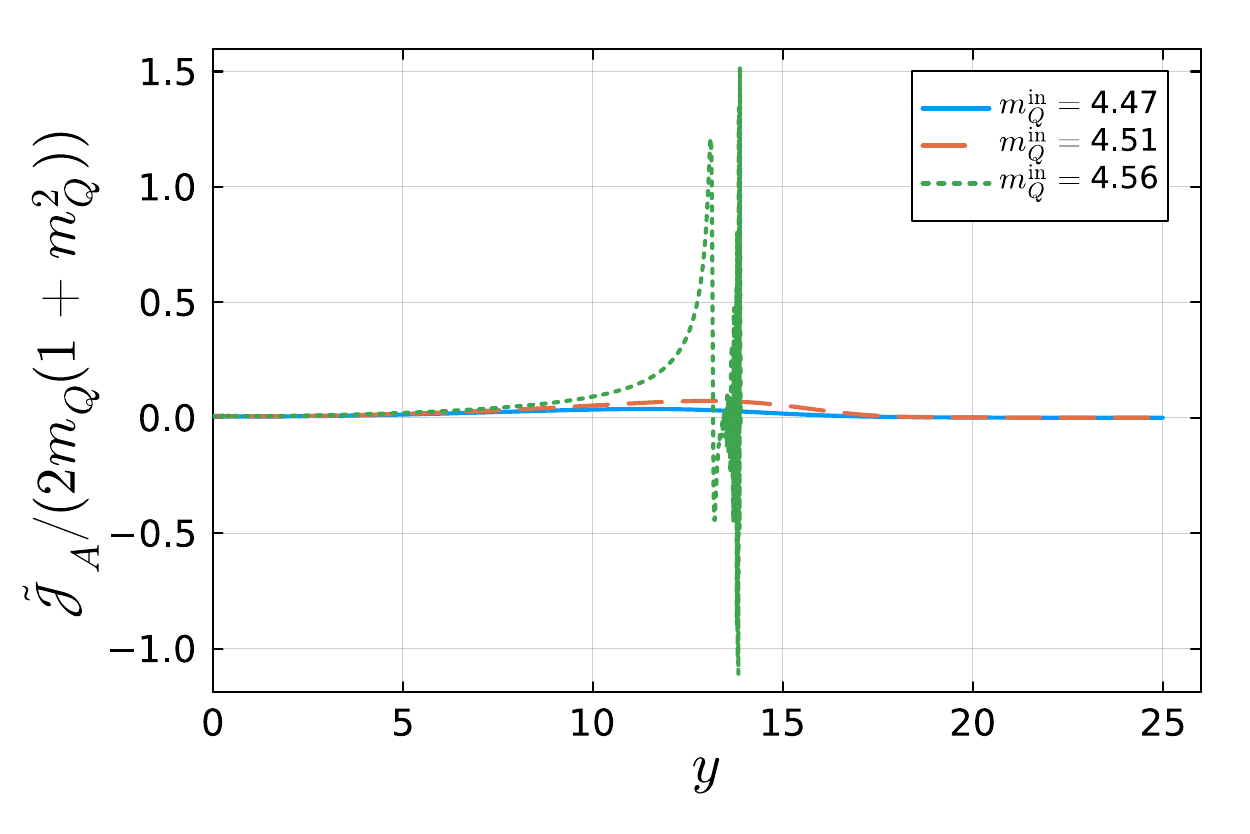}
   \includegraphics[width=7.5cm]{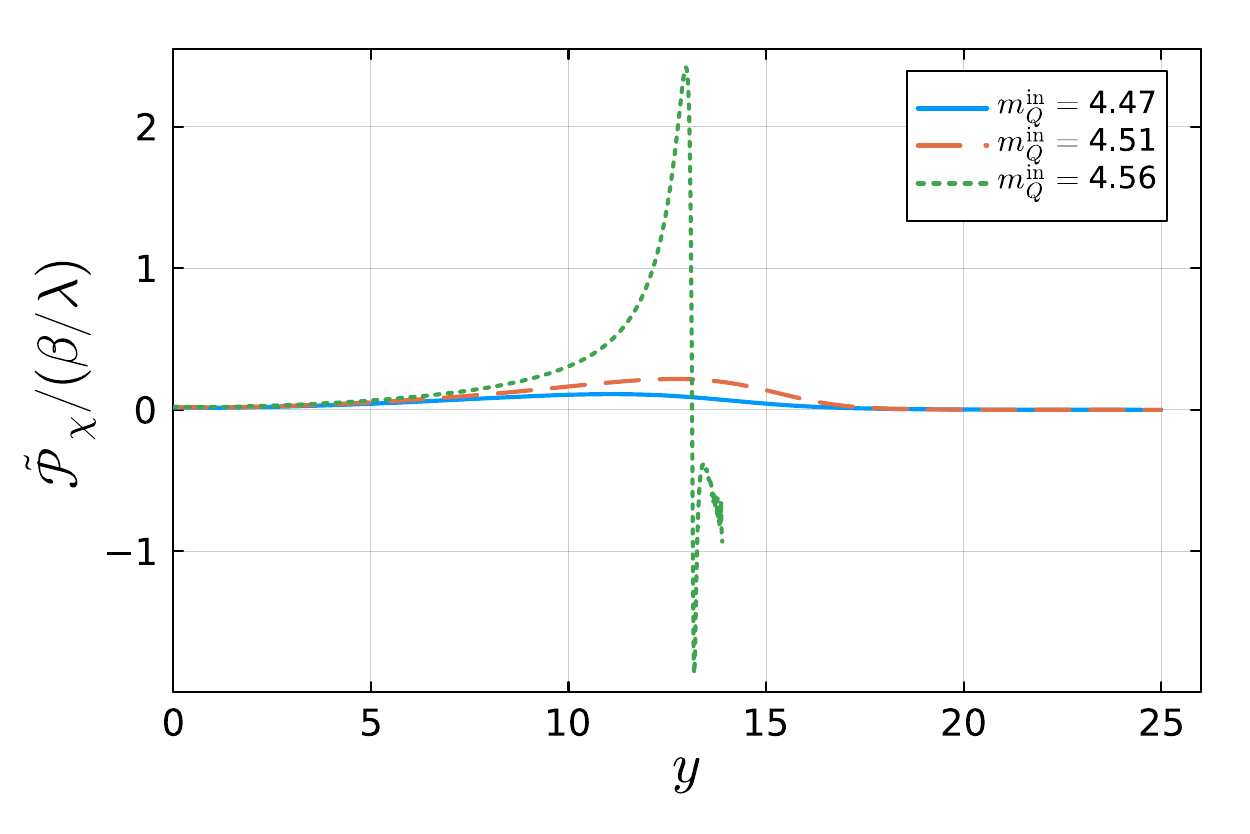}
   \includegraphics[width=7.5cm]{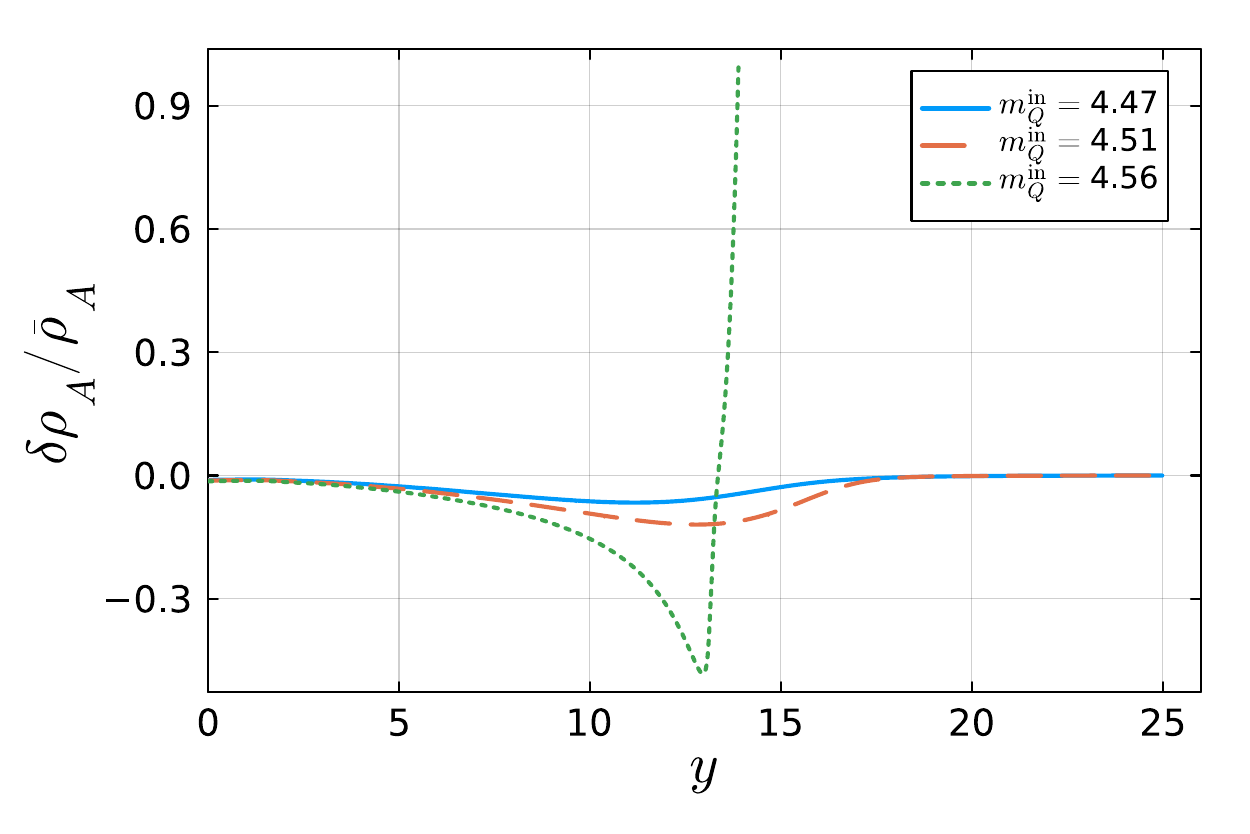}
    \caption{Time evolution of the background fields, $m_Q$
      (top-left), $\tilde{\chi}$ (top-right), the backreaction term
      $\tilde{\cal J}_A$ divided by $2m_Q(1+m^2_Q)$ (middle-left), the
      backreaction term $\tilde{\cal P}_\chi$ divided by
      $\beta/\lambda$ (middle-right), and $\delta \rho_A/\bar{\rho}_A$
      (bottom), as a function of $y=\ln aH$. The values of the parameters are $\lambda=100$,
      $\kappa=10^{-2}$, and $g_A=10^{-3}$. The input values
      of $m_Q^{\rm in}$ are indicated in the legend. The initial
      conditions are given by initial condition (I), in which $m_Q^{\rm in}=m_Q^{\rm st}(\tilde{\chi}=\tilde{\chi}_i)$,
      $\tilde{\chi}_i=0.3\pi$, and their derivatives are given
      by Eqs.\,\eqref{eq:dxiA} and \eqref{eq:dchi}. }
  \label{fig:kappa001_i1_BG}
 \end{center}
\end{figure}

The top two panels of Figure~\ref{fig:kappa001_i1_BG} show the dynamics of the background
fields, $m_Q$ and $\tilde{\chi}$, for $\lambda=10^2$,
$\kappa=10^{-2}$, and $g_A=10^{-3}$ under initial condition (I).
The initial condition (II) yields similar results.
We verified the accuracy of the calculation by confirming that
Eq.\,\eqref{eq:for_check_sol} was satisfied at the
$\order{10^{-2}}$\,\% level. The different lines correspond to various
values of $m_Q^{\rm in}$. The calculation for the largest value of
$m_Q^{\rm in}$ is terminated at around $y\simeq 14$ because the
dynamics reach the perturbative limit, i.e.,
$\delta\rho_A>\bar{\rho}_A$, as shown in the bottom panel. The growth
of $\delta\rho_A$ arises from the enhancement of the right-handed
tensor mode $T_R$, which is caused by its
instability. 

The middle two panels show the
backreaction terms, $\tilde{{\cal J}}_A$ and $\tilde{\cal P}_\chi$, normalized to demonstrate how the backreaction becomes `strong' according to the criterion given in Eq.~\eqref{eq:strong_BR}.  The backreaction terms become
comparable to the largest terms in each differential equation
at around $y\simeq 14$, which coincides with the perturbative limit.
Namely, the growth of $T_R$ enhances both
the backreaction terms and $\delta \rho_A$, thereby breaking the perturbative
expansion of the gauge field.

The slow-roll approximation, $\tilde{\chi}''=m''_Q=\dot{H}=0$ and
$\delta_{Q1}=-(\delta_{Q1}+\delta_{Q2})=\sqrt{\epsilon_{Q_E}}$, holds
for most of the time evolution.  The exceptions are the initial
oscillatory regime in the early stage and the region near the perturbative limit.
Figure~\ref{fig:kappa001_i1_dfields} in Appendix~\ref{app:SR} shows the 
time evolution of $m'_Q$, $\xi$,
$\delta_{Q1}$, and $\delta_{Q1}+\delta_{Q2}$. 
See the discussion there for more details. On the other hand, $-\dot{H}/H^2$ is always small. The behavior remains qualitatively unchanged for $\kappa=10^{-1}$ and
other values of $g_A$. In addition, the choice of initial
conditions has little effect on the dynamics. 
Figure~\ref{fig:kappa001_i0_dfields} in Appendix~\ref{app:SR} demonstrates this with the results for initial condition (II). In this case, while
the amplitudes of the oscillation in $m'_Q$, $\xi$ and the slow-roll
parameters increase, the time evolution
of the background fields remains unaffected.

\subsection{Solutions for case (b): \texorpdfstring{$\kappa=100$}{kappa=100}}
\label{sec:solultionB}

\begin{figure}
 \begin{center}
   \includegraphics[width=7.5cm]{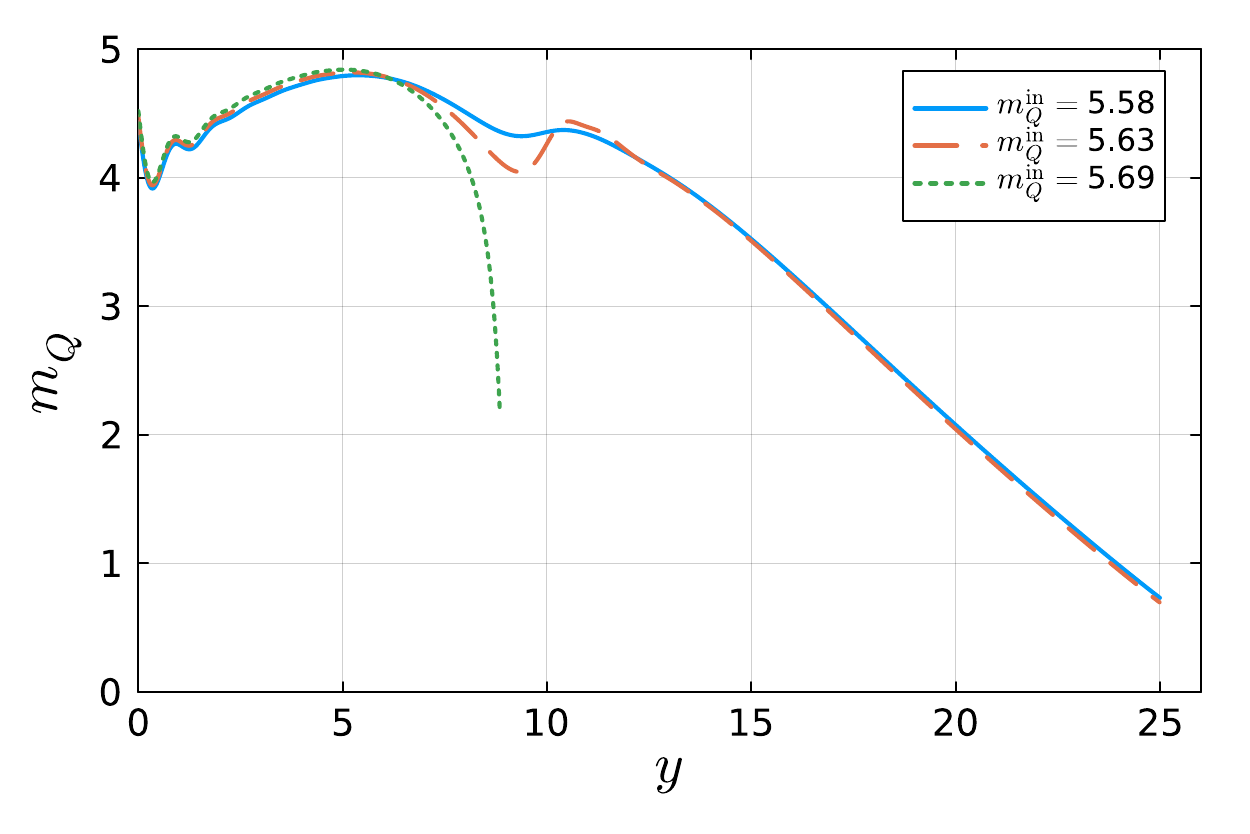}
   \includegraphics[width=7.5cm]{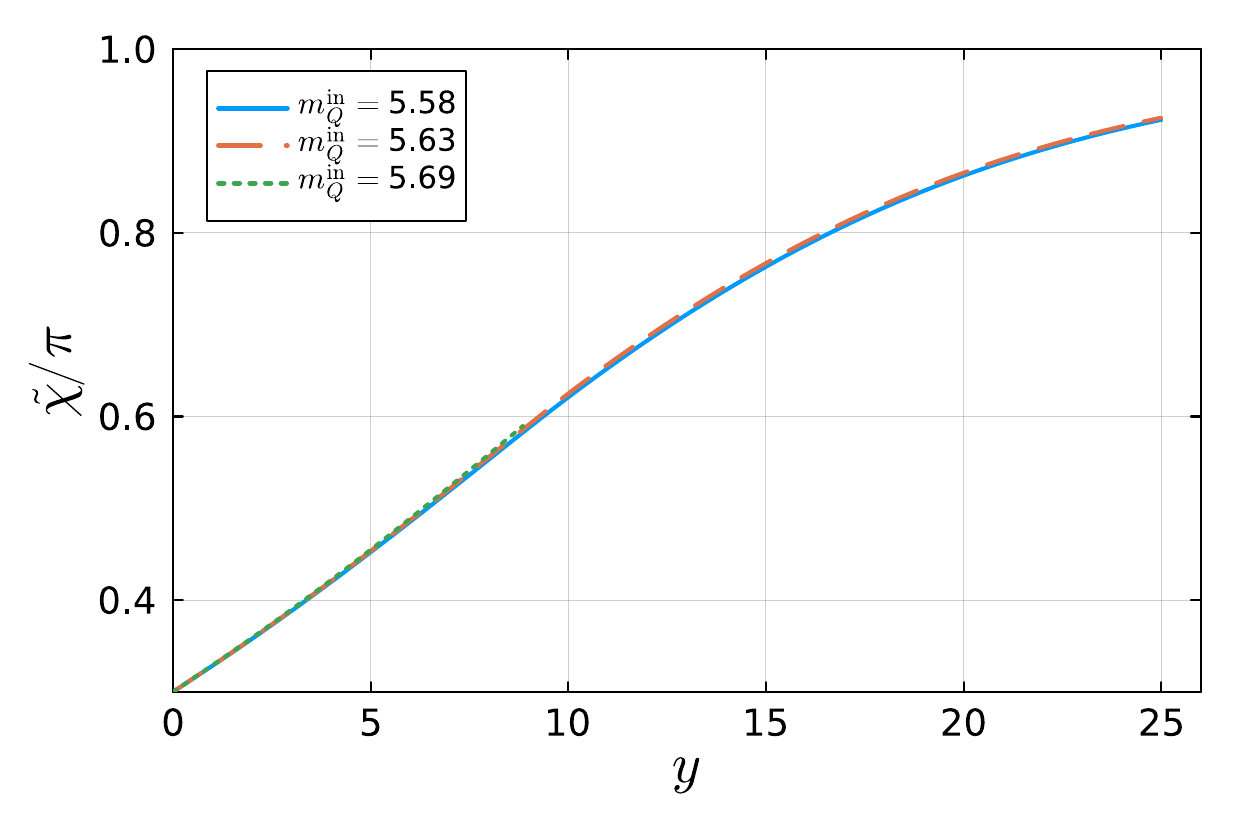}
   \includegraphics[width=7.5cm]{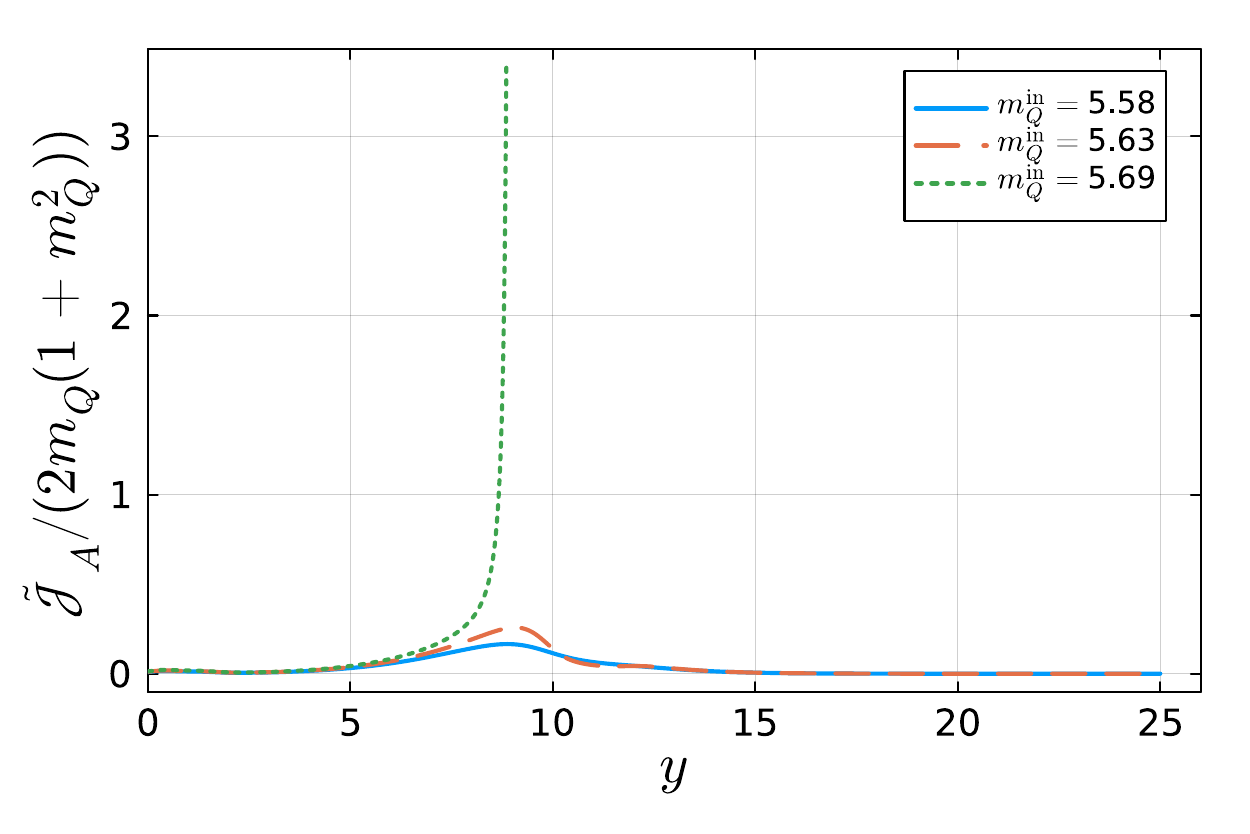}
   \includegraphics[width=7.5cm]{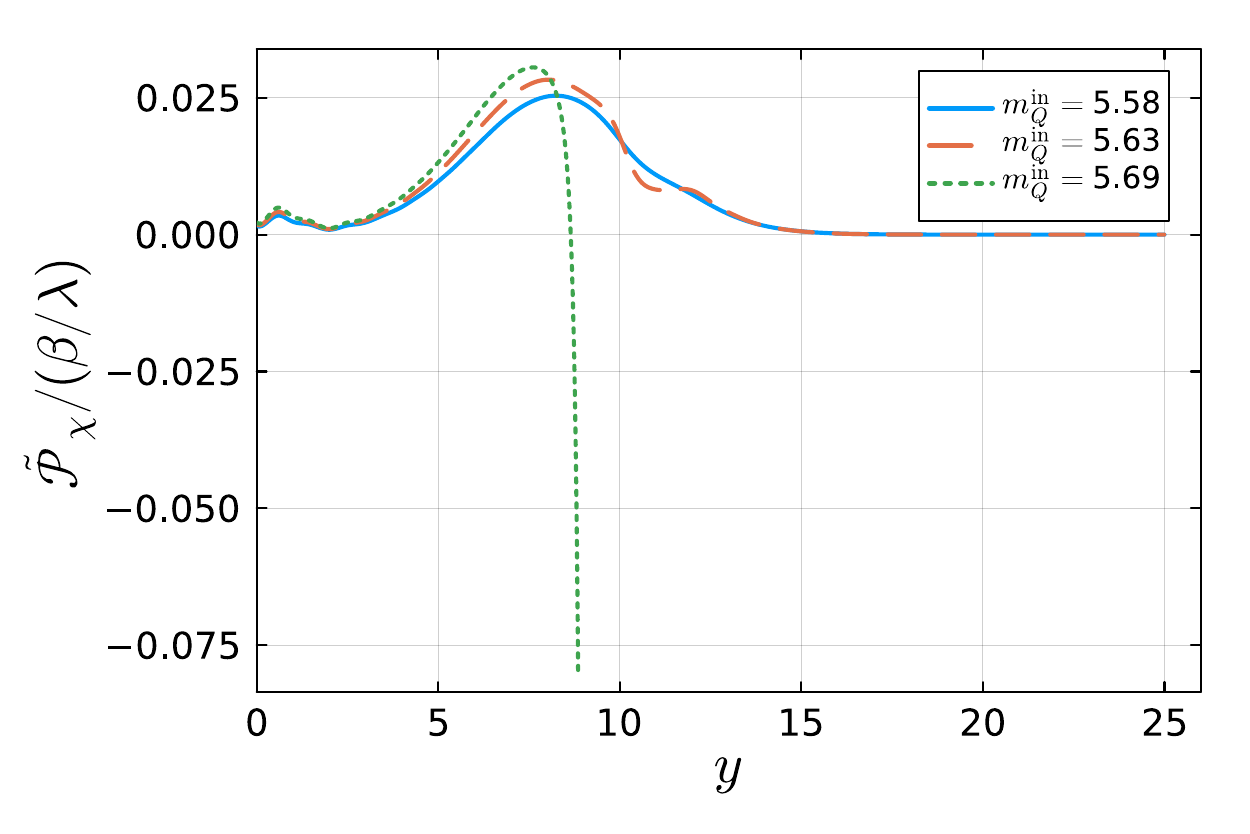}
   \includegraphics[width=7.5cm]{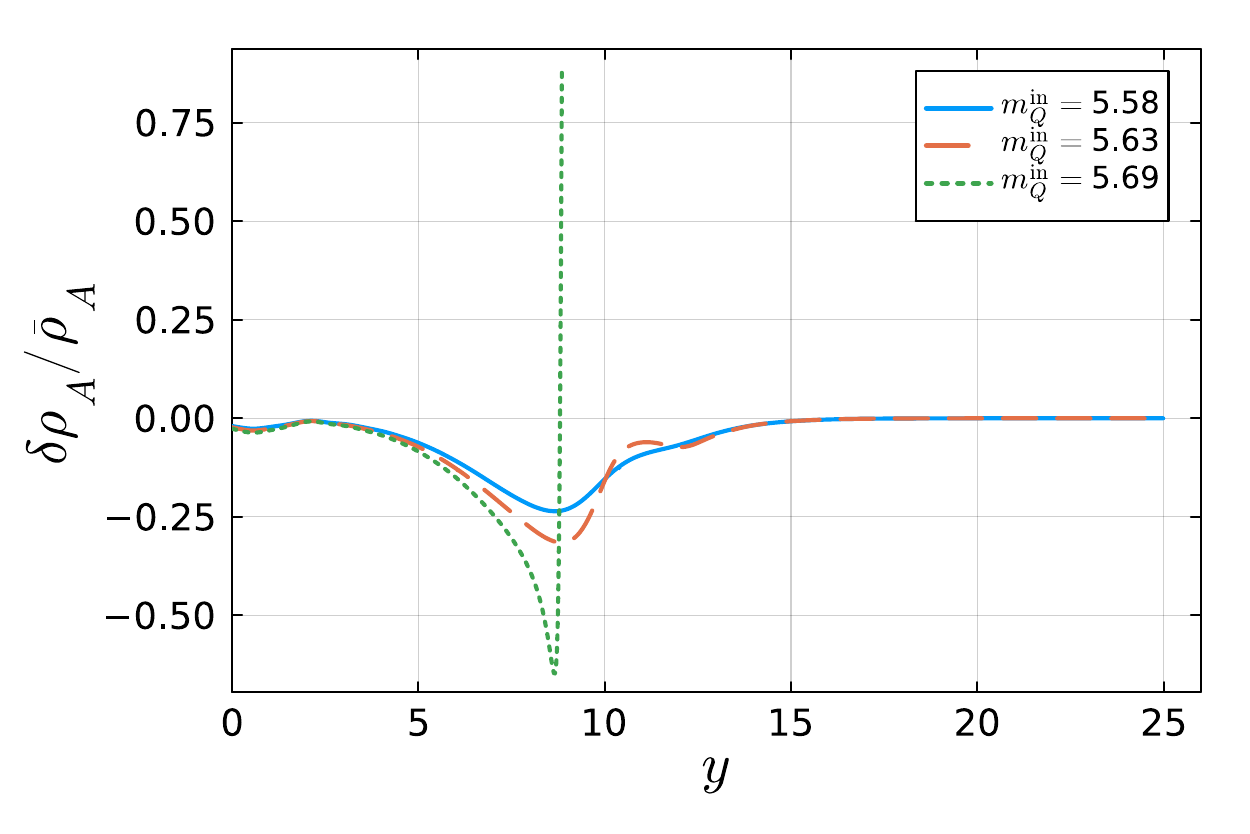}
   \caption{Same as Figure~\ref{fig:kappa001_i1_BG}, but for
     $\kappa=10^{2}$.} 
  \label{fig:kappa100_i1_BG}
 \end{center}
\end{figure}

Figure~\ref{fig:kappa100_i1_BG} shows the results for $\kappa=10^2$ under initial condition (I).
The input
values of $m_Q^{\rm in}$ are adjusted accordingly. 
We verified the accuracy of the calculation again by confirming that
Eq.\,\eqref{eq:for_check_sol} was satisfied at the
$\order{10^{-2}}$\,\% level.
As in case (a), we encountered the perturbative limit and the strong backreaction
almost simultaneously. We find that,
while $\tilde{\cal J}_A$ increases, the contribution of $\tilde{\cal P}_\chi$ to the differential equation is negligible. Consequently,
the backreaction has little effect on the dynamics
of $\tilde{\chi}$, whereas $m_Q$
behaves differently depending on $m_Q^{\rm in}$.

The slow-roll approximation is less valid than in case (a). The slow-roll result of $m_Q'$ assuming Eq.\,\eqref{eq:dxiA} deviates from
the exact numerical solution over a wide range of time
evolution. The approximation of $\delta_{Q1}+\delta_{Q2}\simeq
-\sqrt{\epsilon_{Q_E}}$ is also poorer.  See
Figure~\ref{fig:kappa100_i1_dfields} in Appendix~\ref{app:SR} for
more details.

\begin{figure}
 \begin{center}
   \includegraphics[width=7.5cm]{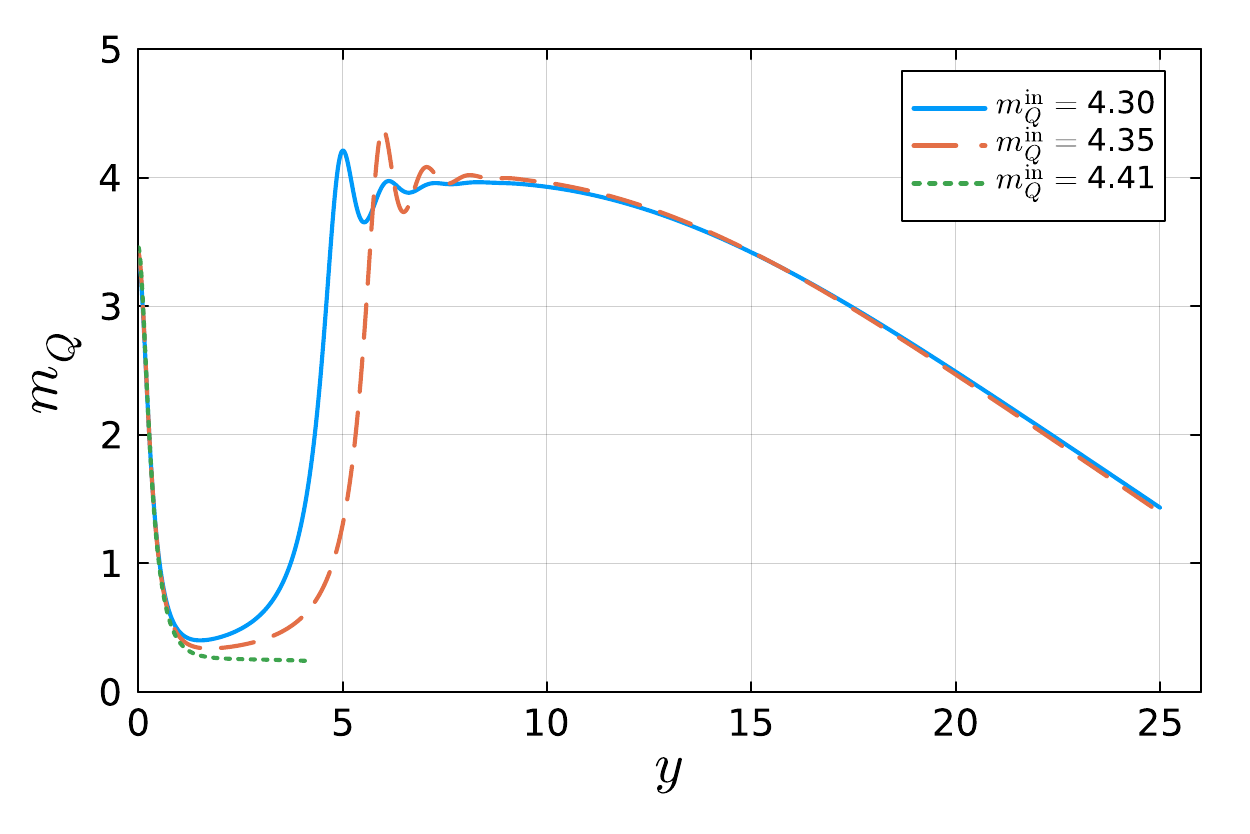}
   \includegraphics[width=7.5cm]{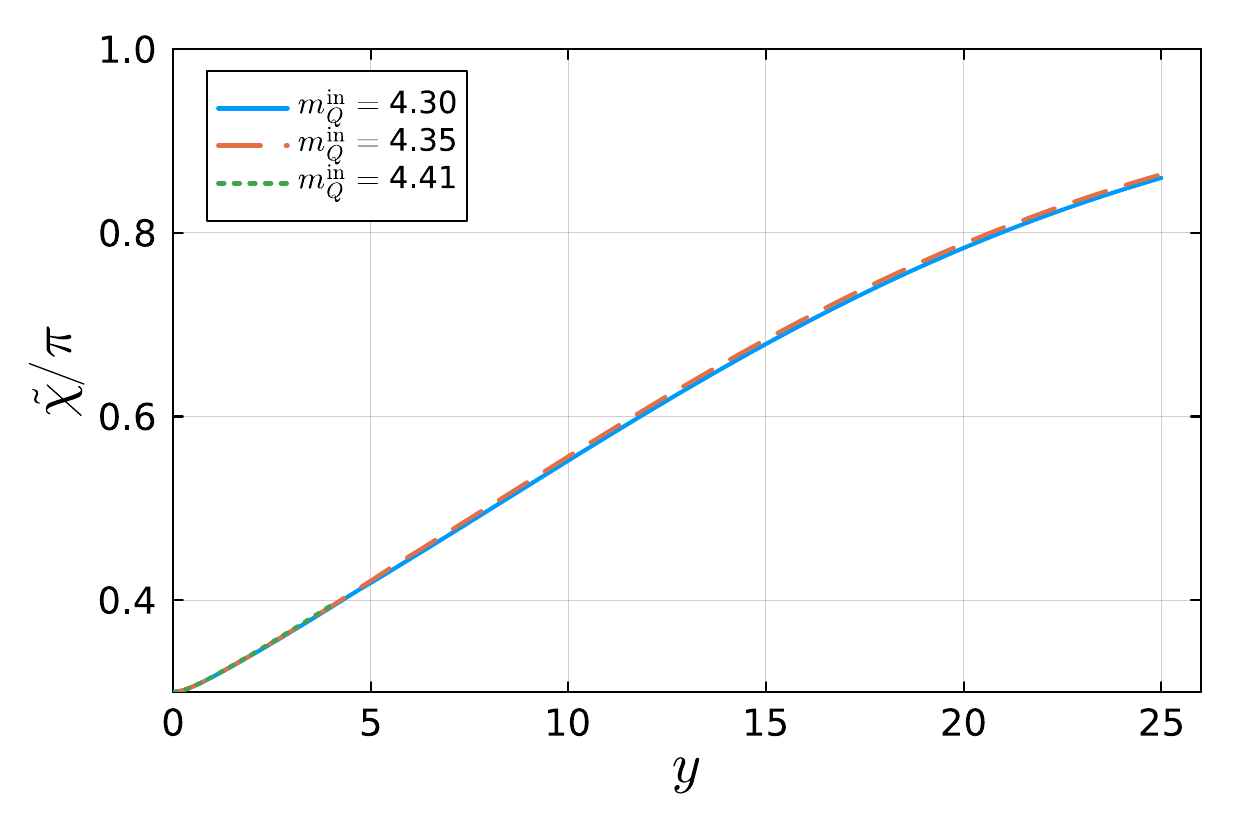}
    \includegraphics[width=7.5cm]{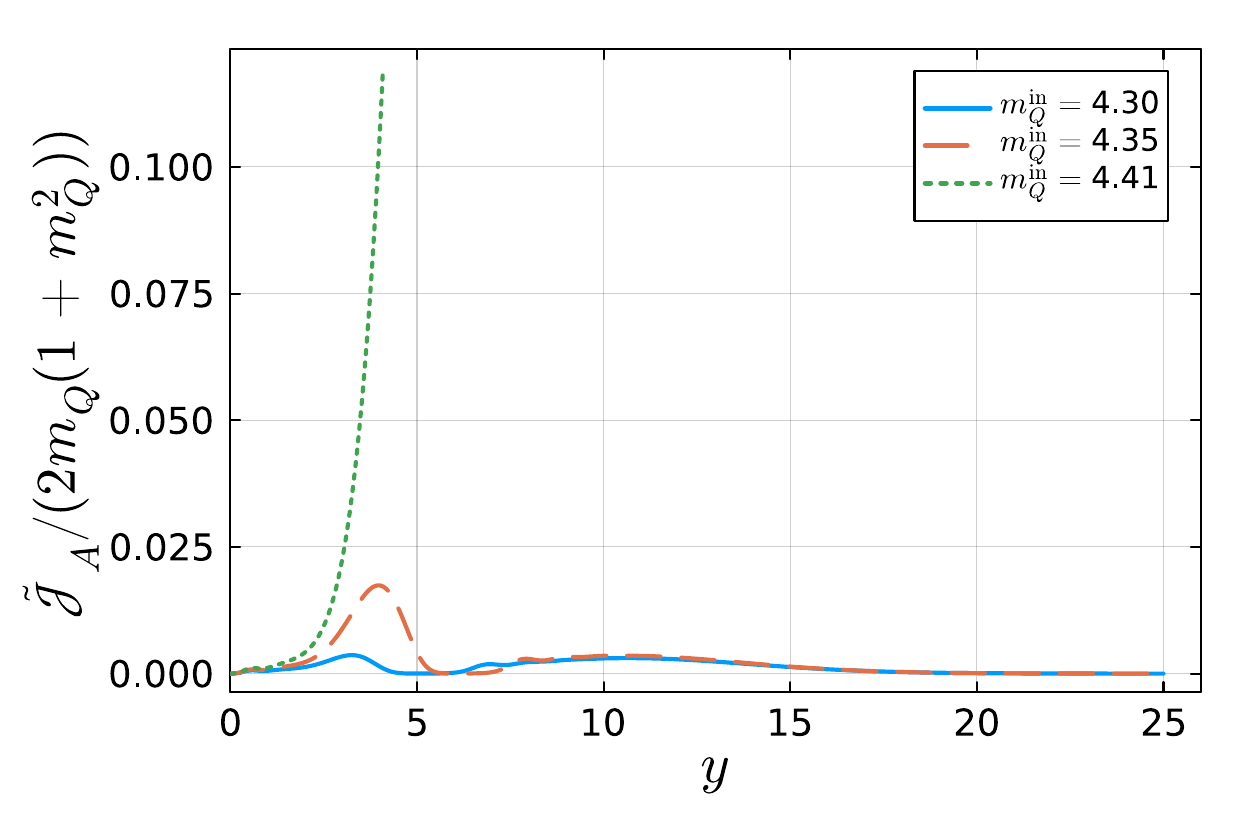}
    \includegraphics[width=7.5cm]{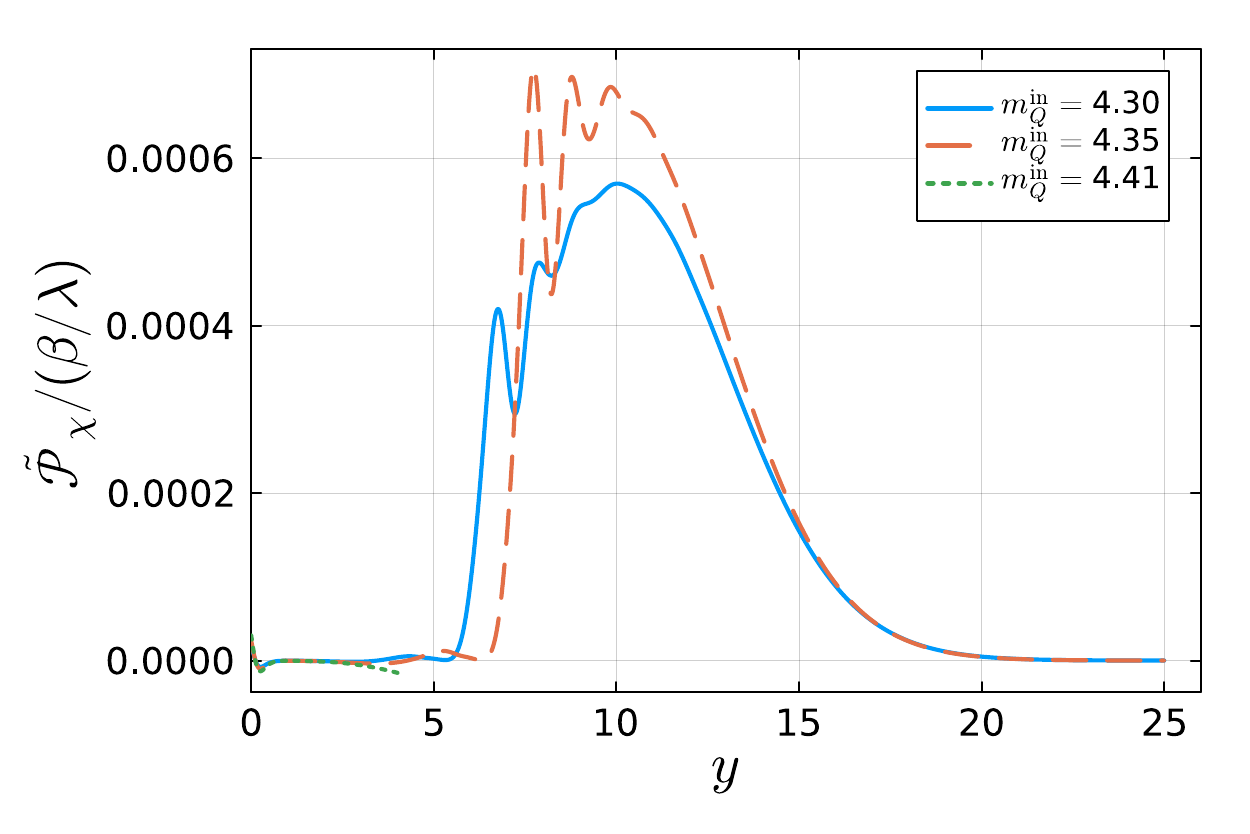}
    \includegraphics[width=7.5cm]{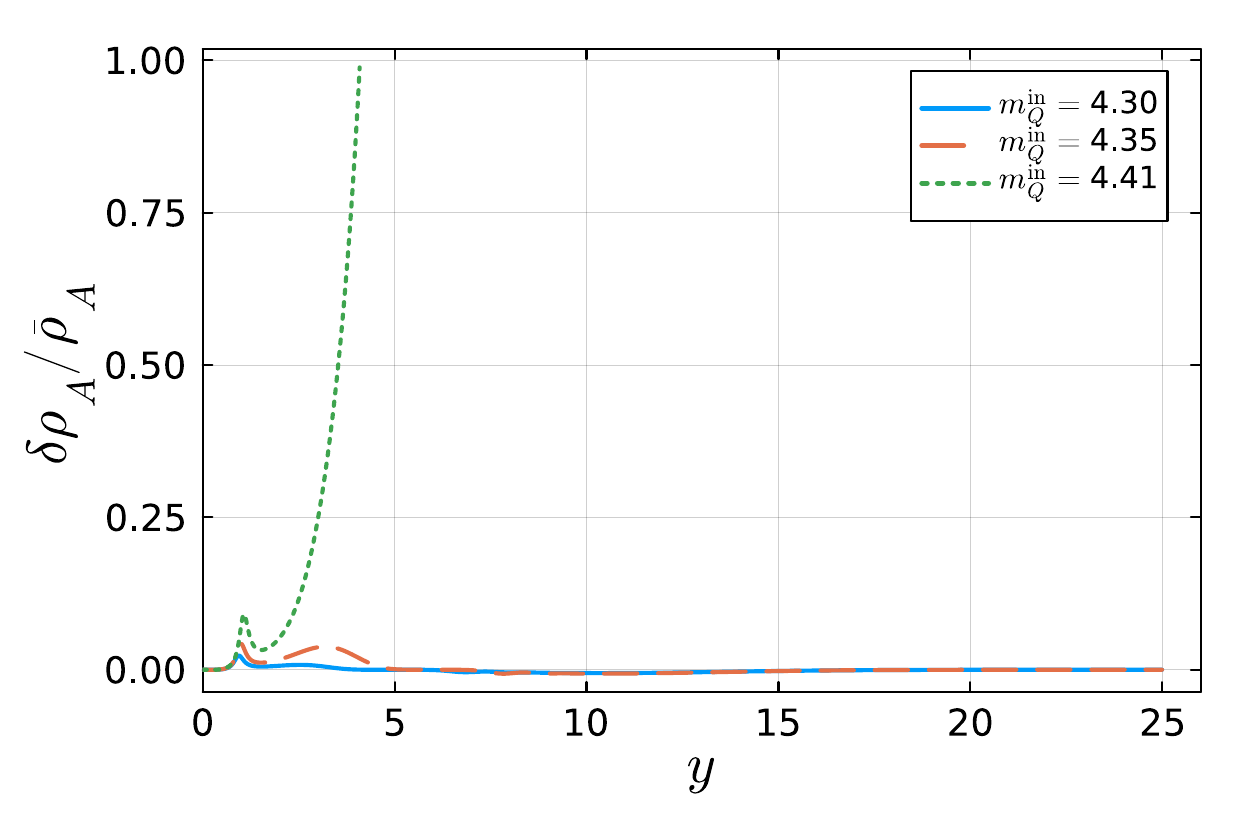}
   \caption{Same as Figure~\ref{fig:kappa100_i1_BG}, but under initial condition (II), in which the
     initial conditions for the derivatives of the background fields are
     given by $m'_Q=\tilde{\chi}'=0$.} 
  \label{fig:kappa100_i0_BG}
 \end{center}
\end{figure}

The behavior of $m_Q$ changes drastically for the initial
condition (II).  The numerical results are shown in
Figure~\ref{fig:kappa100_i0_BG}. We find that $m_Q$ decreases rapidly
at the beginning. This is due to the
$2m_Q(1+m^2_Q)$ term on the left-hand side of Eq.\,\eqref{eq:ELeq_A_3}, which
reduces $m_Q$. On the other hand, $\tilde{\chi}'$ soon becomes positive and the term $\lambda m_Q^2\tilde{\chi}'$ contributes
in Eq.\,\eqref{eq:ELeq_A_3}. After that, $m_Q$ approaches $m_Q^{\rm
  st}$. The suppression of $m_Q$ leads to a smaller
value of $\bar{\rho}_A$, making the perturbative
limit easier to reach.  On the other hand, the backreaction terms are small and the system does not enter the strong backreaction regime.  

The derivatives of the background field and the slow-roll parameters
are presented in Figure~\ref{fig:kappa100_i0_dfields} in
Appendix~\ref{app:SR}, and are qualitatively similar to 
Figure~\ref{fig:kappa100_i1_dfields}.

Regarding the accuracy of the numerical calculation, we find that, for a brief moment, Eq.\,\eqref{eq:for_check_sol} is violated by $\order{10}\%$.
At that moment, both
sides of Eq.\,\eqref{eq:for_check_sol} are suppressed, 
causing the relative deviation between the right- and left-hand sides to increase. 
However, Eq.\,\eqref{eq:for_check_sol} cannot be used for the check in this case because it is derived by neglecting
the right-hand side of the equation of motion for $T_{L/R}$, which is also suppressed
by the slow-roll parameters.  After passing that point, we find that
Eq.\,\eqref{eq:for_check_sol} holds at the $\order{10^{-2}}$\% level.

\subsection{Perturbative limit}

\begin{figure}
 \begin{center}
   \includegraphics[width=7cm]{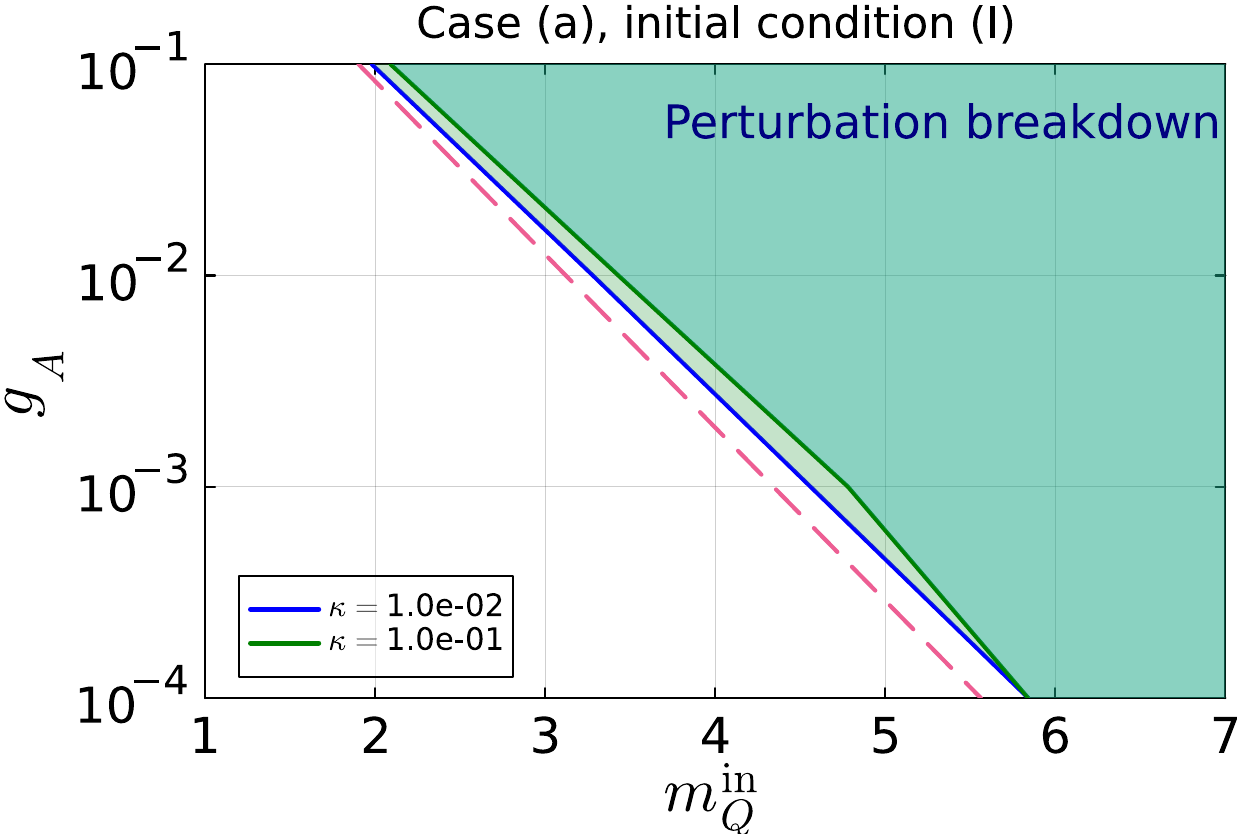}
   \caption{Perturbative limit on the $g_A$-$m_{Q}^{\rm in}$ plane for
     case (a) with $\lambda=10^2$ under initial condition (I). The perturbative calculation breaks down, i.e. $\delta \rho_A>\bar{\rho}_A$, in the shaded regions above the solid lines for $\kappa=10^{-2}$ and $10^{-1}$, as indicated in the legend.  The dashed line shows the upper limit of the stable solution
     found in Ref.\,\cite{Ishiwata:2021yne}. Below this line, the
     backreaction terms have a negligible effect on the differential
     equations.  }
  \label{fig:limit_a_i1}
 \end{center}
\end{figure}

Figure~\ref{fig:limit_a_i1} shows the perturbative limit for case (a) with $\kappa=10^{-2}$ and $10^{-1}$ under initial condition (I). The initial condition (II) yields similar results. The dashed line shows the
stable solution found in Ref.\,\cite{Ishiwata:2021yne}.
Below this line, the backreaction terms have a negligible effect on
the solutions, and the slow-roll approximation remains valid.  We find that the boundary of the perturbative limit nearly coincides with the bound of the strong backreaction. Therefore, the perturbative calculation breaks down when
the backreaction is strong. This is the main conclusion of this paper.

To further investigate the possible dependence on initial conditions, we computed the dynamics using different initial
conditions. For instance, a larger value of initial
$\tilde{\chi}'$, such as $1.5$--$2$ times larger than that given in
Eq.\,\eqref{eq:dchi},  induces strong
backreaction, while perturbativity holds for a certain value of $m_Q^{\rm in}$. However, even in this specific
case, the perturbative calculation breaks down when $m_Q^{\rm in}$ is a few \% larger. Therefore, our main conclusion is valid for case
(a), irrespective of the initial conditions within the percent-level uncertainty.

\begin{figure}
 \begin{center}
   \includegraphics[width=7cm]{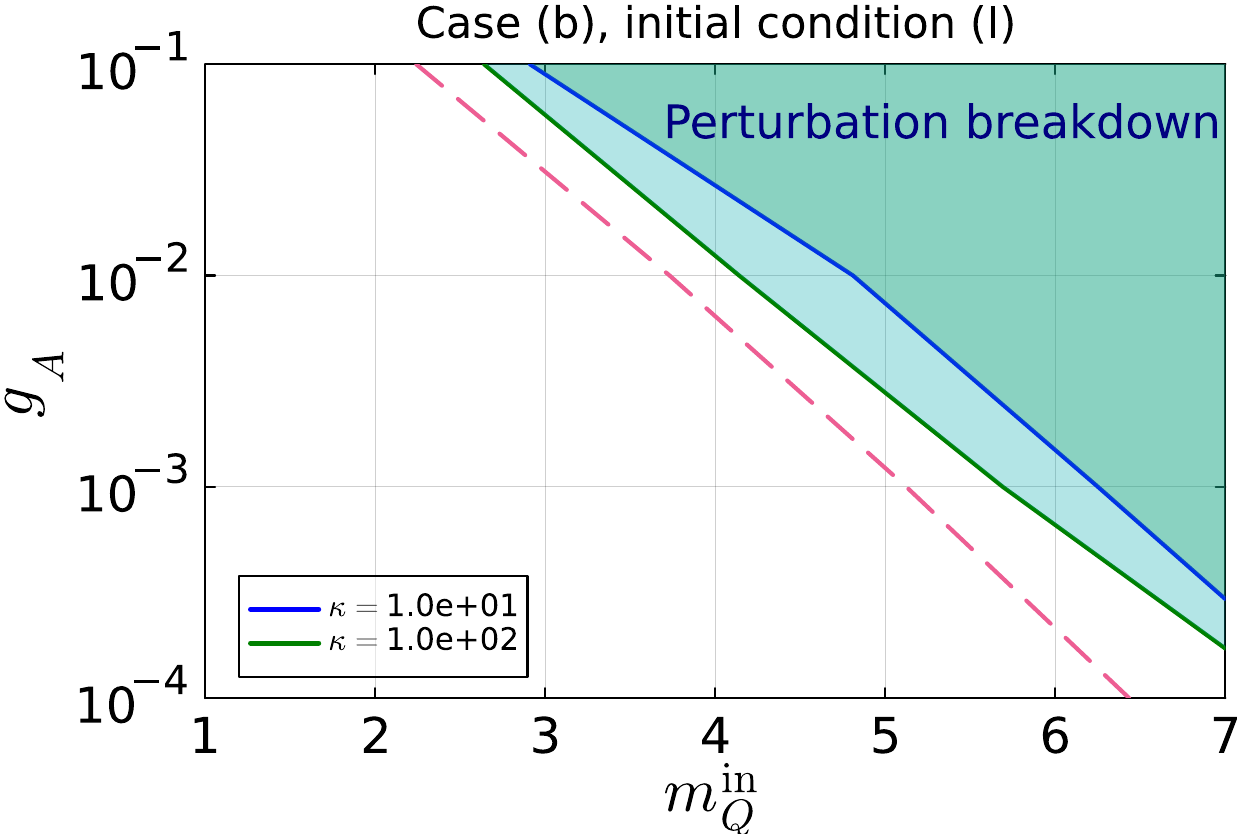}
   \includegraphics[width=7cm]{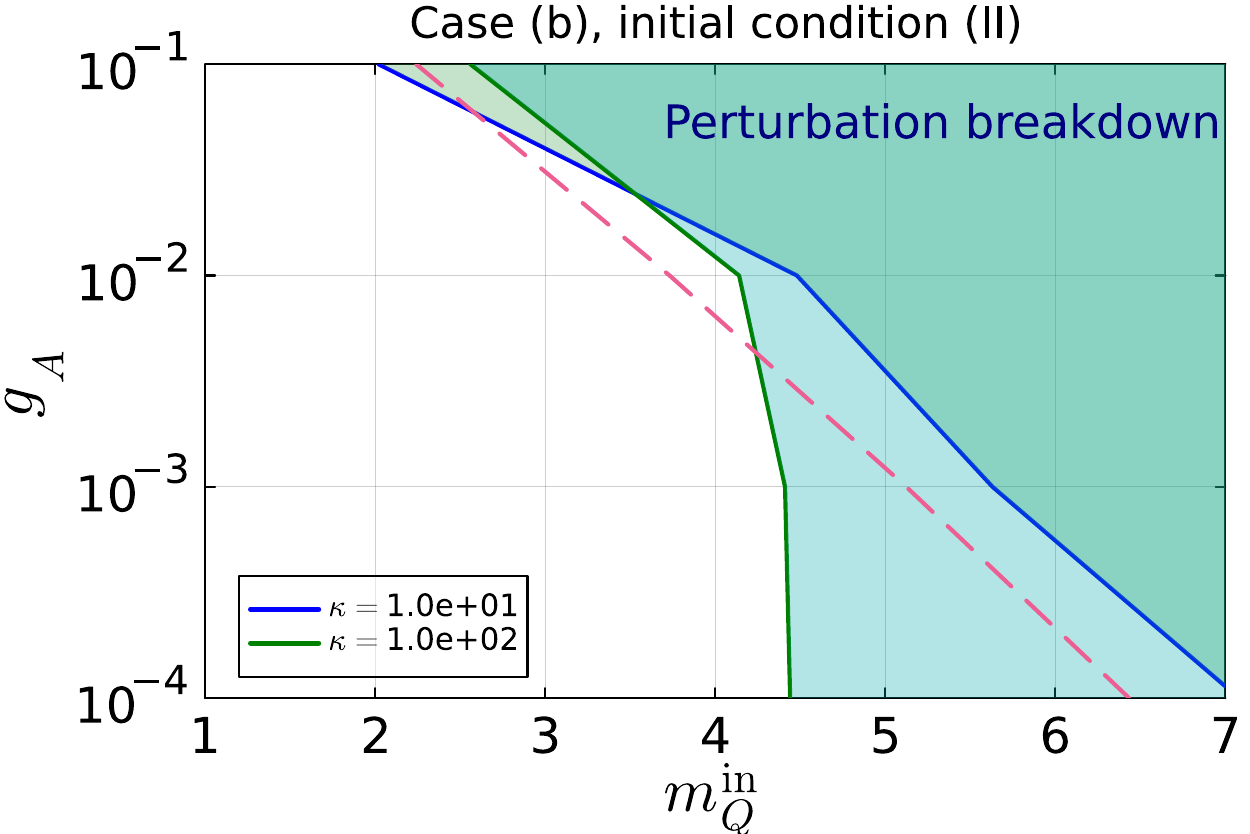}
   \caption{Same as Figure~\ref{fig:limit_a_i1}, but for case (b) under initial conditions (I) (left) and (II) (right). The shaded regions correspond to the breakdown of the perturbative calculation for $\kappa=10$ and $10^2$, as indicated in the legend.} 
  \label{fig:limit_b}
 \end{center}
\end{figure}

Figure~\ref{fig:limit_b} shows the perturbative limit for case
(b). Here we show the results under initial conditions (I) (left panel) and
(II) (right panel).
We also find that the perturbative limit is equivalent to the strong backreaction
regime for initial condition (I). This is because the dynamics of case (b) under initial condition (I) are  similar to those of case (a), as discussed in Section~\ref{sec:solultionB}. 

The situation under initial condition (II) with $\kappa=10$ is similar to that under initial condition (I).\footnote{Upon closer inspection, we find that the bound for $g_A=10^{-1}$ is slightly tighter. Under the initial condition with these specific 
parameters, $m_Q$ never reaches the stationary solution and relaxes to zero. Consequently, $\bar{\rho}_A$ is suppressed and the perturbative limit is reached more easily.} 
On the other hand, the perturbative limit for the case of $\kappa=10^2$
is more stringent when $g_A\le 10^{-2}$. 
This is due to the suppression of $m_Q$ in the
early stage of the dynamics (see the top left panel of Figure~\ref{fig:kappa100_i0_BG}), resulting in a smaller value of $\bar{\rho}_A$.  As a
result, the perturbative limit becomes tighter than the strong
backreaction.  For $g_A=10^{-1}$, however, the dynamics are
similar to those in Figure~\ref{fig:kappa100_i1_BG}, and the perturbative limit and the strong backreaction occur similarly to
the initial condition (I).

\subsection{Gravitational waves}
As seen in Eqs.\,\eqref{eq:T_1} and
\eqref{eq:psi_1}, the tensor modes of the gauge field, $T_{L,R}$, are linearly mixed with the tensor metric perturbations, $\psi_{L,R}$, which gives rise to the
production of primordial gravitational waves~\cite{Adshead:2013qp,Dimastrogiovanni:2012ew,Dimastrogiovanni:2016fuu}. The total gravitational wave power spectrum, ${\cal P}_T$,
is given by ${\cal P}_T={\cal P}_T^{\rm (vac)}+{\cal P}_T^{\rm (src)}$, where
\begin{align}
  {\cal P}_T^{\rm (vac)}&=\lim_{|k\tau|\to 0}\frac{2H^2k^3\tau^2}{\pi^2}
  (|\psi^{\rm int}_L|^2+|\psi^{\rm int}_R|^2)
  \simeq \frac{2H^2}{\pi^2}\,,\\
  {\cal P}_T^{\rm (src)}&=
  \lim_{|k\tau|\to 0}\frac{2H^2k^3\tau^2}{\pi^2}
  (|\psi^{\rm src}_L|^2+|\psi^{\rm src}_R|^2)\,.
\end{align}
Here, ${\cal P}_T^{\rm (vac)}$ is the vacuum contribution, 
while ${\cal P}_T^{\rm (src)}$ is the sourced contribution from the gauge tensor modes.  

\begin{figure}
 \begin{center}
   \includegraphics[width=7cm]{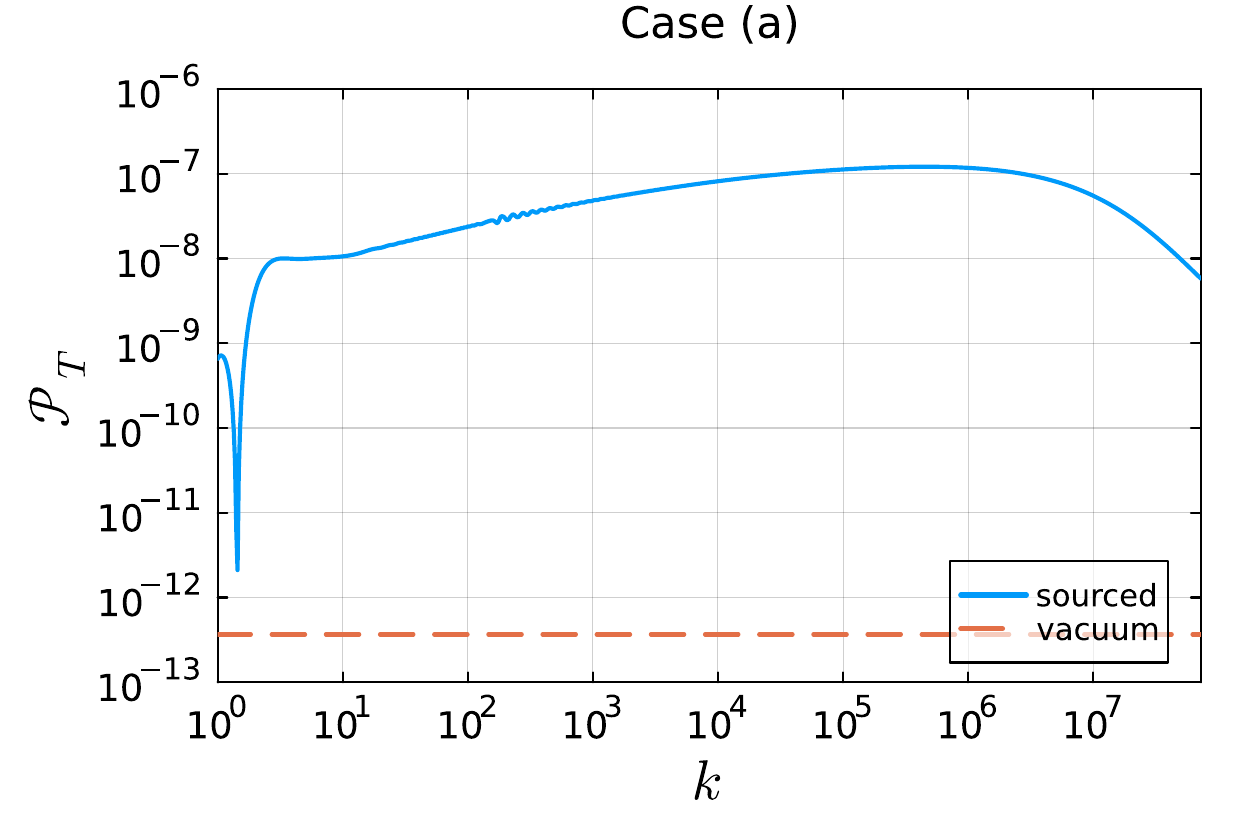}
   \includegraphics[width=7cm]{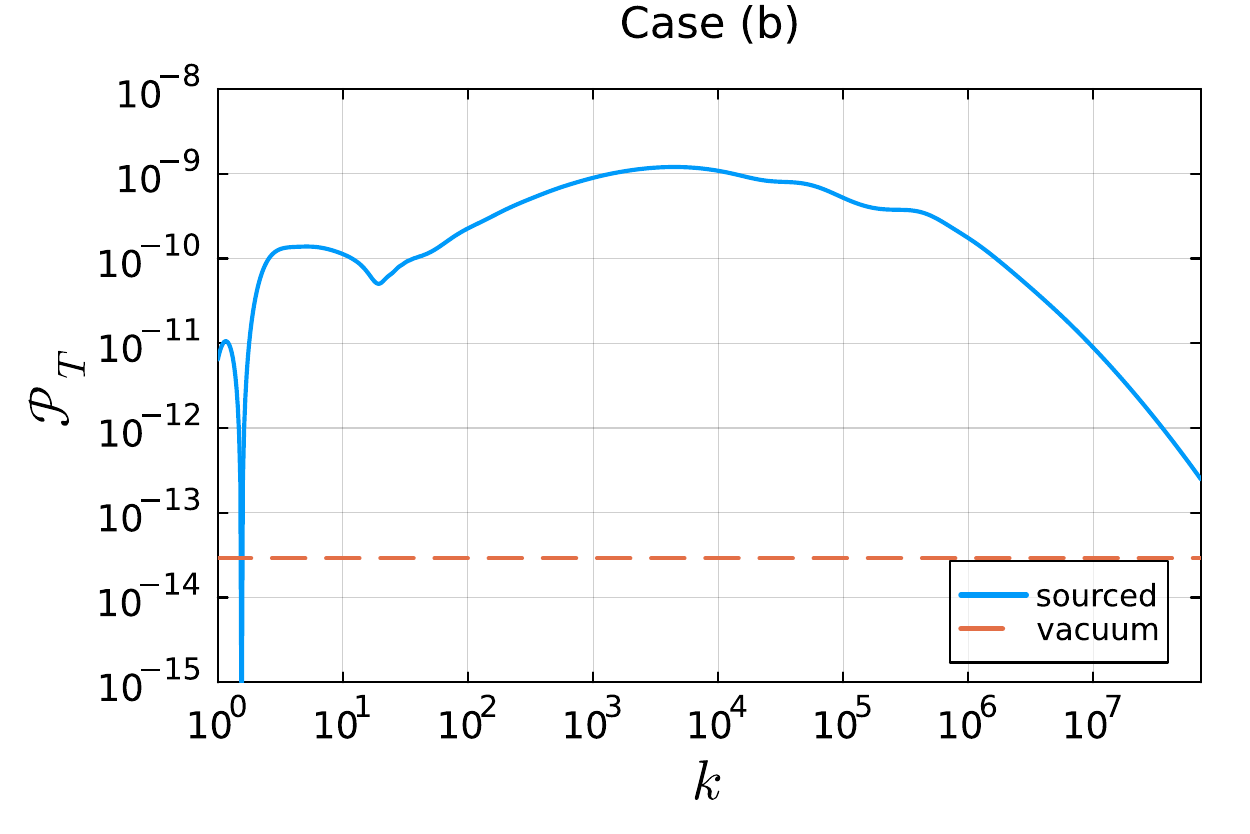}
   \caption{Gravitational waves sourced by the gauge field.  (Left)
     The parameters are the same as in Figure~\ref{fig:kappa001_i1_BG} with $m_Q^{\rm in}=4.51$. (Right) The parameters are the same as in
     Figure~\ref{fig:kappa100_i1_BG} with $m_Q^{\rm in}=5.63$. 
     The dashed lines show the vacuum contributions.}
  \label{fig:i1_GW}
 \end{center}
\end{figure}

The left and right panels of Figure~\ref{fig:i1_GW} show
the power spectra of primordial gravitational waves under initial
condition (I) for cases (a) and (b), respectively. 
The sourced component is dominated by $|\psi_R|^2$ and is much larger than the vacuum
one. We find that the maximum amplitude is $\order{10^{-7}}$ in the
perturbative region of case (a). 
This amplitude is much smaller than the $\order{10^{-3}\,\mathchar`-\,10^{-2}}$ required to explain the recent
observation of Pulsar Timing Array (PTA)~\cite{NANOGrav:2023gor,EPTA:2023fyk,Reardon:2023gzh,Xu:2023wog}, in agreement with Ref.\,\cite{Unal:2023srk}. However, if we could analyze the strong backreaction
of the axion-SU(2) dynamics beyond perturbation theory, the induced
gravitational waves might be amplified enough to explain the amplitude of the  stochastic gravitational waves reported by the PTA observations.

\section{Comparison to previous works}
Ref.\,\cite{Dimastrogiovanni:2024lzj} also studied the perturbative limit of the same system as given in Eq.~\eqref{eq:action}, focusing on the slow-roll dynamics of case (a). Their approach is based on the comparison between the tree-level and one-loop contributions to the two-point correlation function of the gauge field~\cite{Ferreira:2015omg}. In other words, they compared the relative importance of the first- and second-order perturbations in the equations of motion for the gauge field, while all previous work was based on the first-order perturbations. The perturbative limit is reached when the ratio of the one-loop contribution to the tree-level contribution exceeds unity, while in our study the perturbative limit is reached when $\delta\rho_A/\bar\rho_A$ exceeds unity. Their calculations include one-loop contributions from the self-interaction of the gauge fields as well as the non-linear interaction between the inhomogeneous axion and gauge fields. The latter is not addressed in our work, as discussed in Section~\ref{sec:spectator}. Their conclusion is similar to ours for case (a): the perturbative limit is reached when the system enters the strong reaction regime. In this paper, we have studied case (b) in addition to case (a) and found that the perturbative limit is reached even before the system enters the strong backreaction regime in some cases. Therefore, the perturbative limit depends on the configuration of the background fields of the system and it is not universally determined. It remains to be seen whether their approach yields a similar result for case (b).

On the technical side, we have solved the equations of motion,  Eqs.~\eqref{eq:ELeq_chi_3}, \eqref{eq:ELeq_A_3}, \eqref{eq:T_2}, and
\eqref{eq:psi_2}, simultaneously without using the
slow-roll approximation. In previous studies employing the slow-roll approximation (e.g., \cite{Dimastrogiovanni:2016fuu}), several terms were
neglected in the right-hand sides of Eqs.\,\eqref{eq:T_1} and \eqref{eq:psi_1}. 
In the evaluation of the
backreaction terms, we used the cutoff regularization to properly
take into account the contribution due to the tachyonic instability
of the gauge spin-2 components.
Our approach is similar to that of Ref.\,\cite{Garcia-Bellido:2023ser}, in which
the backreaction term was analyzed for the axion-U(1) system.
Ref.\,\cite{Iarygina:2023mtj} regularized the integral for case (a) by 
subtracting the Bunch-Davies-vacuum component from
the tensor modes. Their results are consistent with ours. 

Refs.~\cite{Iarygina:2023mtj} studied the strong backreaction regime of case (a) using perturbation theory. Both our results and those in Ref.\,\cite{Dimastrogiovanni:2024lzj} suggest that the perturbative approach is invalid in this regime. Therefore, it remains to be seen whether their quantitative results hold under non-linear, non-perturbative calculations.

\section{Conclusion}

We have investigated the limits of the perturbative treatment of axion-SU(2) gauge field dynamics during
inflation. We considered the spectator axion-gauge sector as given in Eq.\,\eqref{eq:action}~\cite{Dimastrogiovanni:2016fuu}.
The system's field contents include homogeneous axion and gauge fields, as well as the tensor
modes (spin-2 perturbations) of the gauge field and the metric tensor. 
The instability of the tensor modes of the gauge field results in the copious production of spin-2 particles during inflation, 
which backreacts on the dynamics of the background
fields. We find that the energy density of the tensor
modes, $\delta \rho_A$, grows to be as large as the background gauge field energy density, $\bar{\rho}_A$,
 leading to the breakdown of the perturbative calculation.

The perturbative limits depend on the initial configuration of the
background fields. 
We studied two background configurations, cases (a) [Eq.\,\eqref{eq:mQ_a}] and (b) [Eq.\,\eqref{eq:mQ_b}], based on the
slow-roll stationary solutions with negligible backreaction.  
Cases (a) and (b) were found in Refs.\,\cite{Adshead:2012kp} and \cite{Ishiwata:2021yne}, respectively.
For each case, 
we considered
two initial conditions, (I) and (II), for the derivatives of the
background fields: the slow-roll values and zero, respectively.  The dynamics are similar for both initial conditions in case (a), except for the
initial oscillatory behavior, and we found similar perturbative
limits for both. In contrast, the behavior of the background
gauge field changes significantly depending on the initial conditions in case (b). The
value of the background gauge field is suppressed under initial condition (II), which reduces its energy density. As a result, the perturbative limit becomes more stringent
than in (I).

We studied the correlation between the perturbative limits and the strong
backreaction regime, in which 
the absolute values of the backreaction terms,
$\tilde{\cal J}_A$ and $\tilde{\cal P}_\chi$, become larger than the
other dominant terms in the equations of motion. We found that the perturbative limit is reached in the strong
backreaction regime, 
except for certain
 initial configurations of the background fields. This can be qualitatively
understood as follows: the tachyonic instability of the tensor
modes enhances $\delta \rho_A$ and the backreaction terms almost
simultaneously. The exception is case (b) under initial condition
(II). In this case, while the backreaction terms are much smaller than
the other terms in the equations of motion, $\bar{\rho}_A$ is suppressed due to a smaller value of
the background gauge field, leading to $\delta\rho_A/\bar\rho_A>1$.

To make progress, we need non-perturbative approaches,
such as three-dimensional lattice simulations carried out for axion-U(1) systems \cite{Caravano:2022epk,Figueroa:2023oxc,Figueroa:2024rkr,Sharma:2024nfu,Jamieson:2025ngu}, which will help us understand the full dynamics
of the axion-SU(2) system during inflation and their observational consequences.

\section*{Acknowledgments}

We thank Takashi Hiramatsu, Ippei Obata, and Caner \"Unal for discussions in
the early stage of this project. We also thank Oksana Iarygina for
discussions regarding the backraction. This work was supported in
part by the Excellence Cluster ORIGINS which is funded by the Deutsche
Forschungsgemeinschaft (DFG, German Research Foundation) under
Germany’s Excellence Strategy: Grant No.~EXC-2094 - 390783311. The
Kavli IPMU is supported by World Premier International Research Center
Initiative (WPI), MEXT, Japan. The work of KI was supported by JSPS
KAKENHI Grant Number JP20H01894, JP25K07317, and JSPS Core-to-Core
Program Grant No. JPJSCCA20200002.

\appendix

\section{Energy momentum tensor}
\label{sec:T_munu}

We use the conservation of the energy-momentum tensor to verify the accuracy of the numerical solution of the differential
equations. 
We do not include the metric perturbations in this verification.
The result is given in Eq.~\eqref{eq:for_check_sol}.

The energy density is given by
\begin{align}
  \rho_\chi &= -T^\chi_{00}
  = \frac{\dot{\chi}^2}{2}+V_\chi\,,
  \\
  \rho_A &= -T^A_{00}
  = \frac{1}{2}a^{-2}(F^a_{0i})^2+\frac{1}{4}a^{-4}(F^a_{ij})^2\,,
\end{align}
where $T^\chi_{\mu\nu}$ and $T^A_{\mu\nu}$ are given in Eqs.~\eqref{eq:Tmunuchi} and \eqref{eq:TmunuA}, respectively, and
\begin{align}
  (F^a_{0i})^2 &=3\dot{{\cal Q}}^2+\dot{B}^2_{ij}\,,
  \\
  (F^a_{ij})^2 &=6g_A^2M^4_Q-2g_A^2{\cal Q}^2B^2_{ij}+(\delta F^a_{ij})^2\,.
\end{align}
 We do not distinguish between upper and lower Latin indices ($i,j,a$) and repeated indices are summed regardless of their location.
Here, we have introduced ${\cal Q}\equiv a Q$ and
\begin{align}
  (\delta F^a_{ij})^2 = (\partial_iB_{jk}-\partial_jB_{ik})^2
  +2g_A^2{\cal Q}^2B^2_{ij}-4g_A{\cal Q}\epsilon^{iqp}(\partial_iB_{jq})B_{jp}
  +\order{B_{ij}^3}\,.
\end{align}
We will omit the terms of order ${\cal O}(B_{ij}^3)$ and separate $T^A_{00}$
into contributions from the background field and the perturbations:
\begin{align}
  \bar{\rho}_A &= -T^A_{00}|_{\rm BG} = \frac{3}{2}a^{-2}\dot{{\cal Q}}^2
  +\frac{3}{2}a^{-4}g_A^2{\cal Q}^4\,,
  \label{eq:bar_rho_A}
  \\
  \delta \rho_A &= -T^A_{00}|_{B^2} = \frac{1}{2}a^{-2}\dot{B}^2_{ij}
  +\frac{1}{4}a^{-4}\left[(\partial_iB_{jk}-\partial_jB_{ik})^2
  -4g_A{\cal Q}\epsilon^{iqp}(\partial_iB_{jq})B_{jp}
  \right]\,.
  \label{eq:delta_rho_A}
\end{align}
With $\nabla_\mu
T^{A\,\mu}_{~~~0}=-a^{-4}\partial_0(a^4T_{00})+a^{-2}\partial_iT_{i0}$
and neglecting the total derivative term, we obtain
\begin{align}
  &\nabla_\mu T^{\chi\,\mu}_{~~~0} =
  \dot{\chi}(\ddot{\chi}+3H\dot{\chi}+V'_\chi)\,,
    \\
  & \nabla_\mu T^{A\,\mu}_{~~~\,0}|_{\rm BG} = 3\dot{{\cal Q}}a^{-1}
    \left[a^{-1}\ddot{{\cal Q}}+Ha^{-1}\dot{{\cal Q}}
      +2g_A^2Q^3\right]\,,
    \\
   & \nabla_\mu T^{A\,\mu}_{~~~\,0}|_{B^2} =
    a^{-5}\partial_\tau B_{ij}
    \left[\partial_\tau^2B_{ij}-\partial_i^2 B_{ij}
      +2g_A{\cal Q}\epsilon^{kjp}\partial_kB_{ip}
      \right]
    +a^{-5}g_A \partial_\tau {\cal Q}\epsilon^{iqp}(\partial_iB_{jq})B_{jp}\,.
\end{align}

To check the energy conservation, we rewrite the above equations 
by using the equations of motion. To do so, we need the equation of motion for $B_{ij}$. We derive the
Euler-Lagrange equation from the action without metric
perturbations. For convenience, we use the conformal time, $\tau$,
instead of the physical time, $t$:  
\begin{align}
  S|_{B^2} 
  =\int d\tau d^3x \,a^4({\cal L}_A
    +{\cal L}_{\rm CS})|_{B^2}\,,
\end{align}
where
\begin{align}
  a^4{\cal L}_A|_{B^2} &= \frac{1}{2}(\partial_\tau B_{ij})^2
  -\frac{1}{4}\left[
    (\partial_iB_{jk}-\partial_jB_{ik})^2
  -4g_A{\cal Q}\epsilon^{iqp}(\partial_iB_{jq})B_{jp}
  \right]\,,
  \\
  a^4{\cal L}_{\rm CS}|_{B^2}&=
  -\frac{\lambda \chi}{2f}
  \left[-g_A(\partial_\tau {\cal Q}) B^2_{ij}
    +2(\partial_\tau B_{ka})\epsilon^{ijk}\partial_iB_{ja}
    -2g_A{\cal Q}(\partial_\tau B_{ij}) B_{ij}\right]\,.
\end{align}
Then, the Euler-Lagrange equation is
\begin{align}
  (\partial_\tau^2-\partial_i^2)B_{ij}
  +2g_A{\cal Q}\epsilon^{kjp}\partial_k B_{ip}
  +\frac{\lambda \partial_\tau \chi}{f}(\epsilon^{kip}\partial_k B_{jp}+
  g_A{\cal Q}B_{ij})=0\,.
  \label{eq:ELeq_Bij}
\end{align}
Using the equations of motion, i.e., Eqs.\,\eqref{eq:ELeq_chi_2},
\eqref{eq:ELeq_A_2}, and \eqref{eq:ELeq_Bij}, we find 
\begin{align}
 & \nabla_\mu T^{\chi\,\mu}_{~~~0} =\dot{\chi}
  \left[-3a^{-1}\frac{g_A\lambda}{f}Q^2\dot{\cal Q}
    +{\cal P}_\chi\right]\,,
  \\
  &\nabla_\mu T^{A\,\mu}_{~~~\,0}|_{\rm BG} = 3\dot{{\cal Q}}a^{-1}
  \left[\frac{g_A\lambda}{f} \dot{\chi}Q^2-{\cal J}_A
    \right]\,,
  \label{eq:drho_A_BG}
  \\
  &\nabla_\mu T^{A\,\mu}_{~~~\,0}|_{B^2} = -\dot{\chi}{\cal P}_\chi
  +3a^{-1}\dot{{\cal Q}}{\cal J}_A\,,
  \label{eq:drho_A_B2}
\end{align}
which leads to the conservation of the total energy-momentum tensor:
\begin{align}
  \nabla_\mu T^{\chi\,\mu}_{~~~0}+
  \nabla_\mu T^{A\,\mu}_{~~~\,0}|_{\rm BG}+
  \nabla_\mu T^{A\,\mu}_{~~~\,0}|_{B^2}=0\,.
\end{align}

We use Eq.\,\eqref{eq:drho_A_BG} to verify the accuracy of the numerical solution. We rewrite it using $y=-\ln (-\tau)=\ln aH$ as the time
variable:
\begin{align}
  \frac{3}{2}H^3a^{-4}\dv{y}\left[a^4(\epsilon_{Q_E}+\epsilon_{Q_B})\right]
  =3H^3\sqrt{\epsilon_{Q_E}}\left[
    \lambda  \sqrt{\epsilon_{Q_B}}\tilde{\chi}'
    -\frac{H}{g_A}\tilde{{\cal J}}_A\right]\,,
\end{align}
where $\epsilon_{Q_E}$ and $\epsilon_{Q_B}$ are defined in Eq.~\eqref{eq:miscvariables} and $\tilde{\cal J}$ is defined in Eq.~\eqref{eq:misc}.
Integrating both sides, we get
\begin{align}
  \frac{1}{2}(\epsilon_{Q_E}+\epsilon_{Q_B})(y)
  -\frac{1}{2}(\epsilon_{Q_E}+\epsilon_{Q_B})(0)\frac{a^4(0)}{a^4(y)}
  =\frac{1}{a^4(y)}\int_0^ydy'
  \left[\sqrt{\epsilon_{Q_B}}\tilde{\chi}-\frac{H}{g_A}\tilde{{\cal J}}_A\right]\,.
  \label{eq:for_check_sol}
\end{align}
We verify the accuracy of our numerical solution by computing both sides of Eq.\,\eqref{eq:for_check_sol} at each step of
$y$.

\section{Time derivative of the Hubble parameter}
\label{sec:Hdot}

The time derivative of the Hubble parameter is given by
\begin{align}
  \dot{H} = -\frac{1}{2}(\rho+P)\,,
\end{align}
where $\rho=\rho_\phi+\rho_\chi+\rho_A$ is the total energy density
and $P=P_\phi+P_\chi+P_A$ is the total pressure. Using
\begin{align}
  \rho_\phi=\frac{\dot{\phi}^2}{2}+V_\phi\,,~~
  P_\phi=\frac{\dot{\phi}^2}{2}-V_\phi\,,\\
  \rho_\chi=\frac{\dot{\chi}^2}{2}+V_\chi\,,~~
  P_\chi=\frac{\dot{\chi}^2}{2}-V_\chi\,,\\
  \rho_A=3P_A\,,
\end{align}
we obtain Eq.~\eqref{eq:Hdot}.

However, this result does not agree with that given in 
Refs.\,\cite{Dimastrogiovanni:2024xvc,Dimastrogiovanni:2025snj}, where $\dot{H}$ includes ${\cal J}_A$. 
Therefore, we rederive Eq.~\eqref{eq:Hdot} by taking the
derivative of both sides of $3H^2=\rho_\phi+\rho_\chi+\rho_A$,
\begin{align}
  \dot{H}=\frac{1}{6H}(\dot{\rho}_\phi+\dot{\rho}_\chi+\dot{\rho}_A)\,,
\end{align}
and show that the backreaction terms cancel in the final result.

Using the equation of motion, $\dot{\rho}_\phi$ and $\dot{\rho}_\chi$ are
given by
\begin{align}
  \dot{\rho}_\phi &= -3H\dot{\phi}^2\,,
  \\
  \dot{\rho}_\chi &= -3H\dot{\chi}^2+\dot{\chi}
  \left[-3a^{-1}\frac{g_A\lambda}{f}Q^2\dot{\cal Q}
    +{\cal P}_\chi\right]\,.
\end{align}
Using $\nabla_\mu
T^{A\,\mu}_{~~~0}=-a^{-4}\partial_0(a^4T_{00})+a^{-2}\partial_iT_{i0}$
and Eqs.\,\eqref{eq:drho_A_BG} and \eqref{eq:drho_A_B2}, $\dot{\rho}_A$ is given by
\begin{align}
  \dot{\rho}_A = -4H\rho_A + \dot{\chi}
  \left[3a^{-1}\frac{g_A\lambda}{f}Q^2\dot{\cal Q}
    -{\cal P}_\chi\right]\,.
\end{align}
Combining all of the above, the backreaction terms cancel, and the derivative of the Hubble parameter is obtained as
\begin{align}
  \dot{H} = -\left[
    \frac{1}{2}\dot{\phi}^2+
    \frac{1}{2}\dot{\chi}^2+\frac{2}{3}\rho_A\right]\,,
  \label{eq:Hdot_res}
\end{align}
where $\rho_A=\bar{\rho}_A+\delta\rho_A$, in agreement with Eq.\,\eqref{eq:Hdot}.

In terms of the slow-roll
parameters, it is given by
\begin{align}
\nonumber
  -\frac{\dot{H}}{H^2}&=
  \frac{\dot{\phi}^2}{2H^2}+
  \frac{\dot{\chi}^2}{2H^2}+
  \frac{2}{3H^2}\rho_A
  \\
  &=\epsilon_\phi+\epsilon_\chi+\epsilon_{Q_E}+\epsilon_{Q_B}
  +\frac{2}{3H^2}\delta \rho_A\,,
\end{align}
where $\delta \rho_A$, given in Eq.\,\eqref{eq:delta_rho_A}, is
calculated from the tensor perturbations as [Eq.~\eqref{eq:rhoAint}]
\begin{align}
  \delta \rho_A =
  \frac{1}{2a^2}
  \sum_{h=L,R} \int  \frac{d^3k}{(2\pi)^3}
  \left[|\dot{T}_h|^2+\left(\frac{k^2}{a^2}
+2m_QH\frac{c_hk}{a}\right)|T_h|^2
    \right]\,.
\end{align}

\section{Initial conditions for tensor modes}
\label{app:ELeq_tensor}

To solve the differential equations of the tensor modes, i.e.,
Eqs.\,\eqref{eq:T_2} and \eqref{eq:psi_2}, we adopt the procedure
shown in Ref.\,\cite{Fujita:2018ndp}. Considering the tensor field
as a linear combination of intrinsic and sourced components,
the equations can be decomposed as 
\begin{align}
  & T^{{\rm int}\,\prime\prime}_{L,R}+T^{{\rm int}\,\prime}_{L,R}
  +\tau^2\Omega_T^2T^{{\rm int}}_{L,R}
  =  2\delta_{Q1}\psi^{{\rm src}\,\prime}_{L,R}
  +2\left[\sqrt{\epsilon_{Q_{B}}}(m_Q\mp k\tau)
     -(\delta_{Q1}+\delta_{Q2})
      \right]\psi^{\rm src}_{L,R} \,,
   \label{eq:T_int}
    \\
    & \psi^{{\rm int}\,\prime\prime}_{L,R}+\psi^{{\rm int}\,\prime}_{L,R}
    +\tau^2\Omega_\psi^2\psi^{\rm int}_{L,R}
    =-2\delta_{Q1} T^{{\rm src}\,\prime}_{L,R}
    +2\sqrt{\epsilon_{Q_B}}
    \left[m_Q\mp k\tau \right]T^{\rm src}_{L,R} \,,
    \label{eq:psi_int}\\
     & T^{{\rm src}\,\prime\prime}_{L,R}+T^{{\rm src}\,\prime}_{L,R}
  +\tau^2\Omega_T^2T^{{\rm src}}_{L,R}
  =  2\delta_{Q1}\psi^{{\rm int}\,\prime}_{L,R}
  +2\left[\sqrt{\epsilon_{Q_{B}}}(m_Q\mp k\tau)
     -(\delta_{Q1}+\delta_{Q2})
      \right]\psi^{\rm int}_{L,R} \,,
   \label{eq:T_src}
    \\
    & \psi^{{\rm src}\,\prime\prime}_{L,R}+\psi^{{\rm src}\,\prime}_{L,R}
    +\tau^2\Omega_\psi^2\psi^{\rm src}_{L,R}
    =-2\delta_{Q1} T^{{\rm int}\,\prime}_{L,R}
    +2\sqrt{\epsilon_{Q_B}}
    \left[m_Q\mp k\tau \right]T^{\rm int}_{L,R} \,.
    \label{eq:psi_src}
\end{align}
Here, `int' and `src' stand for the intrinsic and sourced ones,
respectively.  We assume that the sourced components are zero at the
short-wavelength limit and that they are induced via the interactions described
by the right-hand sides. The intrinsic ones, on the other
hand, are assumed to have a Bunch-Davies vacuum at the short
wavelengths. 

The solution for $T_{L,R}$ is given by neglecting
the right-hand side of the equations and assuming that $m_Q$ and $\xi$ are
constant; $(1/\sqrt{2k})i^\beta W_{\beta,\alpha}(2ik\tau)$ where
$W_{\beta,\alpha}$ is the Whittaker function,
$\alpha=-i\sqrt{2m_Q\xi-1/4}$, and $\beta=\pm i(m_Q+\xi)$ for left- and
right-handed fields~\cite{Adshead:2013qp,Maleknejad:2013npa,Bhattacharjee:2014toa,Obata:2014loa,Obata:2016tmo}. Here we use the WKB solution in the
sub-horizon limit, i.e. $T_{L,R}(-k\tau \to
\infty)=(1/\sqrt{2k})(-2k\tau)^\beta e^{-ik\tau}$. On the other
hand, the analytic solutions for $\psi_{L,R}$ are well-known and are given by
$(1/\sqrt{2k})(1-i/k\tau )e^{-ik\tau}$ \cite{Starobinsky:1979ty}, neglecting the
right-hand sides. Using these solutions, we determine the initial conditions
 for the differential equation at $y=0$ as
\begin{align}
  T_{L,R}^{\rm int}(0)&=
  \frac{1}{\sqrt{2k}}i^\beta W_{\beta,\alpha}(2ik\tau)\,,\\
  \psi_{L,R}^{\rm int }(0)&=\frac{1}{\sqrt{2k}}
  \left(1-\frac{i}{k\tau}\right)e^{-ik\tau}\,,\\
  T_{L,R}^{\rm src}(0)&=0\,,\\
  \psi_{L,R}^{\rm src}(0)&=0\,,
\end{align}
for each $k$. The values of $\alpha$ and $\beta$ are determined by the initial
values of the dynamical variables, $m_Q$ and $\xi$.

\section{Cutoff regularization}
\label{app:cutoff}

\begin{figure}
  \begin{center}
    \includegraphics[width=7.5cm]{
      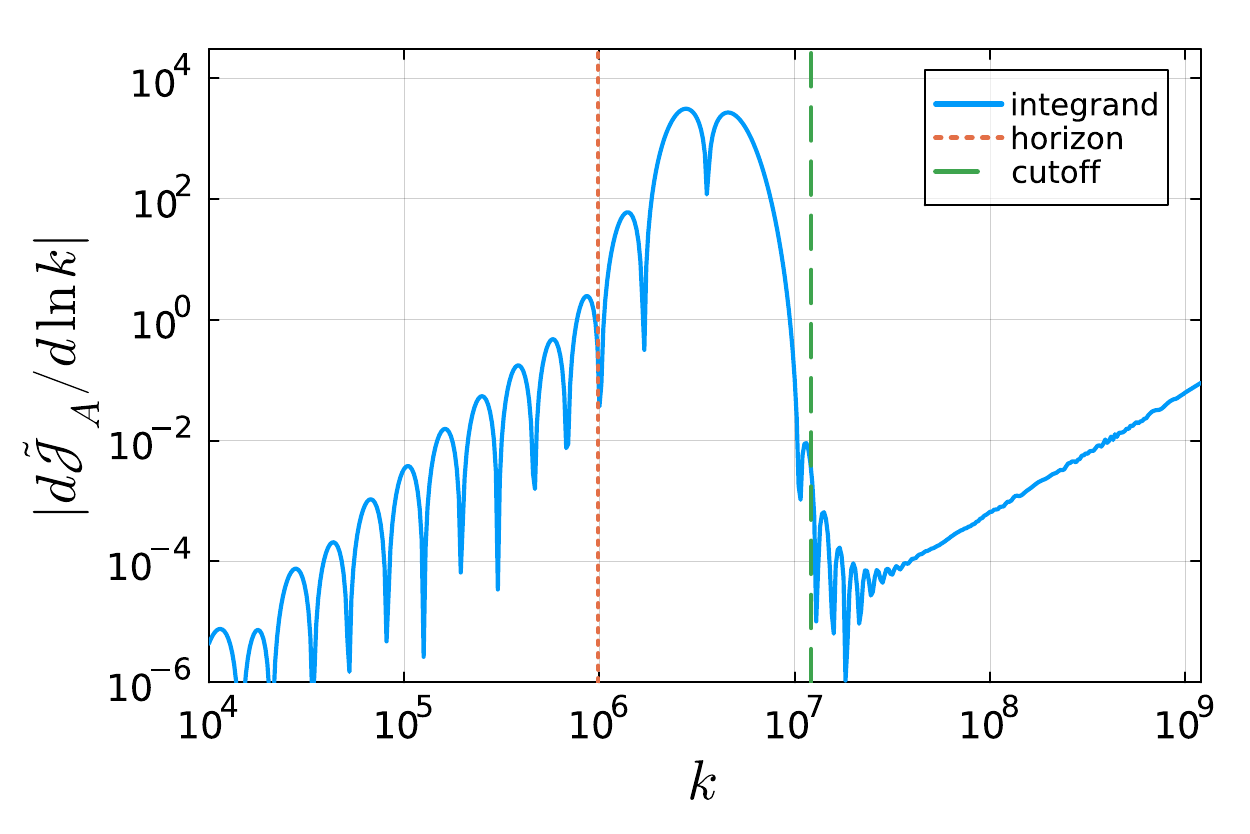}
    \includegraphics[width=7.5cm]{
      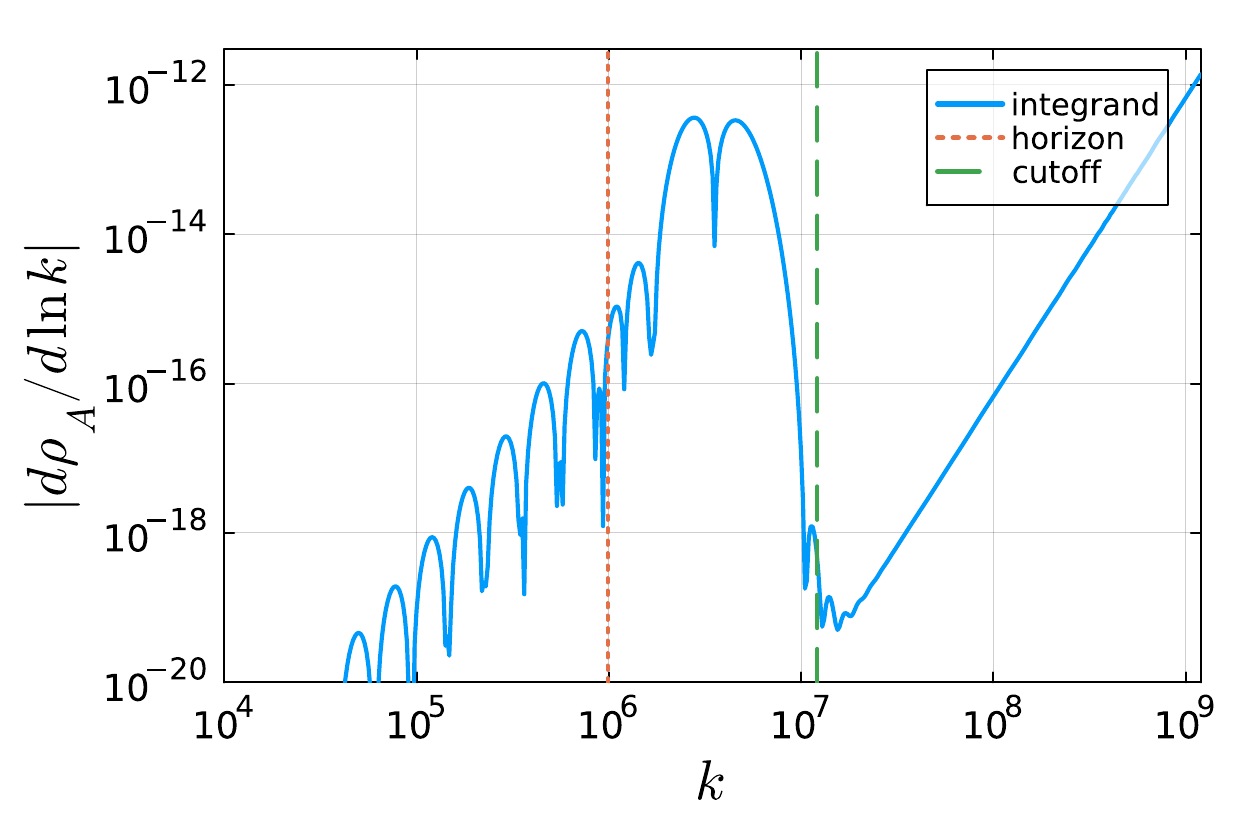}
    \caption{Absolute values of the integrand of
      $\tilde{\cal J}_A$, $d\tilde{\cal J}_A/d\ln k$ (left), and the integrand of $\delta
      \rho_A$, $d\delta\rho_A/d\ln k$ (right), as a function of $k$ in units of $aH$ at $y=0$. The input parameters are the
      same as in Figure~\ref{fig:kappa001_i1_BG}, but $m_Q^{\rm in}=4.56$ and
      $y=13.8$ are taken. The comoving scale that exits
      the horizon (dotted, `horizon') and the cutoff scale (dashed,
      `cutoff') are also shown.  }
  \label{fig:kappa001_i1_dJd-rho}
 \end{center}
\end{figure}

To verify that the $k$-integral is executed properly with the cutoff,
$k_{\rm cut}$, given in Eq.\,\eqref{eq:kcut}, we present the integrands of
$\tilde{\cal J}_A$ and $\delta \rho_A$. The
integrands are defined in the logarithmic scale as
\begin{align}
  \tilde{\cal J}_A &= \int d\ln k \frac{d\tilde{\cal J}_A}{d\ln k}\,, \\
  \delta \rho_A &= \int d\ln k \frac{d\delta\rho_A}{d\ln k}\,.
\end{align}
Figure~\ref{fig:kappa001_i1_dJd-rho} shows $|d\tilde{\cal J}_A/d\ln k|$
and $|d\delta\rho_A/d\ln k|$ for case (a) under initial condition
(I).  The parameters are the same as in Figure~\ref{fig:kappa001_i1_BG}
and we choose $m_Q=4.56$. We show the results at $y=13.8$, which is just before the system reaches the perturbative limit. The
cutoff scale, $k_{\rm cut}$, is shown by the vertical dashed lines. For
reference, the comoving scale that exits the horizon at
that time, $k=aH\,(=e^y)$, is shown by the vertical dotted lines; the modes with $k$ smaller than this scale are outside
the horizon.  

We find that the modes between $aH$ and $k_{\rm cut}$ dominate the integral. 
The modes with $k>k_{\rm cut}$ correspond to the Bunch-Davies
vacuum and should not be included in $\tilde{\cal J}_A$ or $\delta \rho_A$. The figure
clearly shows that $k_{\rm cut}$ properly eliminates the
Bunch-Davies vacuum contribution. Note that Ref.\,\cite{Iarygina:2023mtj} adopts a
subtraction scheme in which the Bunch-Davies
vacuum contribution is subtracted from the integrand. 
Both approaches yield similar results.

\section{Slow-roll approximation}
\label{app:SR}

In this section, we verify the accuracy of the slow-roll approximation, i.e.,
$\tilde{\chi}''=m''_Q=\dot{H}=0$ and
$\delta_{Q1}=-(\delta_{Q1}+\delta_{Q2})=\sqrt{\epsilon_{Q_E}}$.

\begin{figure}
  \begin{center}
    \includegraphics[width=7.5cm]{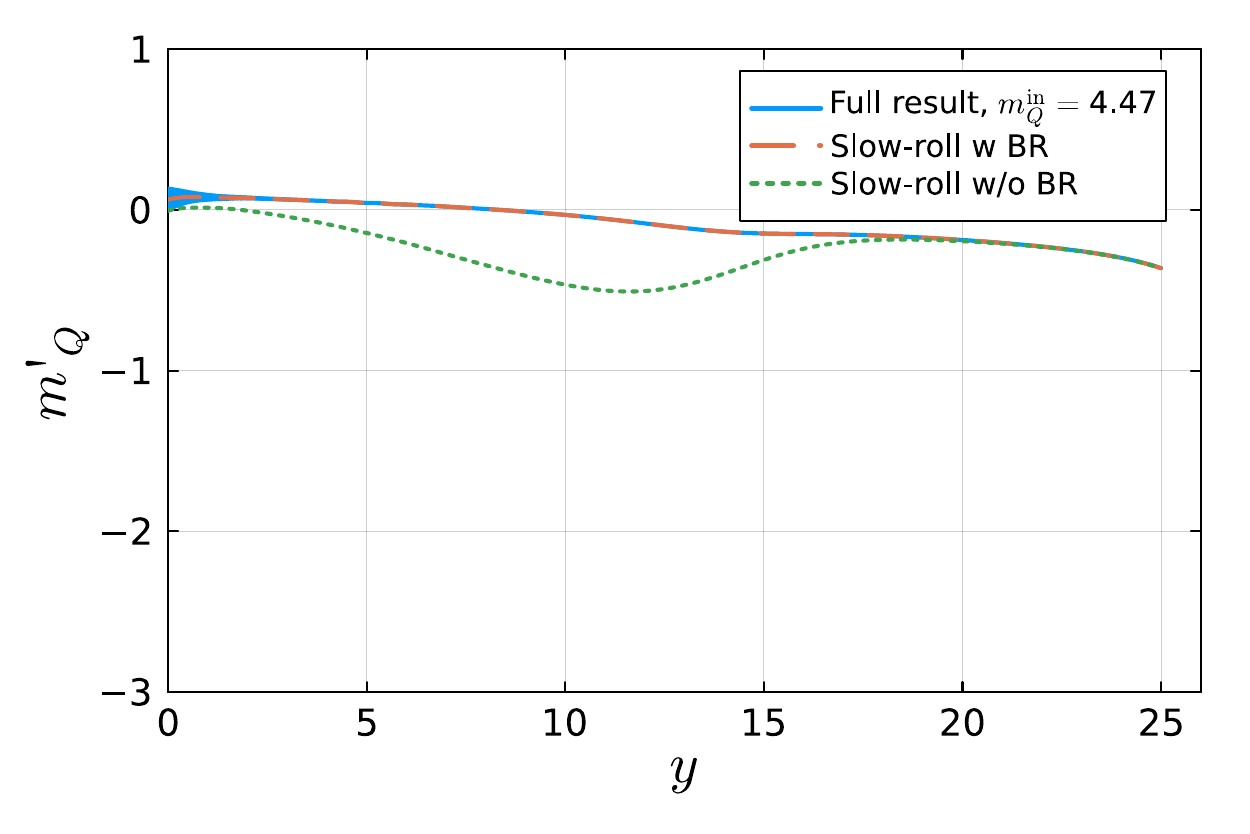}
    \includegraphics[width=7.5cm]{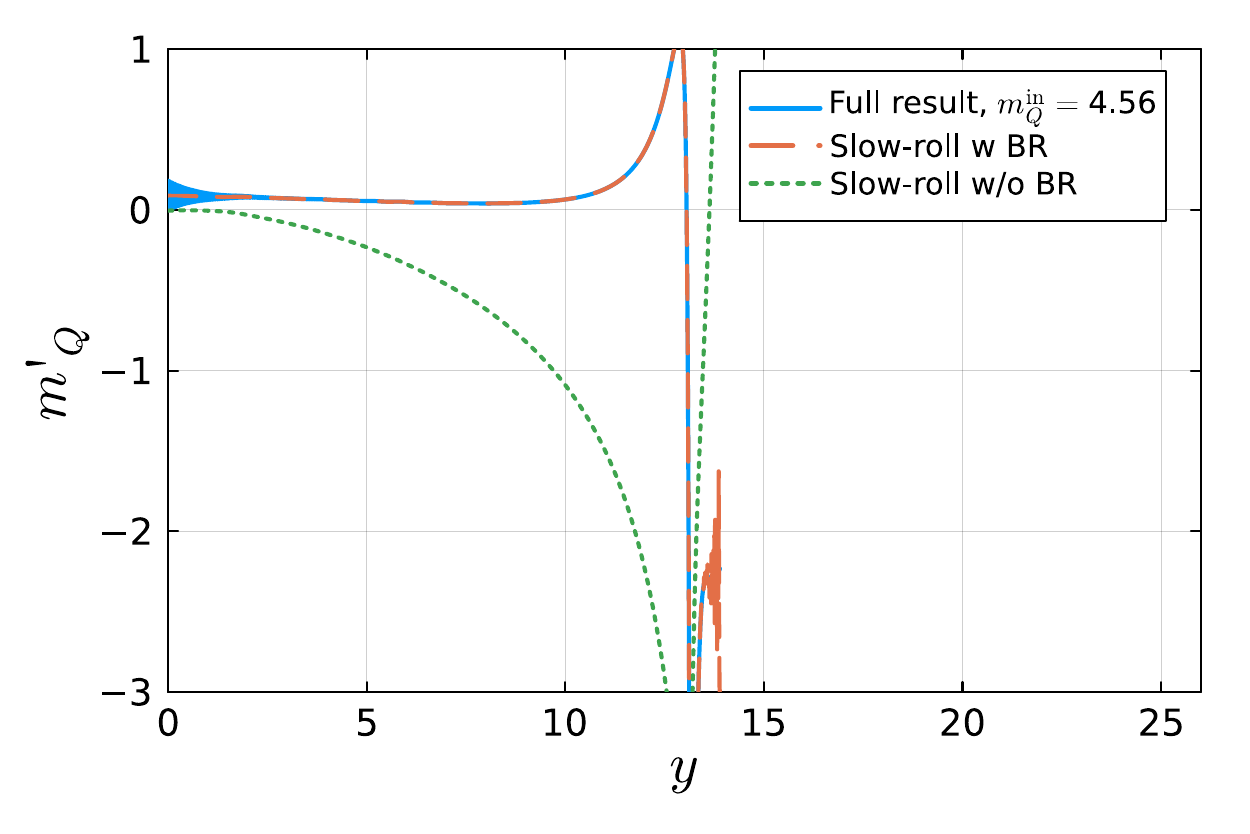}
    \includegraphics[width=7.5cm]{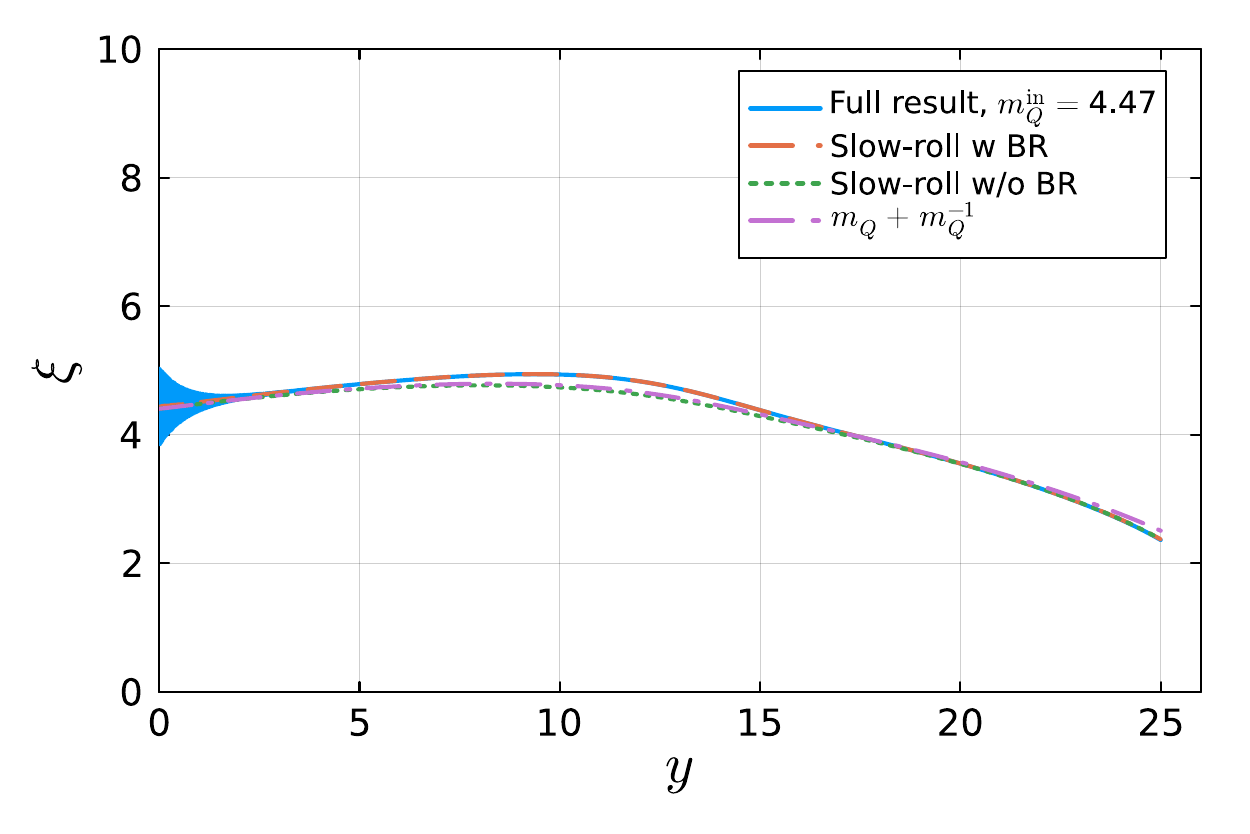}
    \includegraphics[width=7.5cm]{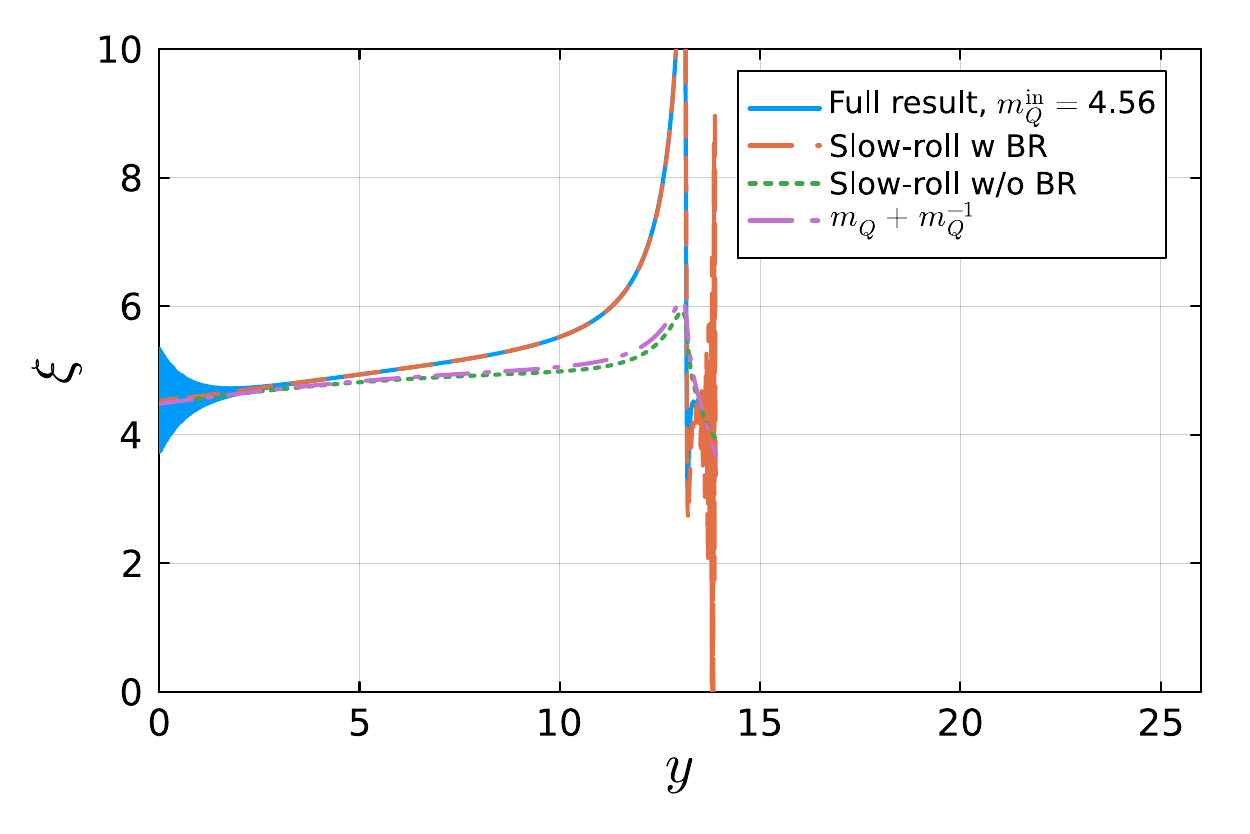}
    \includegraphics[width=7.5cm]{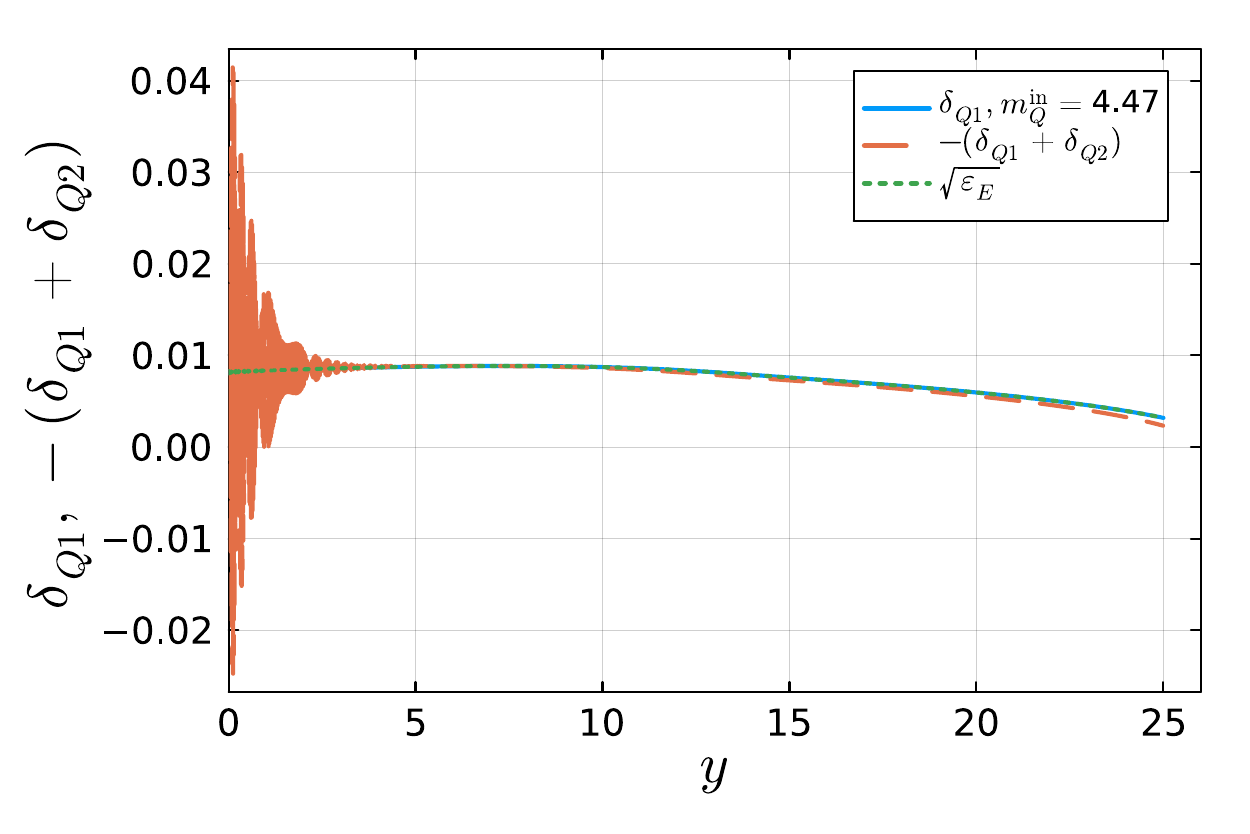}
    \includegraphics[width=7.5cm]{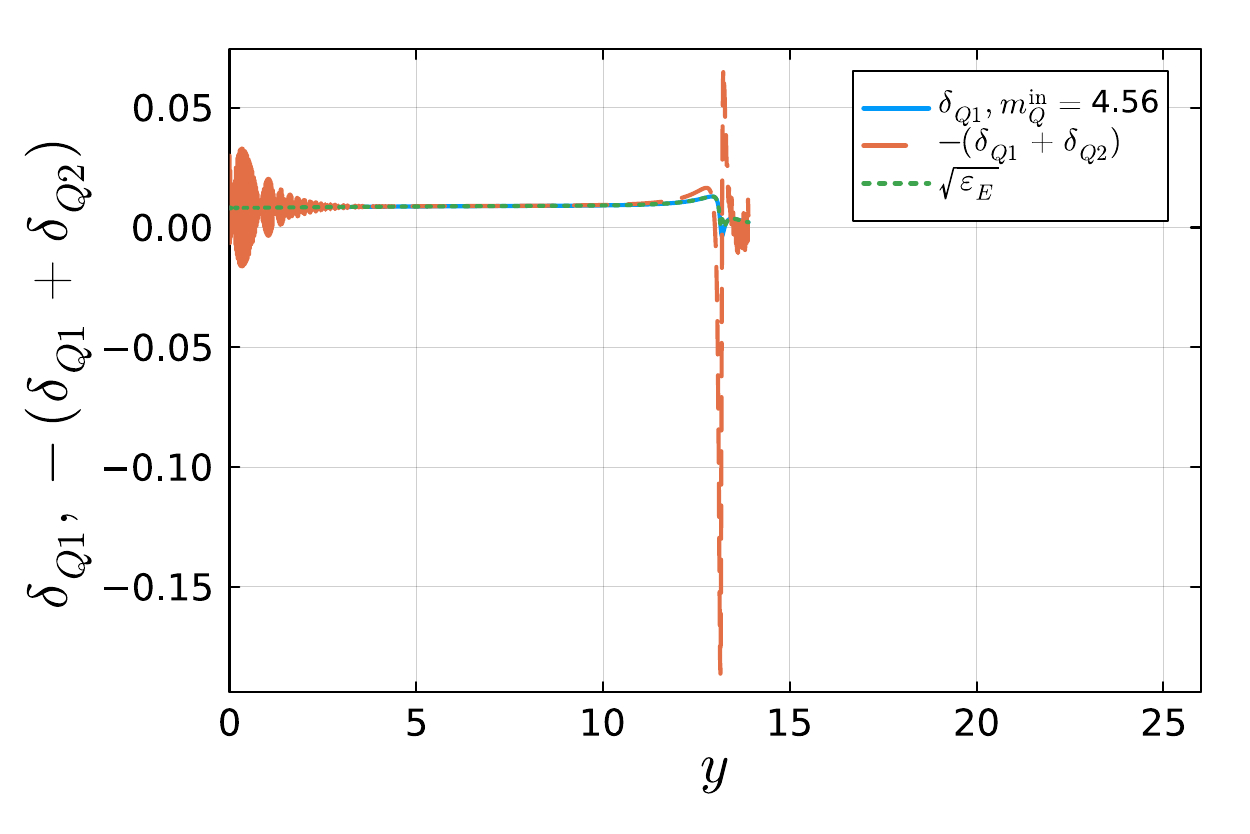}
    \caption{Time evolution of $m'_Q$ (top), $\xi$ (middle), and
      $\delta_{Q1}$ and -($\delta_{Q1}+\delta_{Q2}$) (bottom), 
      as a function of $y$. The parameters
      are the same as in Figure~\ref{fig:kappa001_i1_BG}, i.e., case (a) 
      under initial condition (I), but taking $m_Q^{\rm
        in}=4.47$ (left) and $4.56$ (right). `Full result' shows the
      numerical solutions to the integro-differential equations. `Slow-roll w BR' shows Eqs.\,\eqref{eq:dxiA} and \eqref{eq:dchi} with $m_Q$, $\tilde{\chi}$, and the tensor
      modes on the right-hand sides from the numerical solutions. 
      `Slow-roll w/o BR' shows the results obtained by
      setting $\tilde{\cal J}_A=\tilde{\cal P}_\chi=0$. The dot-dashed lines in the middle panels show $\xi\simeq m_Q+m_Q^{-1}$, while the dotted lines in the bottom panels show $\delta_{Q1}+\delta_{Q2}\simeq-\sqrt{\epsilon_{Q_E}}$.}
  \label{fig:kappa001_i1_dfields}
 \end{center}
\end{figure}

Figure~\ref{fig:kappa001_i1_dfields} shows the results for $m_Q^{\rm
  in}=4.47$ and $4.56$ in the left and right panels, respectively. 
The solid lines in the top four panels (`Full result') show the derivatives of the
background fields, $m'_Q$, $\xi \,(=\lambda \tilde{\chi}'/2)$,
$\delta_{Q1}$, and $-(\delta_{Q1}+\delta_{Q2})$, for case (a) under initial
condition (I), which is associated with the dynamics shown in
Figure~\ref{fig:kappa001_i1_BG}. We have verified that $|\dot{H}|/H^2\lesssim
\order{10^{-3}}$ is satisfied.
The dashed lines (`Slow-roll w BR') and the dotted lines (`Slow-roll w/o BR') show $m'_Q$ and $\xi$ given in Eqs.\,\eqref{eq:dxiA} and \eqref{eq:dchi}, in which the numerical solutions are used for the quantities on the right-hand sides 
with and without the backreaction (BR)
terms, respectively. We find that the full results and the slow-roll results with BR agree well for $m_Q^{\rm
  in}=4.47$ (left panels), 
except for the earlier stage of the
dynamics, where the $m_Q''$ and $\tilde{\chi}''$ terms give the
oscillatory behavior. However, the slow-roll results without
BR deviate from them. This means that the
BR terms contribute to $m'_Q$ and $\tilde{\chi}'$ even
before the strong BR regime; thus, the BR terms should always be
included. We find similar results for $m_Q^{\rm
  in}=4.56$ (right panels). The slow-roll results with BR deviate from the full results 
  near the perturbative limit. 

\begin{figure}
  \begin{center}
    \includegraphics[width=7.5cm]{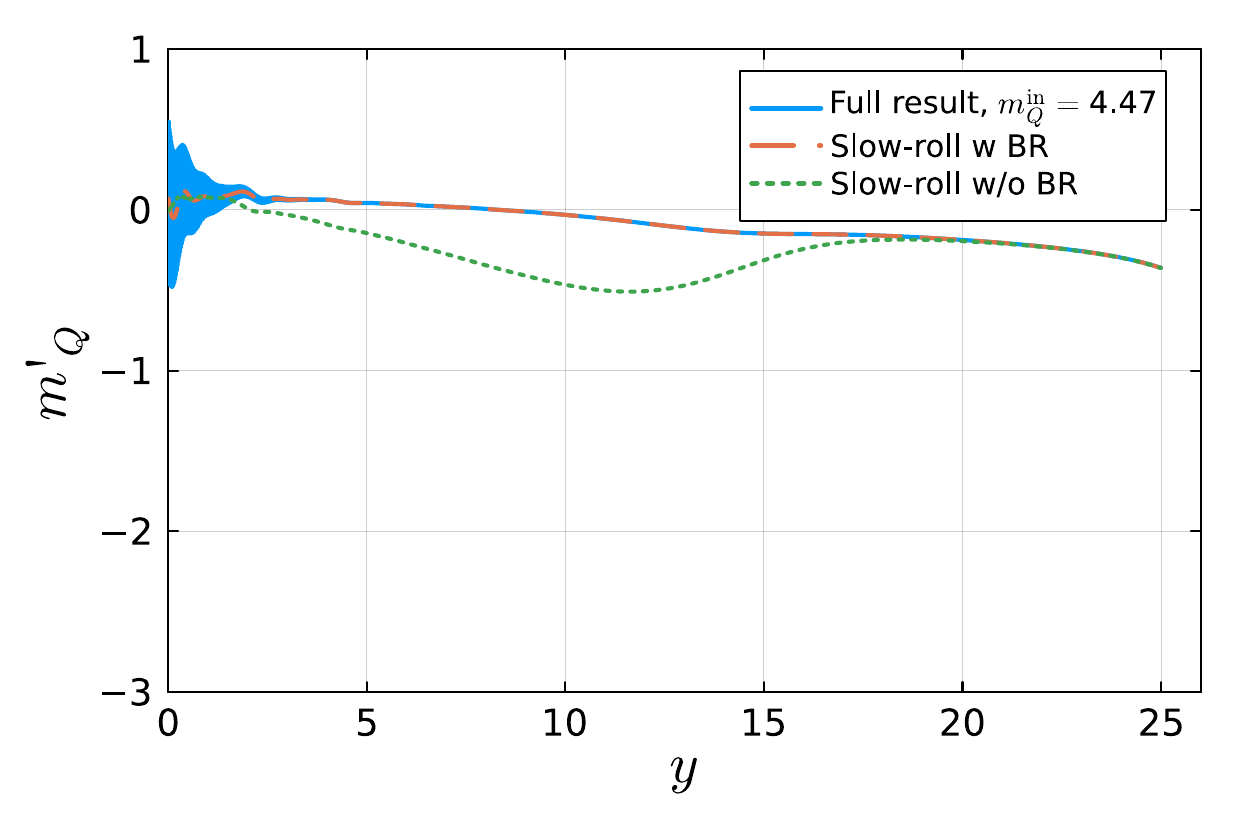}
    \includegraphics[width=7.5cm]{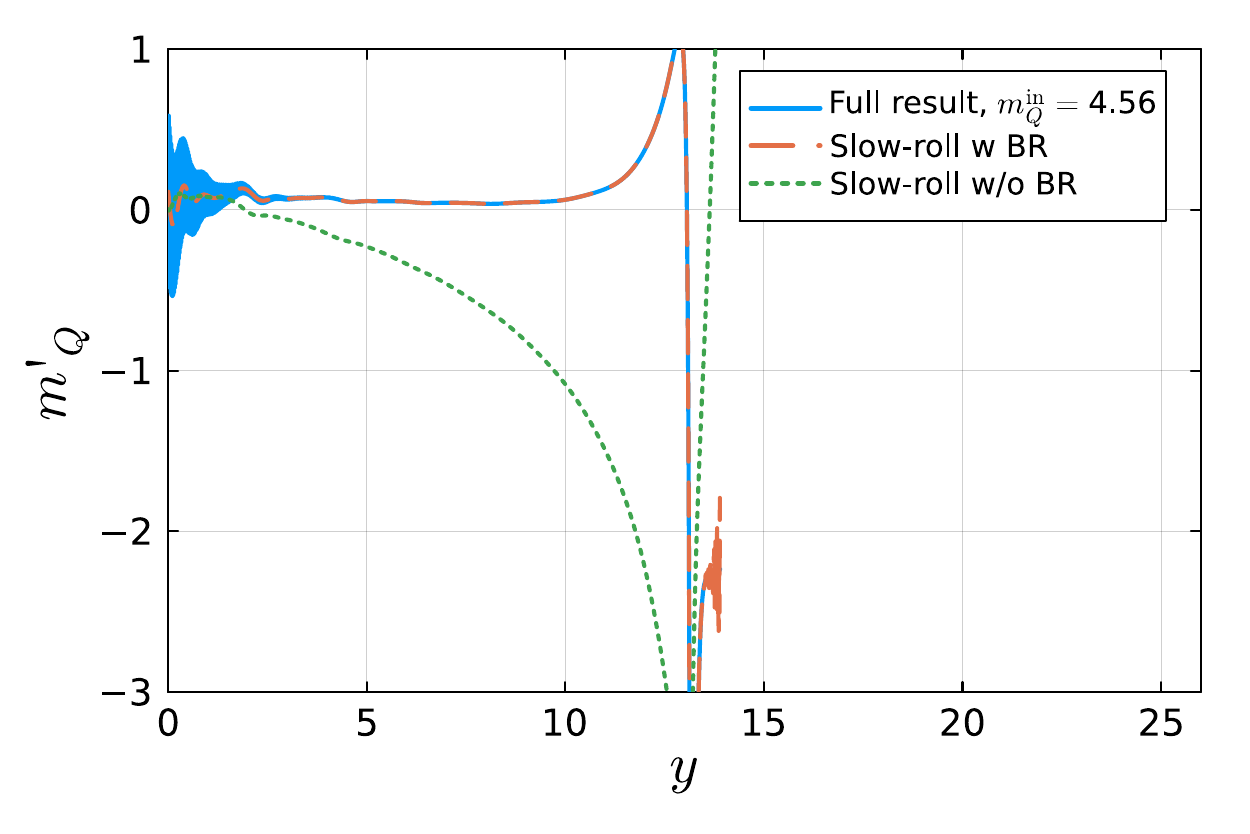}
    \includegraphics[width=7.5cm]{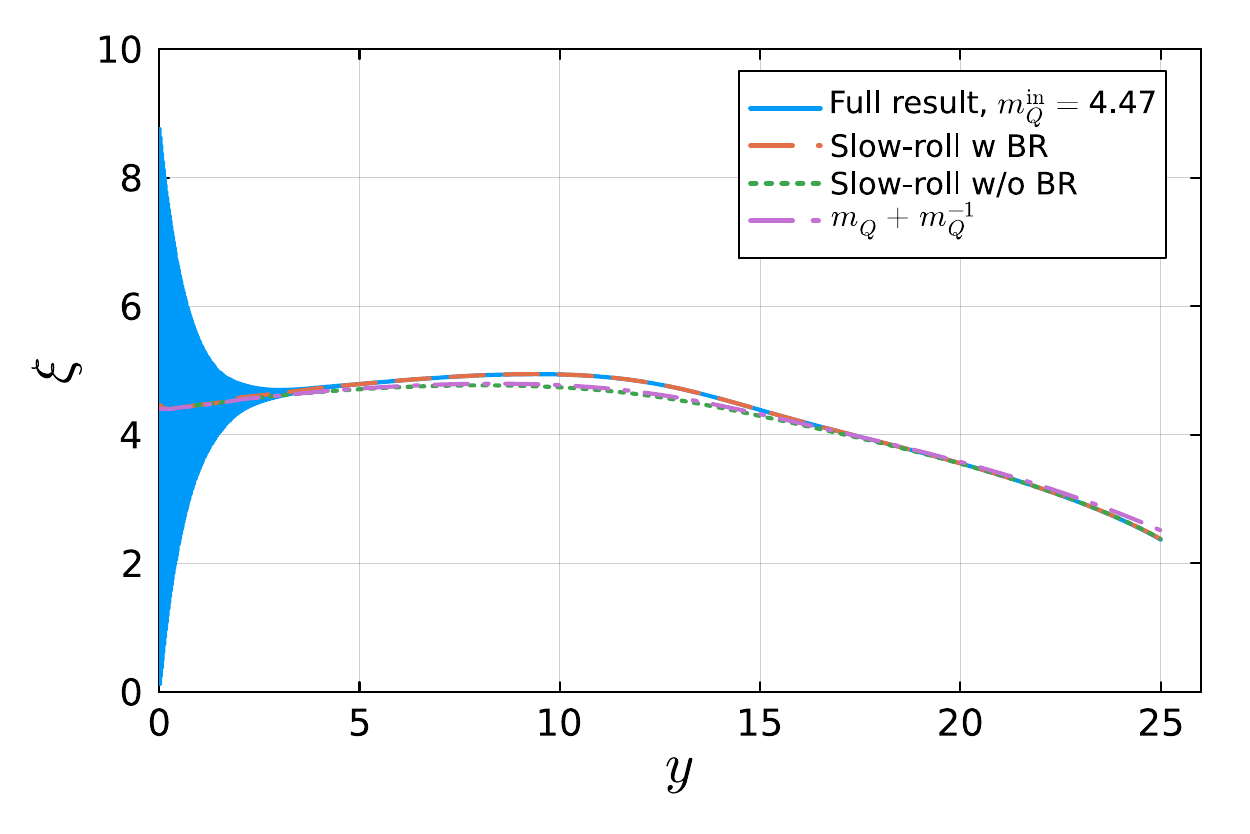}
    \includegraphics[width=7.5cm]{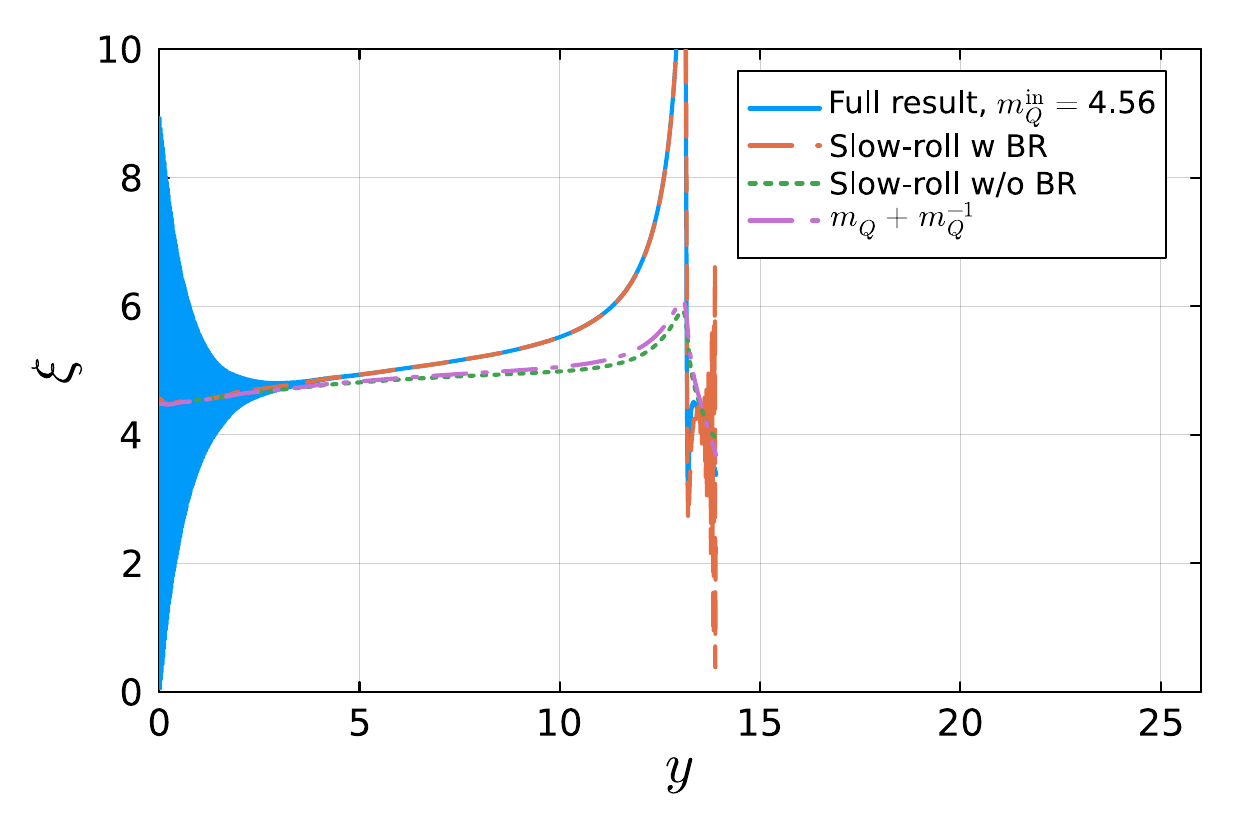}
    \includegraphics[width=7.5cm]{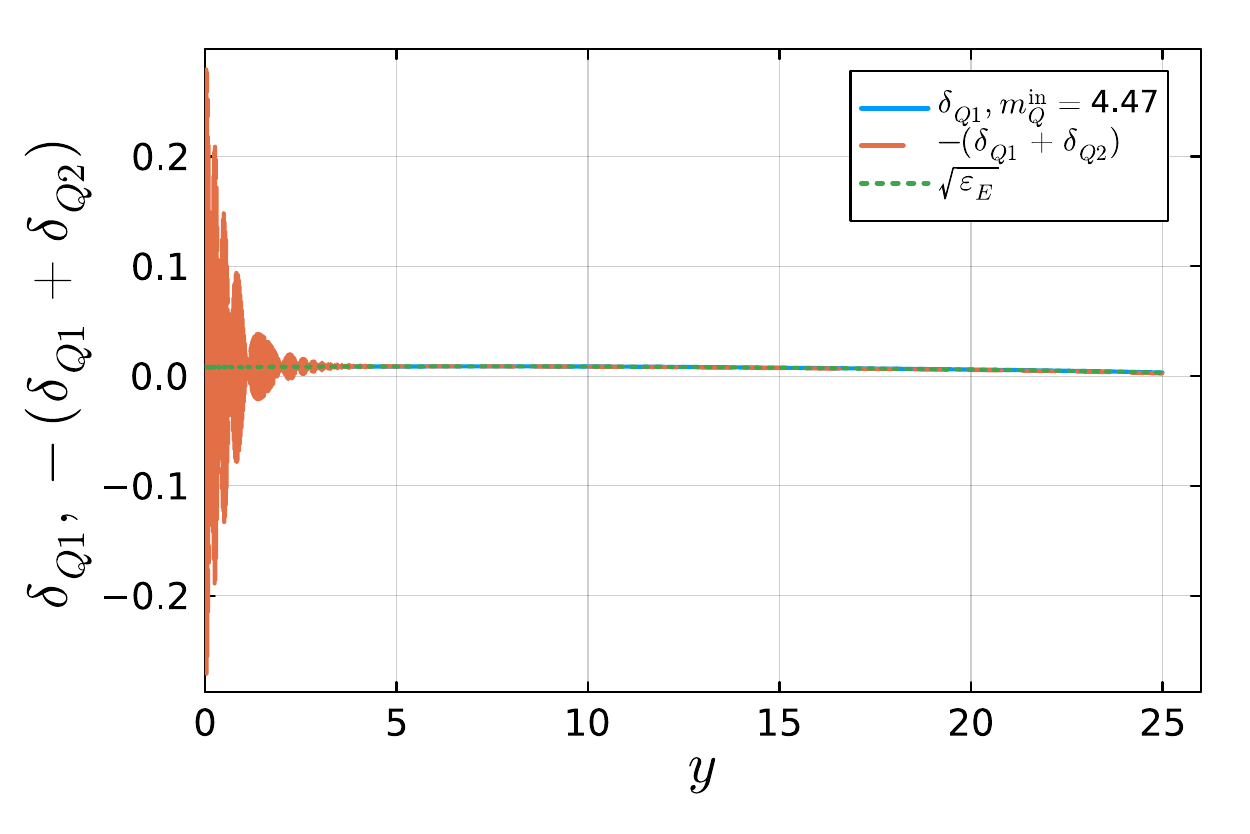}
    \includegraphics[width=7.5cm]{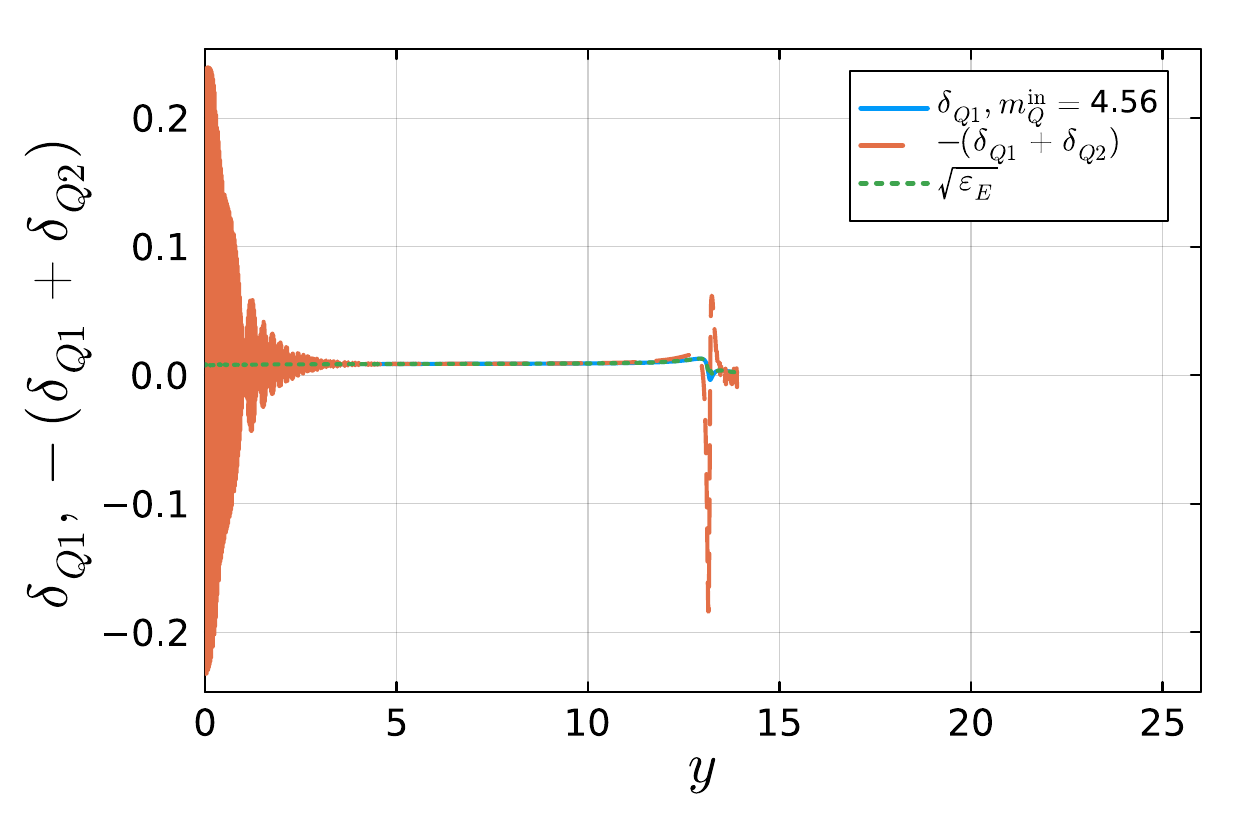}
    \caption{Same as Figure~\ref{fig:kappa001_i1_dfields}, but under
       initial condition (II), in which the
     initial conditions for the derivatives of the background fields are
     given by $m'_Q=\tilde{\chi}'=0$.  }
  \label{fig:kappa001_i0_dfields}
 \end{center}
\end{figure}

The bottom two panels of Figure~\ref{fig:kappa001_i1_dfields} show $\delta_{Q1}$ (solid) and $-(\delta_{Q1}+\delta_{Q2})$ (dashed). We find that
$-(\delta_{Q1}+\delta_{Q2}) \simeq \sqrt{\epsilon_{Q_E}}$ (dotted) is a good
approximation, except for the earlier stage of the dynamics and near
the perturbative limit. In summary, the slow-roll
approximation, $m_Q''=\tilde{\chi}''=0$ and
$\delta_{Q1}+\delta_{Q2}=-\sqrt{\epsilon_{Q_E}}$, cannot describe the oscillatory behavior and the dynamics 
near the perturbative limit.
We find similar results under initial condition (II) (Figure~\ref{fig:kappa001_i0_dfields}).

\begin{figure}
  \begin{center}
    \includegraphics[width=7.5cm]{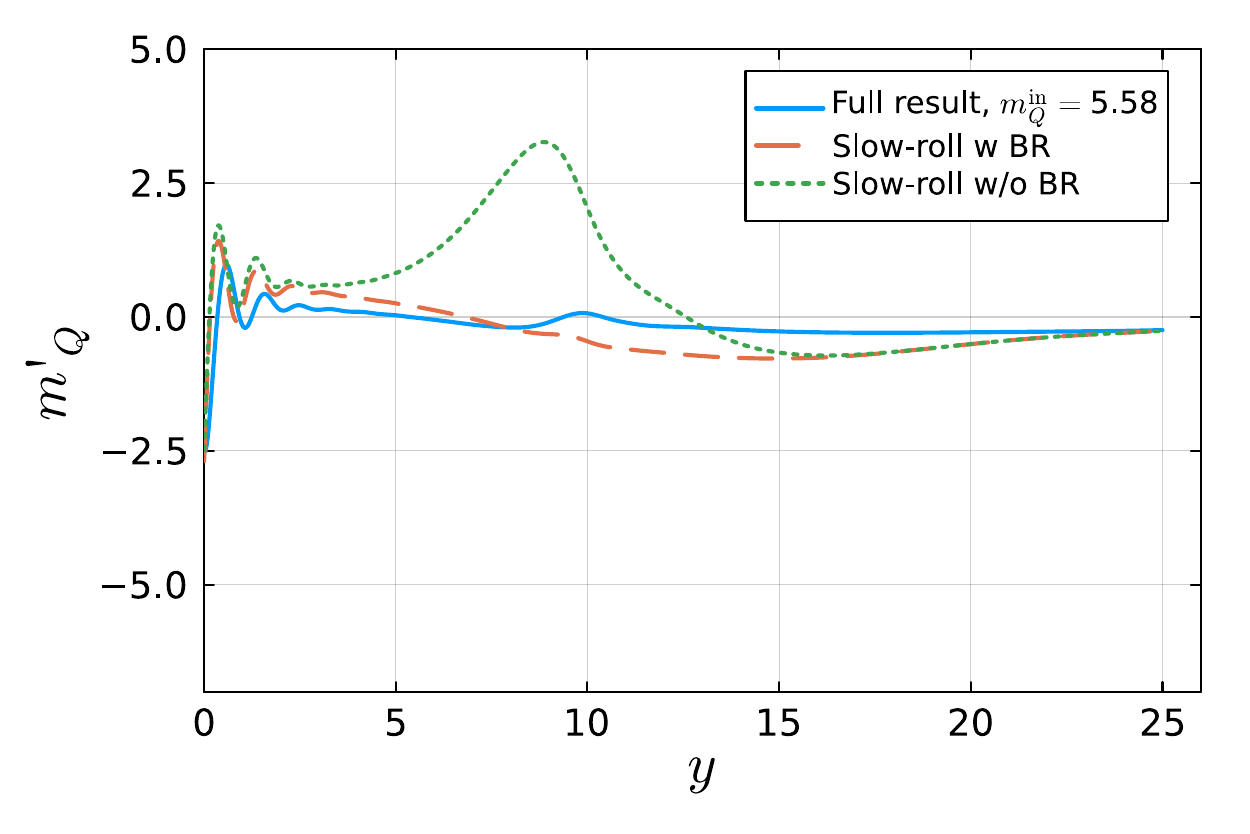}
    \includegraphics[width=7.5cm]{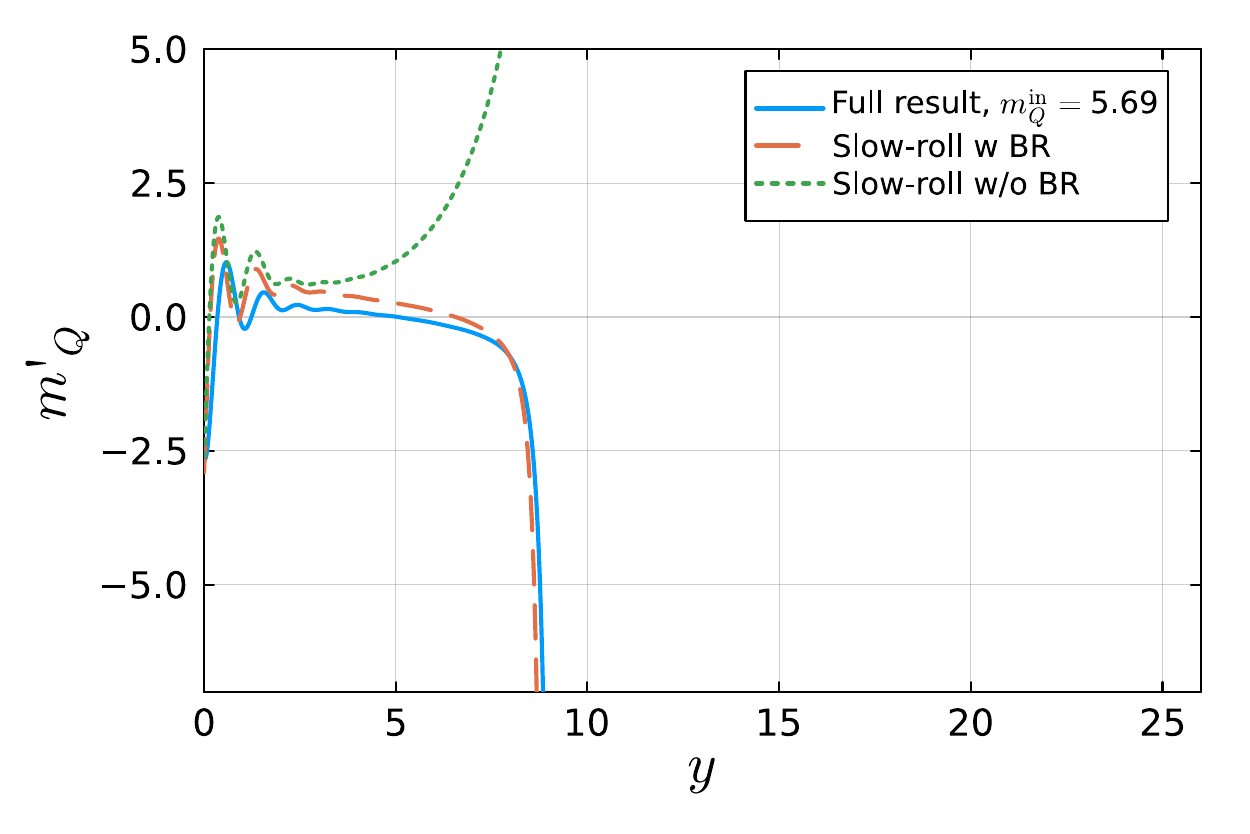}
    \includegraphics[width=7.5cm]{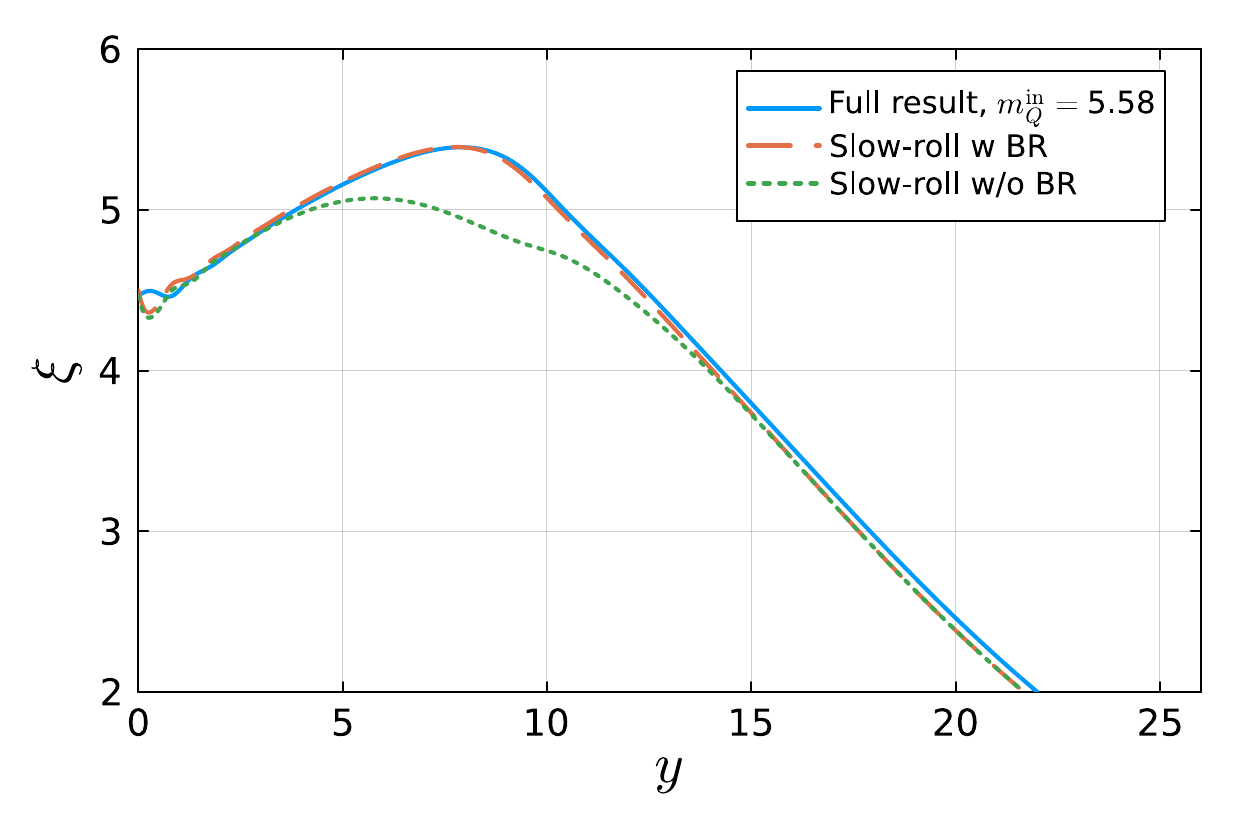}
    \includegraphics[width=7.5cm]{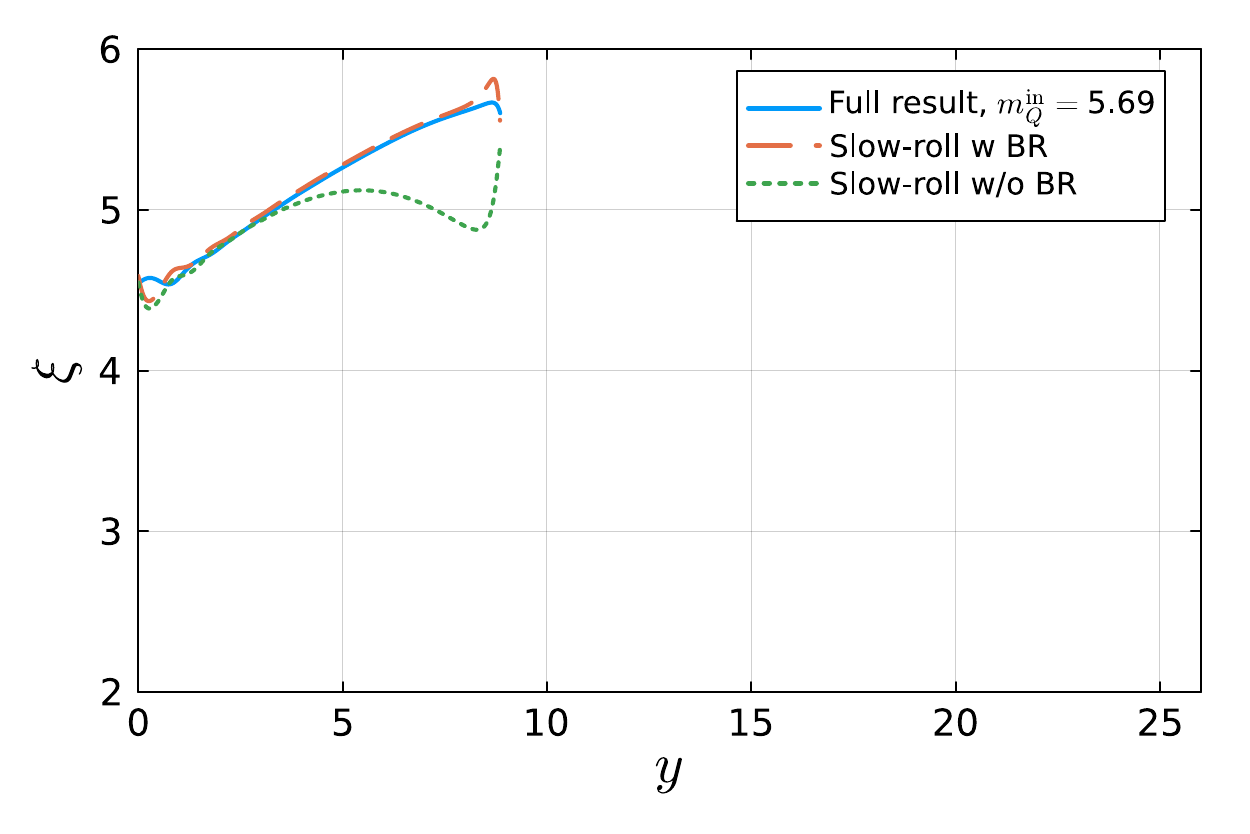}
    \includegraphics[width=7.5cm]{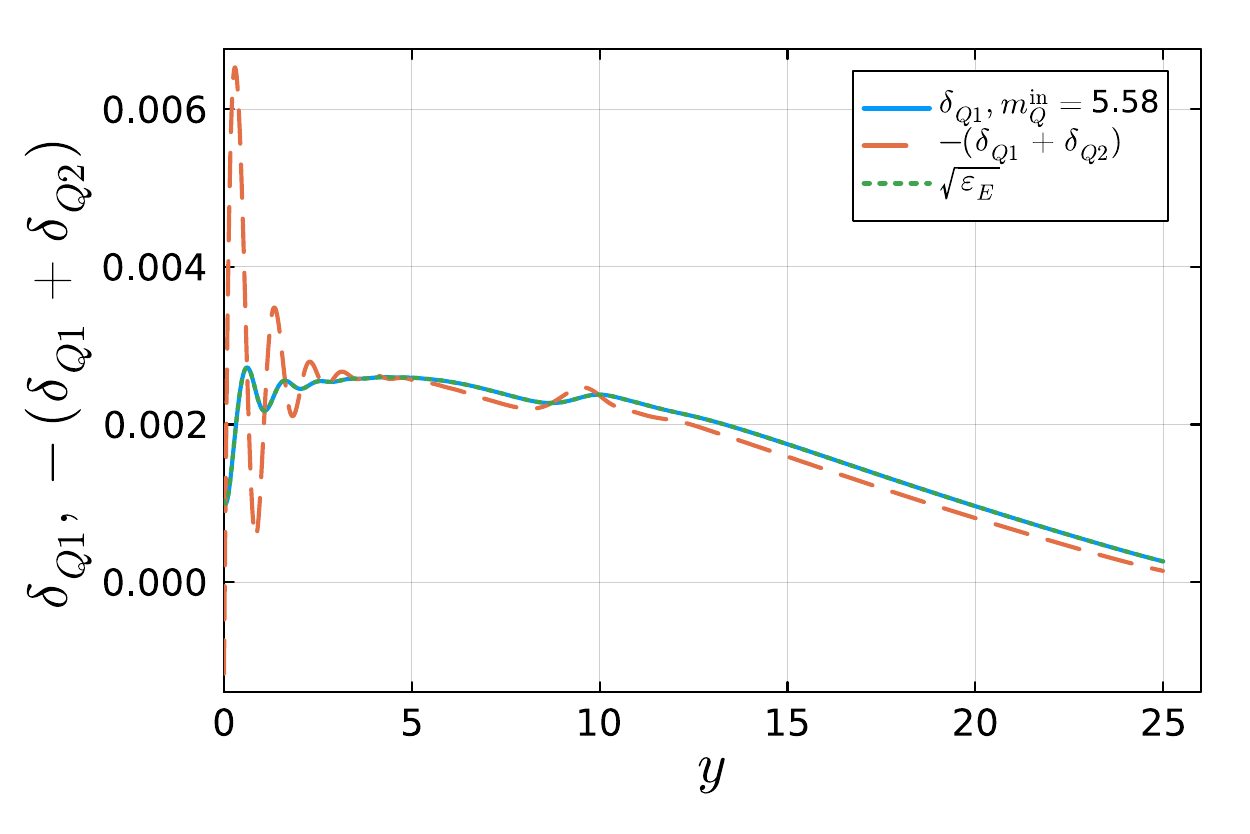}
    \includegraphics[width=7.5cm]{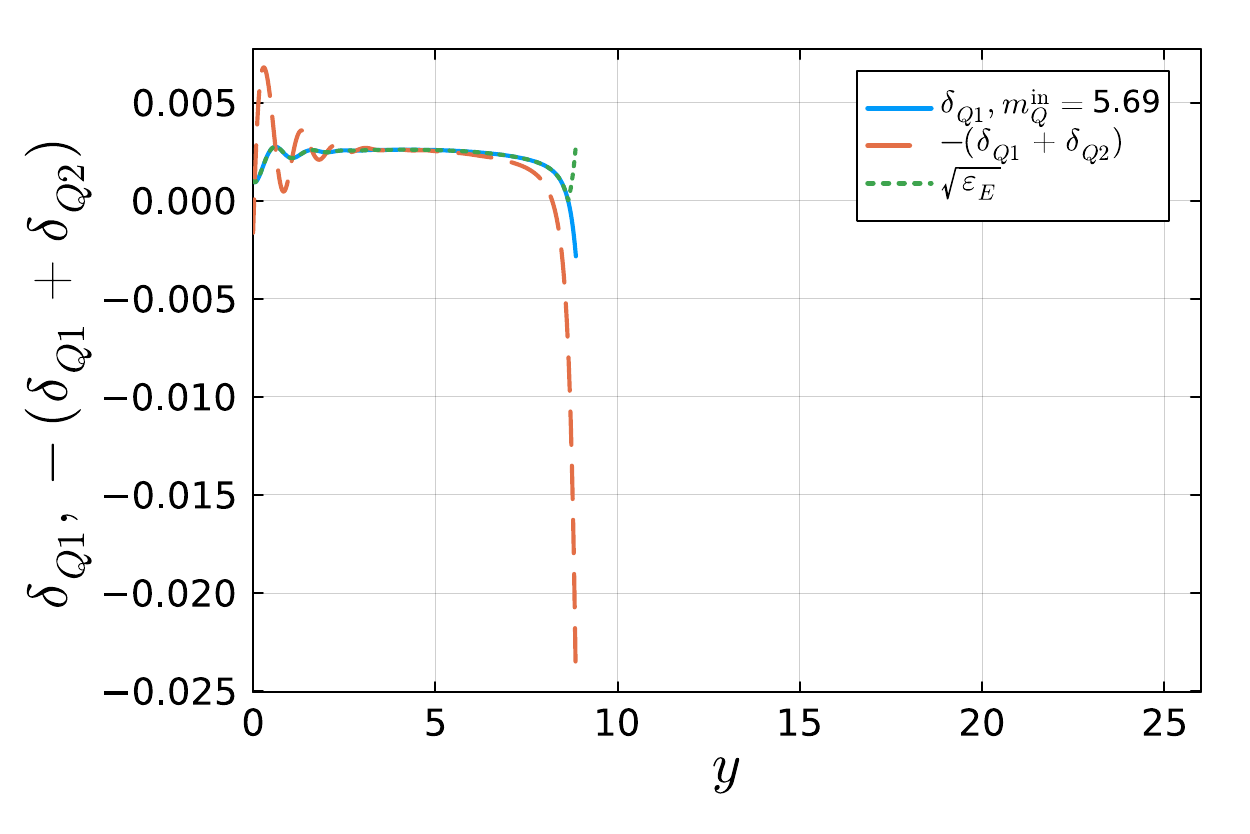}
    \caption{Same as Figure~\ref{fig:kappa001_i1_dfields}, but for
    case (b) under initial condition (I). The
      parameters are the same as in Figure~\ref{fig:kappa100_i1_BG}, but
      taking $m_Q^{\rm in}=5.58$ (left) and $5.69$ (right).}
  \label{fig:kappa100_i1_dfields}
 \end{center}
\end{figure}

\begin{figure}
  \begin{center}
    \includegraphics[width=7.5cm]{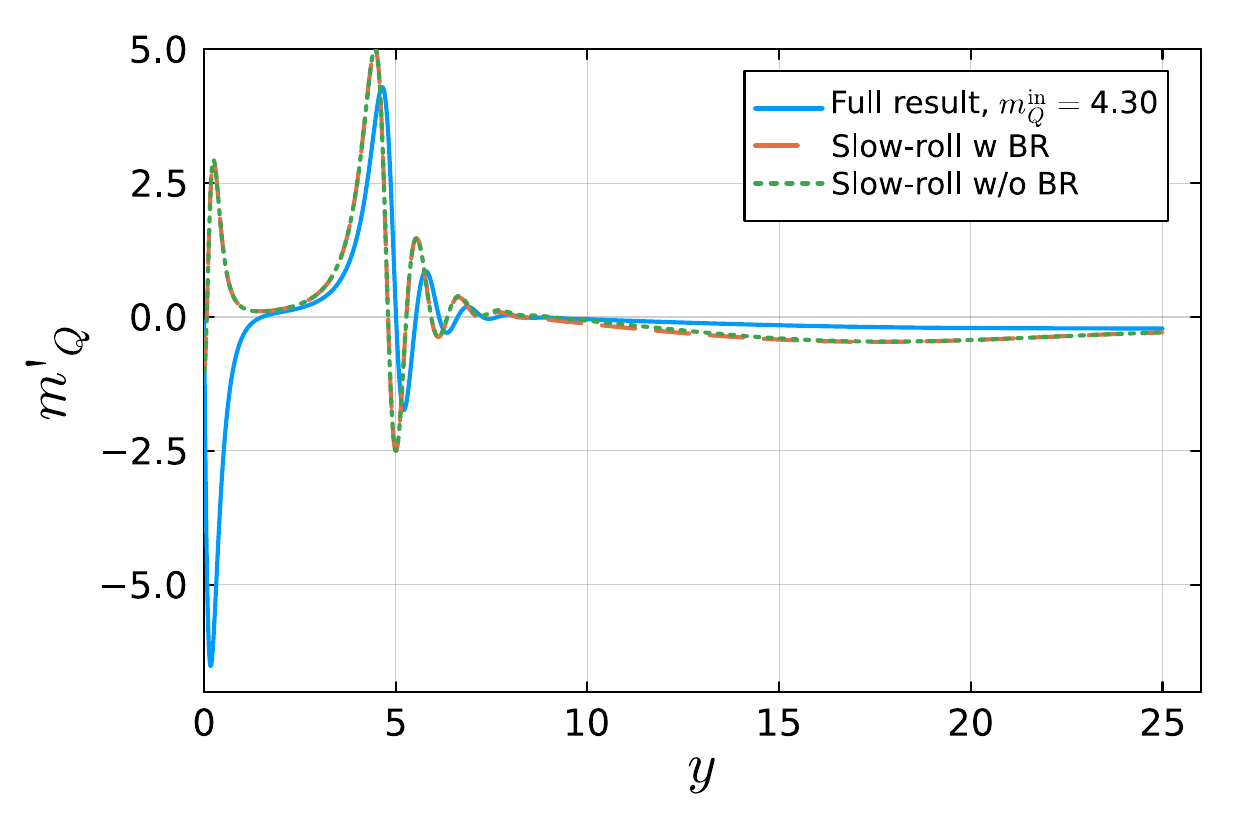}
    \includegraphics[width=7.5cm]{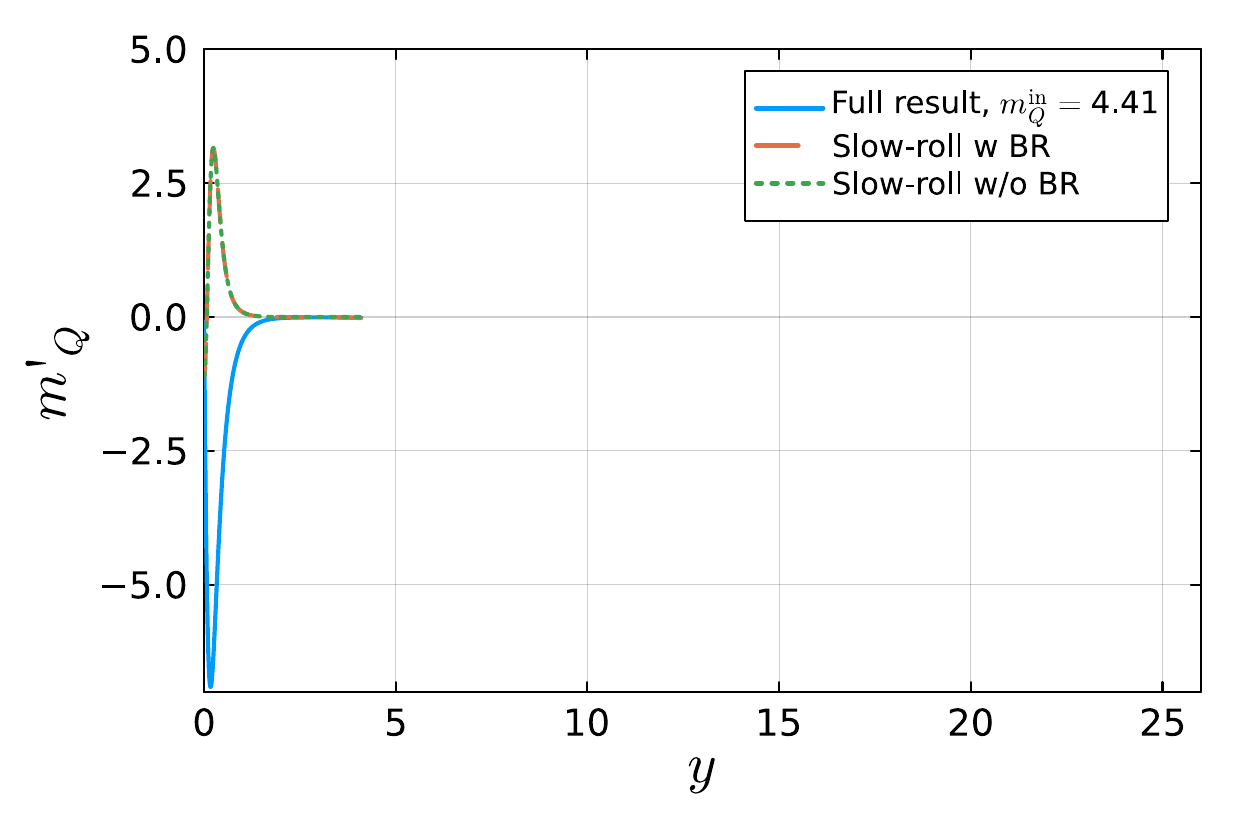}
    \includegraphics[width=7.5cm]{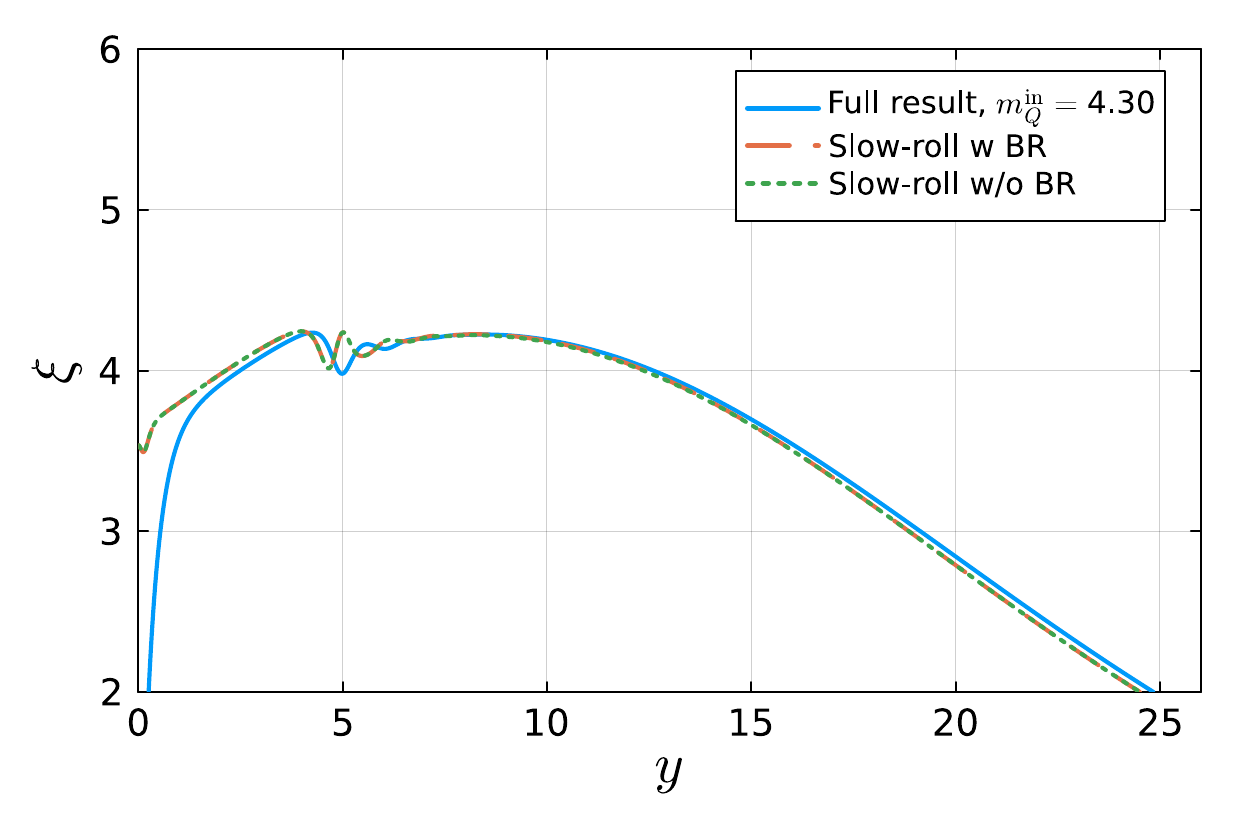}
    \includegraphics[width=7.5cm]{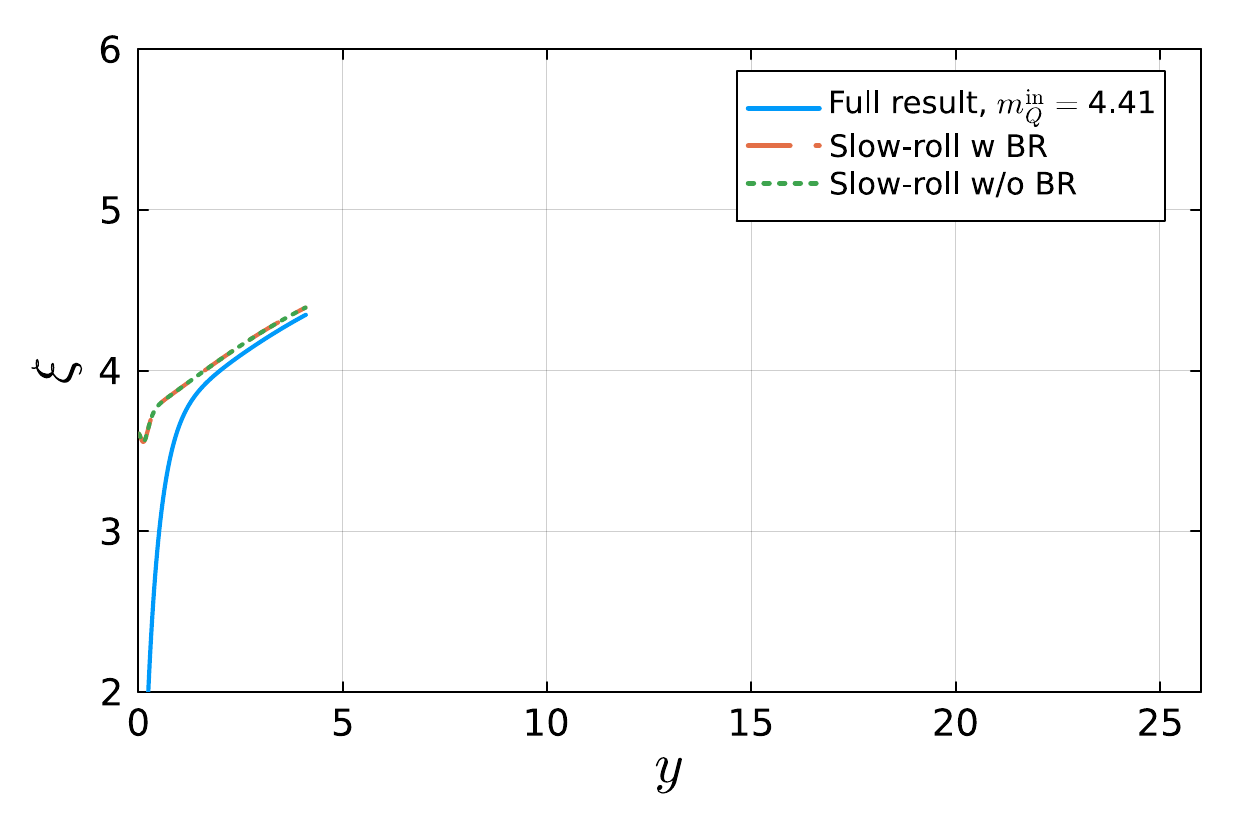}
    \includegraphics[width=7.5cm]{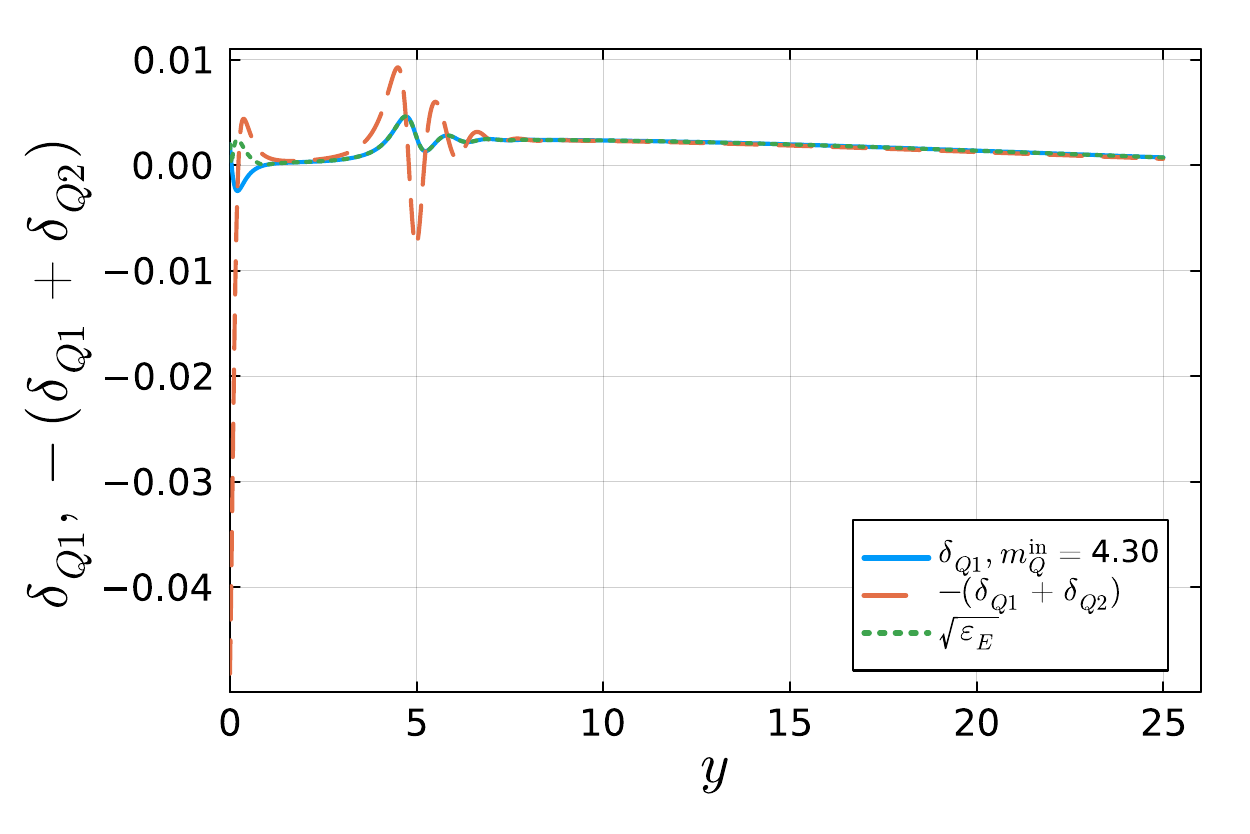}
    \includegraphics[width=7.5cm]{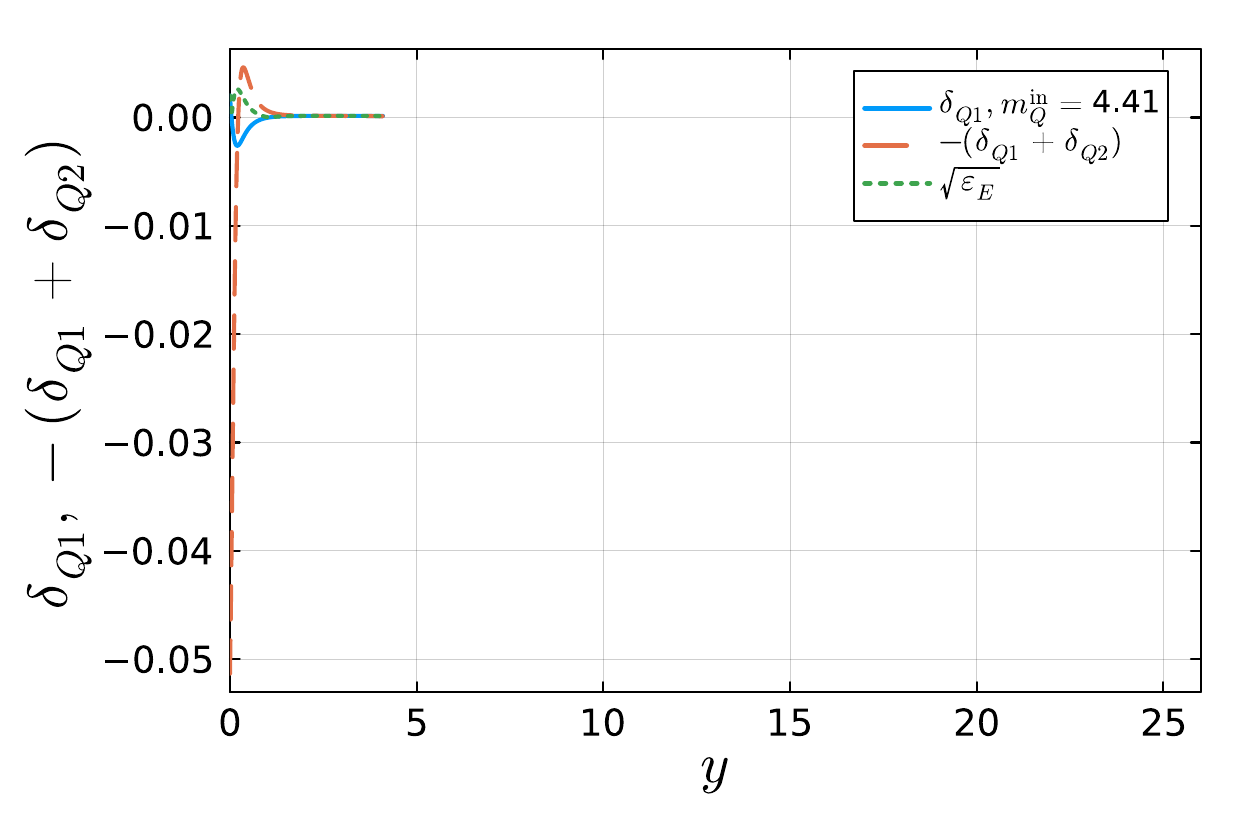}
    \caption{Same as Figure~\ref{fig:kappa100_i1_dfields}, but under
    initial condition (II). The
      parameters are the same as in Figure~\ref{fig:kappa100_i0_BG}, 
      but taking 
      $m_Q^{\rm in}=4.30$ (left) and 4.41 (right). }
  \label{fig:kappa100_i0_dfields}
 \end{center}
\end{figure}

Figures~\ref{fig:kappa100_i1_dfields} and
\ref{fig:kappa100_i0_dfields} show the results for case (b) under initial conditions (I)
and (II), respectively. We have verified that $|\dot{H}|/H^2\lesssim \order{10^{-3}}$ is satisfied.
We find that the
slow-roll results for $m'_Q$ deviate from the full result. For
instance, the result with $m_Q^{\rm
  in}=5.58$ under initial condition (I) shows that the slow-roll approximation, i.e., neglecting
$m''_Q$ and $\tilde{\chi}''$, is invalid.
On the other hand, the slow-roll approximation is relatively
accurate for $\xi$. This is because the right-hand side of the equation of motion for
$\tilde{\chi}$ is suppressed with a large $\kappa$ and
$3\tilde{\chi}'-\beta(\chi)/\lambda\simeq 0$ tends to hold.
  Regarding $\delta_{Q1}$ and
$\delta_{Q2}$, the slow-roll approximation,
$-(\delta_{Q1}+\delta_{Q2})\simeq \sqrt{\epsilon_{Q_E}}$, does not hold
in the early stage of the dynamics, i.e., the oscillatory behavior cannot be captured by the approximation. Otherwise, the approximation
roughly holds, except near the perturbative limit.

\bibliographystyle{JHEP}
\bibliography{refs.bib}

\end{document}